\documentclass[letterpaper,11pt]{JHEP3}
\usepackage{amsmath}
\usepackage{amssymb}
\usepackage{bbm}
\usepackage{subfigure}
\usepackage{epsfig}
\usepackage{yfonts}
\newcommand{\labell}[1]{\label{#1}}

\newcommand{\be}{\begin{equation}}
\newcommand{\ee}{\end{equation}}
\newcommand{\bea}{\begin{eqnarray}}
\newcommand{\eea}{\end{eqnarray}}
\newcommand{\ba}{\begin{eqnarray}}
\newcommand{\ea}{\end{eqnarray}}

\newcommand{\beq}{\begin{equation}}
\newcommand{\eeq}{\end{equation}}
\newcommand{\beqa}{\begin{eqnarray}}
\newcommand{\eeqa}{\end{eqnarray}}
\newcommand{\beqar}{\begin{eqnarray*}}
\newcommand{\eeqar}{\end{eqnarray*}}
\newcommand{\e}{\epsilon}
\newcommand{\reef}[1]{(\ref{#1})}

\newcommand{\eg}{{\it e.g.,}\ }
\newcommand{\ie}{{\it i.e.,}\ }
\newcommand{\comment}[1]{{\bf [[[#1]]]}}

\newcommand{\mt}[1]{\textrm{\tiny #1}}

\newcommand{\X}{\mathcal{X}}

\newcommand{\tL}{\tilde{L}}

\newcommand{\lp}{\ell_{\mt P}}

\newcommand{\fin}{f_\infty}

\newcommand{\hi}{{\hat \imath}}

\newcommand{\rh}{\rho}

\def\ov{\over}
\def\dr{\dot{r}}
\def\drh{\dot{\rho}}

\def\ri{\right}
\newcommand\te{t_\mt{E}}
\newcommand\rmin{\rho_\mt{min}}
\newcommand{\ads}{a_d^*}

\preprint{arXiv:1206.5225 [hep-th]}

\title{Entanglement Entropy for Singular Surfaces}

\author{Robert C. Myers$^{a}$ and Ajay Singh$^{a,b}$ \\
$^a$ {\it Perimeter Institute for Theoretical Physics, Waterloo,
Ontario N2L 2Y5, Canada}\\
$^b$ {\it Department of Physics \& Astronomy and Guelph-Waterloo Physics Institute,}\\
{\it \ \ University of Waterloo, Waterloo, Ontario N2L 3G1, Canada} \\
}
\abstract{We study entanglement entropy for regions with a singular
boundary in higher dimensions using the AdS/CFT correspondence
and find that various singularities make new universal contributions. When the
boundary CFT has an even spacetime dimension, we find that the entanglement
entropy of a conical surface contains a term quadratic in the logarithm
of the UV cut-off. In four dimensions, the coefficient of this contribution
is proportional to the central charge $c$. A conical
singularity in an odd number of spacetime dimensions contributes a term
proportional to the logarithm of the UV cut-off. We also study the entanglement
entropy for various boundary surfaces with extended singularities. In
these cases, similar universal terms may appear depending on the dimension and
curvature of the singular locus.}

\begin{document}

%%%%%%%%%%%%%%%%%%%%%%%%%%%%%%%%%%%%%%%%%%%%
\section{Introduction} \label{intro}

Entanglement entropy has emerged as an interesting theoretical quantity
for studies of quantum matter. For example, it allows one to
distinguish new topological phases or different critical points
\cite{wenx,cardy0,fradkin}. Entanglement entropy has also been
considered in discussions of holographic descriptions of quantum
gravity, in particular, for the AdS/CFT correspondence
\cite{ryu1,ryu2,ryu3}. In this context, as well as characterizing new
properties of holographic field theories \cite{igor0}, it has been
suggested that entanglement entropy may play a fundamental role in the
quantum structure of spacetime, \eg \cite{mvr,vm}.

In quantum field theory (QFT), the typical calculation of entanglement
entropy begins by choosing a particular spatial region $V$ and
integrating out the degrees of freedom in the complement $\bar V$. Then
with the resulting density matrix $\rho_\mt{V}$, one calculates:
$S_\mt{EE} =-{\rm Tr} \left( \rho_\mt{V} \,\log\rho_\mt{V}\right)$.
Unfortunately, this calculation generically yields a UV divergent
answer because the result is dominated short distance correlations in
the vicinity of the boundary $\Sigma=\partial V$. However, if we
regulate the calculation by introducing a short distance cut-off
$\delta$, the entanglement entropy exhibits an interesting geometric
structure \cite{tarun,liu}. For example, with a QFT in $d$ spacetime
dimensions, this allows us to organize the results as follows:
 \be
 S_\mt{EE}=\frac{c_2}{\delta^{d-2}}+\frac{c_4}{\delta^{d-4}}+\cdots\,,
 \labell{diverg0}
 \ee
where each of the coefficients $c_i$ involves an integration over the
boundary $\Sigma$. In particular, the first two coefficients may be
written as
 \bea
 c_2&=&\int_\Sigma d^{d-2}y\, \sqrt{h}\,h_2 = h_2\ {\cal
 A}_\Sigma\,,
 \labell{coeff0}\\
c_4&=&\int_\Sigma d^{d-2}y\,\sqrt{h} \left[h_{4,1}\, {\cal R}+ h_{4,2}\,
K^\hi_a{}^a K^\hi_b{}^{\,b}\right]\,.
 \labell{coeff2}
 \eea
Of course, the leading term then yields the famous `area law' result
\cite{bomb,mark}. The geometry of the boundary becomes even more
evident in the second coefficient $c_4$ with the appearance of $\cal R$
and $K^\hi_{ab}$, the intrinsic Ricci scalar and the extrinsic
curvature of this surface.\footnote{We are assuming here that the
background geometry is simply $d$-dimensional flat space. Otherwise
additional contributions could appear in $c_4$ involving the background
curvature.} The coefficients $h_{k,a}$ above will depend on the
detailed structure of the underlying QFT. In particular, we note that
they may become functions of the cut-off through their dependence on
mass scales $\mu_i$ in the QFT since the latter will only appear in the
dimensionless combination $\mu_i\delta$, \ie $h_{k,a} =
h_{k,a}(\mu_i\delta)$ \cite{liu}. The geometric character of the
entanglement entropy illustrated here naturally follows from the
implicit choice of a covariant regulator in the QFT and the fact that
the UV divergences are local.

Unfortunately, the coefficients appearing in the expansion above are
scheme dependent. Clearly, if we shift $\delta \to \alpha\delta$, we
find $h_{k,a} \to \hat{h}_{k,a}=\alpha^{k-d}h_{k,a}(\alpha
\mu_i\delta)$. Hence the regulator dependence here comes both from the
implicit dependence on mass scales in the QFT and the `classical'
engineering dimension of the individual coefficients. Of course, the
latter can be avoided by carrying the expansion in eq.~\reef{diverg0}
to sufficiently high orders. In particular, one may find a logarithmic
contribution to the entanglement entropy
 \be
 S_\mt{univ}= c_{d}\,\log(\delta/L)
 \labell{universal}
 \ee
where $L$ is some (macroscopic) scale in the geometry of $\Sigma$. At
this order in the expansion, the coefficient $c_d$ is dimensionless.
Further it may be natural to eliminate any intrinsic scales in the QFT
by focussing on fixed point theories of the RG flow.\footnote{Universal
terms may also appear in other circumstances, either as a finite
contribution, \eg \cite{wenx,fradkin} or even when explicit mass scales
are present \cite{frank,janet}.} In this case, shifting $\delta$ as
above makes a finite shift in the entanglement entropy but $c_{d}$ is
left unchanged. Hence one expects that this coefficient contains
universal information that characterizes the underlying CFT. For
example, in four dimensions, this universal coefficient is simply given
by \cite{solodukhin}\footnote{Again, we are assuming that the
background geometry is flat.}
 \be
c_{d} = \frac{1}{2\pi} \int_{\Sigma} d^{2}y\sqrt{h}\,\left[\, c\,
\left( K^\hi_a{}^b K^\hi_b{}^a-\frac12 K^\hi_a{}^a
K^\hi_b{}^{\,b}\right) +\,a\, {\cal R}\,\right]\,,
 \labell{solo}
 \ee
where $a$ and $c$ are the two central charges of the CFT. In principle
then, this provides an approach to determine these central charges. For
example, calculating the entanglement entropy for a sphere yields
$c_{d}\propto a$, while only $c$ appears for a cylinder
\cite{solodukhin}.

The preceding discussion implicitly assumes that the boundary of the
region in question is smooth. The purpose of this paper is to explore
the effects of geometric singularities in this `entangling surface'
$\Sigma$. In part, our motivation comes from the observation that in
three dimensions, if the boundary contains corners or kinks, the
corresponding entanglement entropy will contain additional logarithmic
contributions \cite{log1,log3,hirata,log2}
 \be
 S_\mt{kink}= q_3(\Delta\theta)\,\log(\delta/L)\,,
 \labell{bend}
 \ee
where $\Delta\theta$ is the opening angle of the kink -- see figure
\ref{sing}. As a function of $\Delta\theta$, the coefficient
$q_3(\Delta\theta)$ satisfies certain simple properties
\cite{log1,log3,hirata}. In particular, $q_3(\Delta\theta=\pi)=0$ since
the $\Sigma$ becomes a smooth surface when the angle is set to $\pi$.
Strong subadditivity can be used to argue that in general
$q_3(\Delta\theta)$ must satisfy certain inequalities, \eg
$q_3(\Delta\theta)\ge 0$ and $q_3'(\Delta\theta)\leq 0$
\cite{log3,hirata}. If the QFT is in a pure state, we have
$S_\mt{EE}(V) = S_\mt{EE}(\bar{V})$ and so $q_3(\Delta\theta) =
q_3(2\pi-\Delta\theta)$ in this case. Further, in examples
\cite{log1,log3,log2,hirata}, one finds for a small opening angle:
$q_3(\Delta\theta\to0)\propto 1/\Delta\theta$. However, we must add
that the precise universal information contained in $q_3(\Delta\theta)$
remains to be understood.

A natural question is to ask whether similar contributions arise for
singular entangling surfaces in higher dimensions. If yes, we can ask
what the geometric dependence of these new terms is. Further, if we
focus on CFT's, we might ask if the coefficients in these contributions
have a simple dependence on the central charges, analogous to that in
eq.~\reef{solo}. The AdS/CFT correspondence \cite{adscft} provides a
simple framework with which we may begin to address these questions. In
fact, using the standard calculation of holographic entanglement
entropy \cite{ryu1,ryu2}, the logarithmic contribution \reef{bend}
associated with a kink in three dimensions has already been identified
in \cite{hirata}. More generally, this approach allows us to easily
study boundary CFT's in a variety of dimensions and further the
geometric structure of the entanglement entropy becomes readily evident
in holographic calculations \cite{janet,ent1}. In the following then,
we use holography to study some simple singular entangling surfaces in
higher dimensions and we find a variety of new universal contributions.
While these are just first steps towards a full understanding of these
universal terms, our results indicate a rich geometric structure for
the entanglement entropy of singular surfaces.

The remainder of this paper is organized as follows: In the next
section, we give a brief overview of our calculations and summarize our
main results. In section \ref{singular}, we consider entangling
surfaces with a conical singularity for boundary CFT's with $d=4$, 5
and 6. In these cases, the singularity in the geometry of the
entangling surface is confined to a single point and so we broaden our
calculations to consider extended singularities in section
\ref{crease}. There we find that the appearance of universal terms in
the entanglement entropy depends on the dimension and the curvature of
the singular locus. Section \ref{central1} presents calculations of
holographic entanglement entropy for singular surfaces in boundary CFTs
which are dual to the Gauss-Bonnet gravity. These calculations allow us
to examine the dependence of the universal terms on the central charges
of the underlying CFT. In section \ref{dis}, we briefly discuss our
results and consider future directions. Appendix \ref{conformal}
describes an alternate calculation of the entanglement entropy
associated with a conical singularity. In particular, we make a
conformal transformation from $R^d$ to $R\times S^{d-1}$ and so the
conical entangling surface becomes a cylinder in the latter background.
In appendix \ref{appa}, we give certain details for lengthy
calculations presented in sections \ref{singular}, \ref{crease} and
\ref{central1}.

While we were in the final stages of preparing this paper,
ref.~\cite{new} appeared which contains some results which overlap with
ours. Particularly, authors have calculated the holographic
entanglement entropy for a cone and a crease for a four-dimensional
CFT. We also learned of an upcoming paper \cite{ben} where the
entanglement entropy for a cone in a six-dimensional CFT is studied.

%%%%%%%%%%%%%%%%%%%%%%%%%%%%%%%%%%%%%%%%%%%%%%%%%%%%%%%%
\section{Singular entangling surfaces and summary of results}
\label{results}

In the sections \ref{singular}, \ref{crease} and \ref{central1}, we
will describe in detail various holographic calculations of the
entanglement entropy for certain singular surfaces. Each of these
calculations is quite lengthy and individually they are not very
enlightening. Hence in this section, we provide an overview of these
calculations and a summary of our results. We begin by describing the
kinds of singular entangling surfaces which we will consider.

Let us go back to eq.~\reef{bend} for three dimensions. In this case,
the entangling surface is a one-dimensional curve and the `singular
surface' would be one containing a kink or a corner where the direction
of the tangent vector changes discontinuously at a point. We can
characterize this behaviour by saying that the geodesic curvature of
the curve contains a $\delta$-function singularity. In higher
dimensions, the entangling surface is a ($d-2$)-dimensional submanifold
which divides into two a time slice of the $d$-dimensional background
spacetime. In this case, the natural extension of geodesic curvature is
the extrinsic curvature of these surfaces. However, a distinct feature
characterizing the geometry of these higher dimensional surfaces is now
their intrinsic curvature. Of course, for a fixed background, these two
curvatures will be related to each other (and the background curvature)
through the Gauss-Codazzi equations. However, it is worth noting that
in discussing singular surfaces, we might consider singularities in
either the extrinsic curvature or the intrinsic curvature. In
particular, as we show with simple examples below, it is possible to
construct surfaces where the intrinsic curvature is everywhere smooth
while the extrinsic curvature is singular. The other possibility which
we consider is when both the extrinsic and intrinsic curvatures have
singularities.\footnote{We will assume the geometry of the background
is everywhere smooth in the following. Combined with the Gauss-Codazzi
equations, this assumption rules out the possibility that the extrinsic
curvature may be smooth while the intrinsic curvature is singular.} In
either case, the examples which we consider below contain
$\delta$-function singularities. That is, in all of our examples, the
curvatures characterizing the entangling surface are finite and smooth
everywhere, except for a particular locus or subset of points.

\FIGURE[!ht]{
 \begin{tabular}{ccc}
\includegraphics[width=0.42\textwidth]{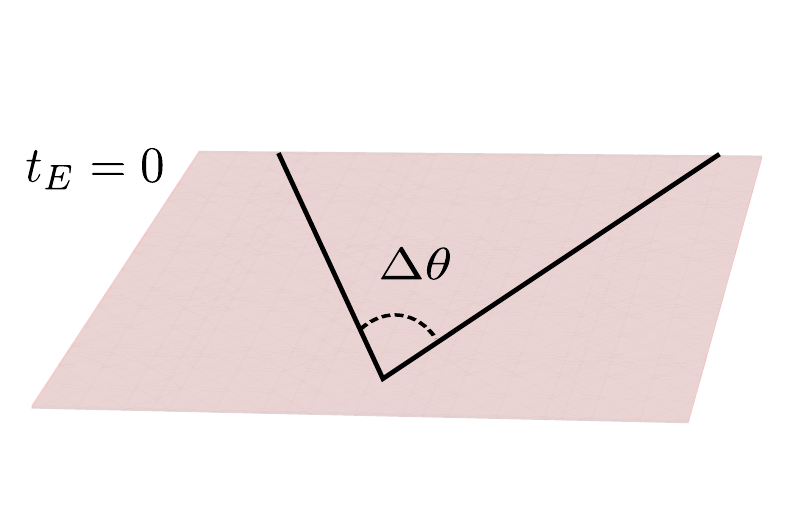}&
\includegraphics[width=0.32\textwidth]{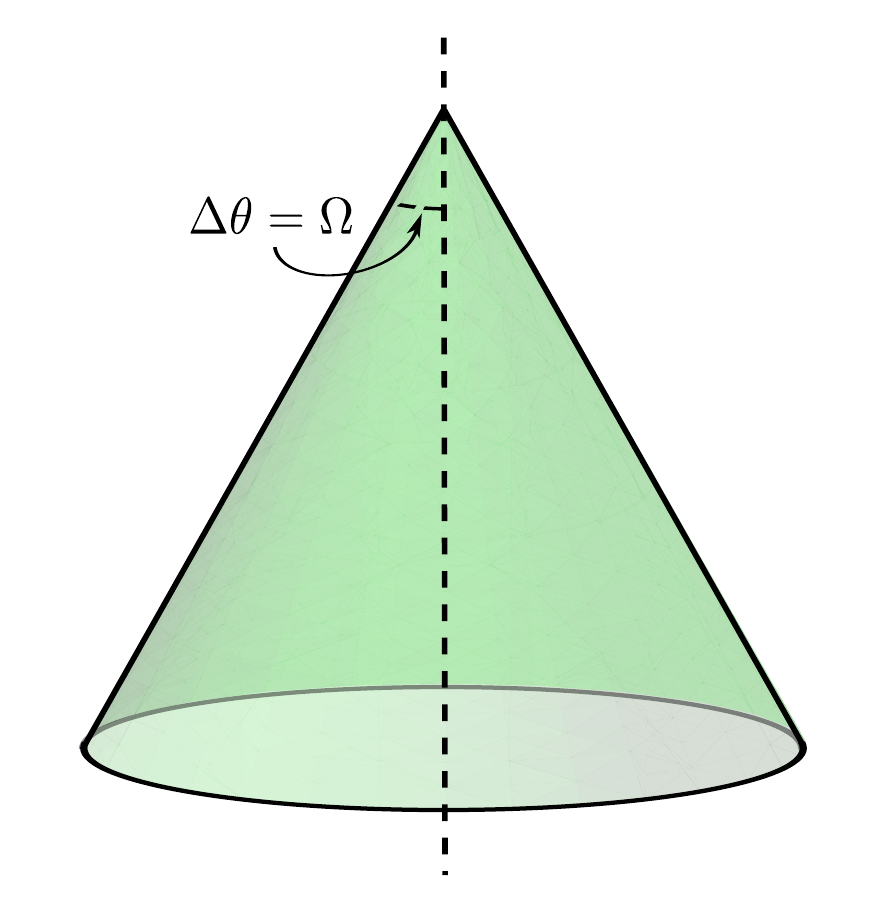}&
\includegraphics[width=0.22\textwidth]{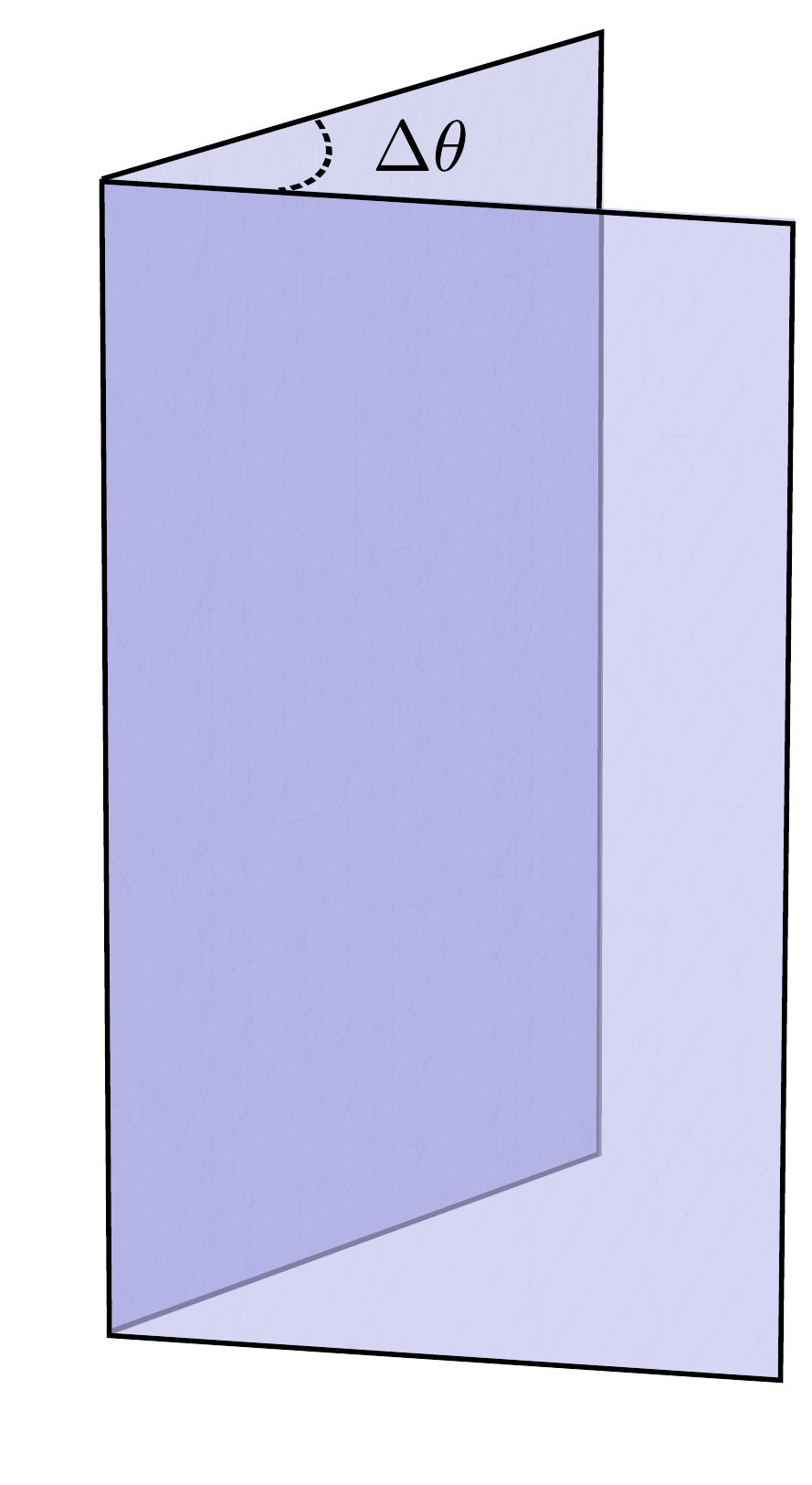}\\
(a) & (b) &(c)
\end{tabular}
\caption{(Colour Online) Panel (a) shows a kink in constant Euclidean
time $t_E$ slice in $d=3$. Panel (b) shows the cone $c_1$. Panel (c)
shows the crease $k\times R^1$.}
 \label{sing}}
In considering such singular entangling surfaces, we introduce an
intuitive nomenclature to simplify the discussion: {\it kink}, {\it
cone} and {\it crease}. Examples of each are illustrated in figure
\ref{sing}. To explain these terms, it suffices to consider evaluating
the entanglement entropy in flat $d$-dimensional background $R^d$. With
Euclidean signature, the metric can then be written in `cylindrical'
coordinates as
 \be
ds^2  \,=\,  d\te^2+d\rho^2+\rho^2 \left(d\theta^2 + \sin^2\theta\,
d\Omega^2_n\right) + \sum_{i=1}^m \left(dx^i\right)^2
  \labell{met0}
 \ee
where $d\Omega^2_n$ denotes the round metric on a unit $n$-sphere.
Hence we have $d=3+n+m$. The entangling surface will be a
($d-2$)-dimensional geometry embedded in this background (on a surface
of constant $\te$).

The natural model of a {\it kink} is then given in $d=3$ ($n=0=m$) as
the union of two rays: $k=\lbrace \te=0, \rho =[0,\infty),
\theta=\pm\Omega\rbrace$. With this choice of angles, the opening angle
between the two rays is $\Delta\theta=2\Omega$. Similarly, our
prototype for a {\it cone} is given in $d=3+n$ as the surface: $c_n=
\lbrace \te=0, \rho =[0,\infty), \theta =\Omega \rbrace$. Hence with
this construction, we are only considering cones with a spherical
cross-section $S^n$. We define a {\it crease} to be the higher
dimensional extension of either of these singular surfaces where we
take the direct product of a kink $k$ or a cone $c_n$ with some other
manifold. With the flat background \reef{met0} above, the natural
extension of the previous constructions gives the crease $k\times R^m$
in $d=3+m$ or a conical crease $c_n\times R^m$ in $d=3+n+m$.

In a certain sense the kink is simply a lower dimensional version of a
cone, \ie $k=c_0$. However, we prefer to distinguish these two classes
to highlight the difference in the singularities noted above. If we
consider the crease $k\times R^m$, the intrinsic geometry of this
submanifold is smooth everywhere including $\rho=0$ and in fact for the
construction above, the intrinsic curvature vanishes everywhere. On the
other hand, the extrinsic curvature contains a $\delta$-function
singularity at the tip of the kink, \ie $\rho=0$. Again for the above
construction, the extrinsic curvature vanishes everywhere away from
this singularity. Now considering the conical crease $c_n\times R^m$,
one finds that both the intrinsic and extrinsic curvatures have
$\delta$-function singularities at the tip of the cone, \ie $\rho=0$.
For the prototype constructed above, the intrinsic curvature vanishes
away from this singular point but the extrinsic curvature is
nonvanishing.

In our calculations of the entanglement entropy, it will also be useful
to consider the CFT in a background of the form $R^{3+n}\times S^m$,
for which we write the metric as
 \be
ds^2  \,=\,  d\te^2+d\rho^2+\rho^2 \left(d\theta^2 + \sin^2\theta
d\Omega^2_n\right) + R^2\,d\Omega_m^2\,.
  \labell{met0x}
 \ee
This allows us to consider creases of the form $k\times S^m$ or
$c_n\times S^m$. With these geometries, the intrinsic curvature of the
entangling surface is nonvanishing everywhere but in particular, the
geometry of the singular locus is now $S^m$.

Now after this description of our singular surfaces, we briefly explain
our holographic calculations and the results. We use the AdS/CFT
correspondence to calculate the holographic entanglement entropy for a
boundary CFT with the above singular entangling surfaces. In different
calculations, we use the following bulk geometries:
 \bea
ds^2 & \,=\, & {L^2 \ov z^2}\left(dz^2+d\te^2+d\rho^2+\rho^2
\left(d\theta^2 + \sin^2\theta d\Omega^2_n\right) + \sum_{i=1}^m
\left(dx^i\right)^2   \ri)\,,
 \labell{metric1} \\
ds^2 & \,=\, & {L^2 \ov z^2}\Big(dz^2+
f_1(z)\left[d\te^2+d\rho^2+\rho^2 \left(d\theta^2 + \sin^2\theta
d\Omega^2_n\right)\right] + f_2(z)\,R^2\,d\Omega_m^2  \Big) \,.
\labell{curvemet}
 \eea
Here $L$ is the AdS curvature scale and $z$ is the radial coordinate
with the asymptotic boundary at $z=0$. Of course, eq.~\reef{metric1} is
just AdS$_{d+1}$ where $d=3+n+m$ as above and with a flat boundary
metric written in the coordinates given in eq.~\reef{met0}. The
geometry in eq.~\reef{curvemet} approaches AdS$_{d+1}$ asymptotically
with the curved boundary metric given in eq.~\reef{met0x}. However,
because this boundary geometry is not conformally flat, the bulk
geometry is not purely AdS$_{d+1}$ and the functions, $f_1$ and $f_2$,
must be determined by evaluating the gravitational equations of motion.
With these bulk metrics, \reef{metric1} and \reef{curvemet}, we can
readily calculate the holographic entanglement entropy for the kink,
cone or crease geometries described above.

Recall the proposal for calculating holographic entanglement entropy
\cite{ryu1,ryu2}: Having selected a particular entangling surface
$\Sigma$ in the $d$-dimensional boundary theory, we must find the
minimal area surface $m$ (of dimension $d-1$) which extends in the bulk
geometry such that the boundary of $m$ matches $\Sigma$ at the
asymptotic AdS$_{d+1}$ boundary. The entanglement entropy is then given
by\footnote{All of our calculations use Euclidean signature for which
the {\it minimal} area surface is appropriate. If the calculation is
done in a Minkowski signature background, one seeks to extremize the
area but this surface will only be a saddle point. In either case, the
area must be suitably regulated to produce a finite answer.}
 \be
S_\mt{EE} \,=\, {2\pi \ov \lp^{d-1}} \,  \underset{\partial m \sim
\Sigma}{\text{min}}[A(m)]\,.
 \labell{c0x1}
 \ee
This conjecture passes a variety of consistency tests, \eg see
\cite{ryu2,ryu3,head1,ent1} and in fact, for spherical entangling
surfaces, there is a derivation of holographic entanglement entropy
which yields this result \reef{c0x1} \cite{casini9}. An implicit
assumption in eq.~\reef{c0x1} is that the bulk physics is described by
(classical) Einstein gravity. However, this proposal was extended
\cite{ent1,ent2} to certain dual gravity theories with higher
derivative interactions, namely, Lovelock theories \cite{lovel}. In
this extension, which we use in section \ref{central1}, one finds the
surface which minimizes a new functional of the geometry of $m$. In
particular, the latter was chosen to match an expression for the black
hole entropy in the Lovelock theories \cite{jacobson}.

The details of our holographic calculations of the entanglement entropy
for kinks, cones and creases are given in the following sections. As
expected, these calculations contain a variety of divergences that must
be regulated. However, we pay particular attention to the question of
whether there are new singular contributions associated with geometric
singularity in the entangling surface. Further, in certain cases, we
find that the singularity produces new universal contributions. Table
\ref{table2} summarizes our results. %rcm
 \begin{table}
 %\begin{center}
\centerline{
 \begin{tabular}{|c|c|c|c|c|c|c|}
 \hline
$d$  & Background &  Geometry of & Crease  & Crease & Expected & New \\
 & spacetime & entangling surface & dimension & curvature & Divergences &Divergences\\
\hline
3 & $R^3$ & $k$   & 0 & -- & ${1/\delta}$ & $\log(\delta)^{\,*}$ \\
\hline
4 & $R^4$ & $c_1$   & 0 & -- & ${1/\delta^2}\,,\;\log(\delta)$ &
$\log^2(\delta)^{\,*}$\\
\hline
5 & $R^5$ & $c_2$   & 0 & -- & ${1/\delta^3}\,,\;{1/\delta}$ &
$\log(\delta)^{\,*}$\\
\hline
6 & $R^6$ & $c_3$   & 0 & -- & ${1/ \delta^4}\,,\;{1/\delta^2}\,,\; \log(\delta)$ &
$\log^2(\delta)^{\,*}$\\
\hline
&&&&&\\[-1.3em]\hline
$>$3 & $R^d$ & $k\times R^{d-3}$ & $d-3$ & flat & ${1/ \delta^{d-2}}$ &
${1/ \delta^{d-3}}$\\
\hline
4 & $R^3\times S^1$ & $k\times S^1$ & 1 & flat & ${1/ \delta^{2}}$ &
${1/\delta}$\\
\hline
5 & $R^3\times S^2$ & $k\times S^2$ & 2 & curved & ${1/ \delta^3}\,,\;{1/ \delta}$
&${1/ \delta^2}\,,\;\log(\delta)^{\,*}$\\
\hline
6 & $R^3\times S^3$ & $k\times S^3$ & 3 & curved & ${1/ \delta^4}\,,\;{1/
 \delta^2}$ & ${1/ \delta^3}\,,\;{1/\delta}$\\
\hline
6 & $R^4\times S^2$ & $k\times (R^1 \times S^2)$ & 3 & curved & ${1/ \delta^4}\,,\;
{1/ \delta^2}$ & ${1/ \delta^3}\,,\;{1/ \delta}$\\
\hline
&&&&&\\[-1.3em]\hline
5 & $R^5$ & $c_1\times R^1$ & 1 & flat & ${1/ \delta^3}\,,\; {1/ \delta}$ &
${\log(\delta)/\delta}$\\
\hline
6 & $R^6$ & $c_1\times R^2$ & 2 & flat & ${1/ \delta^4}\,,\; {1/ \delta^2}\,,\;
 \log(\delta)^{\,*}$ &${\log(\delta)/ \delta}$\\
\hline
7 & $R^7$ & $c_1\times R^3$ & 3 & flat & ${1/ \delta^5}\,,\;  {1/ \delta^3}\,,\;
{1/ \delta}$ &$ {\log(\delta)/\delta^3}$\\
\hline
5 & $R^4\times S^1$ & $c_1\times S^1$ & 1 & flat & ${1/ \delta^3}\,,\; {1/ \delta}$
&${\log(\delta)/ \delta}$\\
\hline
6 & $R^4\times S^2$ & $c_1\times S^2$ & 2 & curved & ${1/\delta^4}\,,\;{1/\delta^2}\,,\;\log\delta$ &
$\log^2(\delta)^{\,*}$\\
\hline
7 & $R^4\times S^3$ & $c_1\times S^3$ & 3 & curved & ${1/ \delta^5}\,,\;
 {1/ \delta^3}\,,\; {1/ \delta}$ &${\log(\delta)/ \delta^3}$\\
\hline
&&&&&\\[-1.3em]\hline
6 & $R^6$ & $c_2\times R^1$ & 1 & flat & ${1/ \delta^4}\,,\;{1/ \delta^2}\,,\;\log(\delta)^{\,*}$
&${1/ \delta}$\\
\hline
7 & $R^7$ & $c_2\times R^2$ & 2 & flat & ${1/ \delta^5}\,,\;{1/ \delta^3}\,,\;{1/ \delta}$
&${1/ \delta^2}$ \\
\hline
6 & $R^5\times S^1$ & $c_2\times S^1$ & 1 & flat & ${1/ \delta^4}\,,\;{1/ \delta^2}\,,\;
\log(\delta)^{\,*}$ &${1/ \delta}$\\
\hline
7 & $R^5\times S^2$ & $c_2\times S^2$ & 2 & curved & ${1/ \delta^5}\,,\;{1/ \delta^3}\,,\;
{1/ \delta}$ &${1/ \delta^2}\,,\;\log(\delta)^{\,*}$\\
\hline
\end{tabular} }
%\end{center}
\caption{Summary of the divergent terms for various singular surfaces
from our holographic calculations. Here $d$ is the spacetime dimension
of the CFT background, which can be both flat or curved. The `expected'
divergences are those which can arise with a smooth entangling surface
-- see discussion in the introduction. The `new' divergences are
produced by the singularity in the surface and vanish when the surface
is smooth, \ie $\Omega=\pi/2$. Any universal terms are marked with a
`$*$' -- see the discussion in the main text. \labell{table2}}
\end{table}

As described above, if the entangling surface has a kink in $d=3$,
there is a new universal contribution \reef{bend} to the entanglement
entropy. We review the holographic calculation of \cite{hirata} which
reveals this result in section \ref{singular}. These calculations are
extended to creases of the form $k\times R^m$ in section \ref{crease}
and there we find no universal contribution, \ie no $\log\delta$ term,
for these cases. However, we note that in this case the locus of the
singularity, \ie the tip of the crease, is flat. The results become
more interesting if this locus is curved. If the singular locus is
curved but odd dimensional, we again find no logarithmic contribution.
However, if the singular locus is both curved and even dimensional, we
find that the entanglement entropy of the crease contains a
$\log\delta$ contribution. While we have a limited number of explicit
examples of this behaviour, \ie the case of $k\times S^2$, it suggests
to us that generally for creases of the form $k\times \Sigma_{2m}$,
there are new universal terms of the form
 \be
S_\mt{univ}\sim \int_\sigma d^{2m}y \, \sqrt{h} \left[{\cal R}^m\right]
\log\delta
 \labell{suggest}
 \ee
where $\sigma$ is the singular locus on the entangling surface and
$\left[{\cal R}^m\right]$ represents some curvature invariant
containing $m$ powers of the curvature on this submanifold.

We also consider entangling surfaces with conical singularities in
section \ref{singular}. In this case, if the boundary CFT lives in an
odd number of spacetime dimensions, \ie $c_n$ with $n$ even, we find
that the singularity contributes a $\log\delta$ term to the
entanglement entropy. However, for an even dimensional boundary theory,
\ie $c_n$ with $n$ odd, we find that the new universal contribution
actually diverges as $\log^2\!\delta$. However, we note, and explain in
detail in section \ref{dis}, that part of this $\log^2\!\delta$
contribution can be identified with the `smooth' contribution given in
eq.~\eqref{solo}. That is, part of this divergence should be associated
with correlations away from the singularity and so depends on details
of the smooth part of the geometry away from the tip of the cone.
However, we also argue part of the contribution is associated with the
singularity itself and so should still be expected to arise for more
general situations, independent of this smooth geometry. Given that our
boundary field theory is a CFT, we might ask if the coefficients of
these new universal contributions are simple functions of the central
charges. As a step in this direction, we work with Gauss-Bonnet gravity
in the bulk in section \ref{central1}, as this allows us to begin to
distinguish the boundary central charges, \eg see \cite{renyi,EtasGB}.
In the case of even dimensions, we see that the coefficient of the
$\log^2\!\delta$ contribution is proportional to a particular central
charge, \ie for $d=4$, it is $c$. However, for odd dimensions, the
$\log\delta$ term does not yield any such simple result.

The holographic calculations are also extended to consider conical
creases of the form $c_n\times R^m$ or $c_n\times S^m$ in section
\ref{crease}. For these cases, we find that the nature of universal
contributions again depends on the dimension of both the full spacetime
and the singular locus, as well as the curvature of the latter. In
particular, if both of these dimensions is even and the locus is
curved, \eg $c_1\times S^2$, then a $\log^2\!\delta$ term arises.
Alternatively, if the background is odd dimensional but the locus is
even dimensional and curved, \eg $c_2\times S^2$, then a $\log\delta$
contribution appears. In any other cases, the singularity does not
contribute any universal terms of this form to the entanglement
entropy. Our results again suggest the appearance of universal
contributions of the form given in eq.~\reef{suggest} for odd
dimensional theories, while similar terms with a $\log^2\!\delta$
divergence seem to be present for even dimensional CFT's.

In discussing these results, we need to be careful about an important
point. As we have illustrated with eq.~\eqref{solo}, in even
dimensional CFT's, the smooth part of the entangling surface will
already produce a universal term proportional to $\log\delta$. Hence,
as noted in our description of the results for cones, we must
distinguish this term from universal contributions associated with the
singularity. Other cases where this issue arises include: $c_1\times
R^2$, $c_2\times R^1$ and $c_2\times S^1$. All of these examples are in
six dimensions where we do not have the analog of the $d=4$ expression
in eq.~\eqref{solo}. So while a detailed comparison is not possible, in
considering the corresponding holographic calculations in detail, we
see that the coefficient of the $\log\delta$ receives contributions at
all values of the radius $\rho$ and that there is no singularity at
$\rho=0$. Hence we can clearly infer that this contribution is coming
from the smooth part of the geometry and the singularity is not making a
universal contribution to the entanglement entropy.

To close this section, let us note that for many of the examples in
table \ref{table2}, there were no logarithmic terms in the entanglement
entropy. In some of those cases, it may still be that the finite
contribution exhibits some universal behaviour but we did not
investigate this possibility here.

%%%%%%%%%%%%%%%%%%%%%%%%%%%%%%%%%%%%%%%%
\section{EE for singular embeddings}  \label{singular}

In this section, we will study entanglement entropy (EE) with singular
entangling surfaces in a flat background for holographic CFTs which are
dual to Einstein gravity. The simplest case in this category is the
kink in $d=3$. It is already known that EE for a kink has a logarithmic
divergence \cite{log1,log3,log2}. This calculation for holographic EE
was first done by Hirata and Takayanagi in \cite{hirata}. So before
calculating EE for cones in higher dimensions, we briefly review this
case.

We begin with $\{d,n,m\}=\{3,0,0\}$ in metric \eqref{metric1}.
The kink in the boundary is defined by $\rho \in [0,H]$ and
$\theta \in [-\Omega,\Omega]$, where $H$ is an IR cut-off. The
holographic entanglement entropy for this geometry is given by
\eqref{c0x1}, that is the area of the minimal area surface which hangs
in the bulk and is homologous to the kink on the boundary. We assume
that the induced coordinate for the minimal area surface are
$(\rho,\theta)$ and the radial coordinate $z=z(\rho,\theta)$. Now we
can find the induced metric $h_{\mu\nu}$ over the surface and the
entanglement entropy is given by
 \be
S_3 \big|_{k} \,=\,{2\pi \ov \lp^2}\int d\rho\, d\theta \, \sqrt{h} \,=\,{2\pi
\ov \lp^2} \int d\rho \, d\theta \, {L^2 \ov z^2}\sqrt{\rh^2+\rh^2
z'{}^2 + \dot{z}^2}\,,
 \labell{cuspx1}
 \ee
where $z'=\partial_{\rho}z$ and $\dot{z}=\partial_{\theta}z$. Here we
point out that the EE for entangling surface $\Sigma$ in
$d$-dimensional field theory will be represented by
$S_d\big|_{\Sigma}$. Now we can easily find the equation of motion for
$z(\rho,\theta)$. From scaling symmetry of the AdS space and the fact
that there is no other scale in the problem, we can say that
 \be
z\,=\, \rh \, h(\theta)\,,
 \labell{cuspx2}
 \ee
where $h(\theta)$ is such that $h\to 0$ as $\theta \to \pm \Omega$.
After using this ansatz, the entropy functional reduces to
 \be
S_3 \big|_{k} \,=\,{4 \pi L^2 \ov \lp^2} \int_{\delta/h_0}^{H} {d\rh \ov \rh}
\int_{0}^{\Omega-\epsilon} d\theta  {\sqrt{1+h^2 + \dot{h}{}^2} \ov
h^2}\,,
 \labell{cuspx3}
 \ee
where we have introduced the UV cut-off at $z=\delta$,
$\dot{h}=dh/d\theta$ and defined $h_0$ such that at $\theta=0$,
$h(0)=h_0$. Note that $\dot{h}(0)=0$ and $h_0$ is the maximum value of
$h(\theta)$. Also, $\e$ is a function of $\rh$ defined using
\eqref{cuspx2}, such that at $z=\delta$, $h(\Omega-\e)=\delta/\rh$.
Further, the substitution of ansatz \eqref{cuspx2} in equation of
motion for $z(\rho,\theta)$ gives
 \be
h(1+h^2)\ddot{h} + 2 \dot{h}{}^2 + (h^2+1)(h^2+2)\,=\,0\,.
 \ee
For this equation of motion, we can easily see that there exists a
quantity $K$ which is conserved along $\theta$ translation and is given
by
 \be
K \,=\, \frac{1+h^2}{h^2 \sqrt{1+h^2+\dot{h}{}^2}} \,=\,
{\sqrt{1+h_0^2} \ov h_0^2 }\,,
 \labell{cuspx4}
 \ee
where we have used $\dot{h}(0)=0$ and $h(0)=h_0$ to get the expression
on the right hand side. Now plan is to convert $\theta$ integral in
\eqref{cuspx3} to integral over $h$ and then separate the divergent
part in the integral. We also make the coordinate transformation to
$y=\sqrt{1/h^2-1/h_0^2}$, where $y\to\infty$ as we approach the
boundary. After this coordinate transformation, the integrand has
following divergence in the limit $y\to \infty$:
 \be
\sqrt{{1+h_0^2(1+y^2) \ov 2+h_0^2(1+y^2)}} \sim 1 + \mathcal{O}\left(
{1\ov y^3} \right)\,.
 \labell{cuspx4x1}
 \ee
So now we can write EE as
 \bea
S_3 \big|_{k} &\,=\,&{4 \pi L^2 \ov \lp^2} \int_{\delta/h_0}^{H} {d\rh \ov \rh} \int_{0}^{\sqrt{(\rho/\delta)^2-1/h_{01}^2}} dy \left(  \sqrt{{1+h_0^2(1+y^2) \ov 2+h_0^2(1+y^2)}} - 1 \right) \nonumber \\
&&\qquad \qquad \qquad \qquad + {4 \pi L^2 \ov \lp^2}
\int_{\delta/h_0}^{H} {d\rh \ov \rh} \sqrt{{\rho^2 \ov \delta^2}-{1 \ov
h_{01}^2}}\,,
 \labell{cuspx5}
 \eea
which can be simplified to give
 \be
S_3 \big|_{k} \,=\,{4 \pi L^2 \ov \lp^2 } \left( { H \ov \delta} +
q_3(\Omega)\log \left( {\delta \ov H} \right) + \mathcal{O}(\delta)
\right)\,,
 \labell{cuspx6}
 \ee
where $q_3(\Omega)$ is
 \be
q_3(\Omega) \,=\, \int_{0}^{\infty} dy  \left[1 -\sqrt{{1+h_0^2(1+y^2)
\ov 2+h_0^2(1+y^2)}} \right]\,.
 \labell{cuspx7}
 \ee
Note that $q_3(\Omega)$ is the cut-off independent term in the EE for
the kink. After this quick review of the calculations by Hirata and
Takaynagi, we turn towards the cone in higher dimensions where we will
see $\log^2\!\delta$ divergence for even $d$.

%%%%%%%%%%%%%%%%%%%%%%%%%%%%%%%%%%%%%%%
\subsection{Cone in $d=4,5$ and $6$ CFT}  \label{cone}

In this section, we will calculate EE for cone $c_n$ in some even and
odd dimensional spacetime. We will give detailed calculations for
$c_1$ in $d=4$ and discuss the final results for $c_2$ and $c_3$ in
$d=5,6$ dimensional CFTs\footnote{The cone in $d=4$ CFT is also discussed in
\cite{new}.}. However, in the beginning of the calculations, we will
keep the discussion general for arbitrary $d$.

With $m=0$ in the bulk metric \eqref{metric1}, we can define the cone
geometry by $\theta\in [0,\Omega]$ and $\rho\in[0,H]$, where $H$ is the
IR cut-off for the geometry. Now, we define the minimal area surface,
that gives the entanglement entropy, by coordinates
$(\rho,\theta,\xi^i)$ where $\xi^i$'s are coordinates on sphere
$S^n$ in \eqref{metric1}. As for the cone $c_n$, we have a rotational
symmetry $SO(n+1)$ along the sphere $S^n$, the radial coordinate $z$
will depend only on $\theta$, \ie $z=z(\rho,\theta)$. Then, the induced
metric is given by
 \be
h \,=\, \left[
 \begin {array} {c c c }
{L^2 \ov z^2}\left( 1 + z'{}^2 \right) & {L^2 \ov z^2} z' \, \dot{z}  &   \\
{L^2 \ov z^2} z' \dot{z}  & {L^2 \ov z^2}\left( \rho^2 + \dot{z}^2 \right) &    \\
  &   &  {L^2 \, \rho^2 \ov z^2} \sin^2\theta \, g_{ab}(S^n)
% & & & \ddots & & \\
%  &  &  &  & & {L^2 \, \rho^2 \ov z^2} \Pi_{i=0}^{d-4} \sin^2\theta_i
\end {array}
\right]\,,
 \labell{conex2}
 \ee
where $\dot{z}=\partial_{\theta}z$, $z'=\partial_{\rho}z$ and
$g_{ab}(S^n)$ represents the metric over unit $S^n$. We also regulate the
minimal area surface by stopping it at the UV cut-off $z=\delta$. Now
the entropy functional becomes
 \bea
& S_d \big|_{c_{d-3}} &\,=\, {2 \pi \,L^{d-1} \Omega_{d-3} \ov \lp^{d-1}} \int d\rho \,
d\theta \, {\rho^{d-3} \sin^{d-3}(\theta) \ov z^{d-1}} \sqrt{\left(
\rho^2 + (\partial_\theta z)^2 +\rho^2 (\partial_\rho z)^2 \right)} \,,
 \labell{conex3}
 \eea
where $\Omega_{d-3}$ is the surface area of the unit $(d-3)$-sphere. Note
that here we have used $n=d-3$. From this entropy functional, we can
find the equation of motion for $z(\rho,\theta)$.
 \bea
0&\,=\,& \rho ^2 \sin(\theta) z  \left(\rho ^2+\dot{z}^2\right) z'' + \rho^2 \sin(\theta) z  \left(1+z'{}^2\right)\ddot{z}- 2 \rho^2 \sin(\theta) z \dot{z} z' \dot{z}' \nonumber \\
& & + (d-1) \rho ^2 \sin(\theta) \left(\dot{z}^2+ \rho^2 \left(1+z'{}^2\right)\right)  + z (d-3)\cos(\theta) \dot{z} \left(\dot{z}^2 + \rho ^2 \left(1+z'{}^2\right)\right)	\nonumber \\
& & +\rho z  \sin(\theta) z' \left((d-1) \dot{z}^2 + (d-2) \rho^2
\left(1+z'{}^2\right)\right)
 \labell{conex4}
 \eea	
To proceed further, we use the scaling symmetry of the AdS background.
As the background geometry is scale invariant and the only scale in our
discussion is $\rho$,  solution of $z(\rho,\theta)$ should be of the
following form
 \be
z(\rho,\theta)\,=\,\rho \, h(\theta)\,,
 \labell{conex5}
 \ee
where $h(\theta)$ is a function of $\theta$ such that as $\theta \to
\Omega$, $h \to 0$. Also at $\theta=0$, $z$ achieves its maximum value
and we have $\dot{h}(0)=dh/d\theta|_{\theta=0}=0$. We also define $h_0$
such that $h_0=h(\theta=0)$. Now to extract the cut-off independent
term, first we change the integration over $\theta$ to integration over
$h$:
 \bea
& S_d \big|_{c_{d-3}}  &\,=\, {2 \pi \,L^{d-1} \Omega_{d-3} \ov \lp^{d-1}}
\int_{\delta/h_0}^H {d \rho \ov \rho} \int_{h_0}^{\delta/\rho} dh \, {
\sin^{d-3}(\theta) \ov \dot{h} h^{d-1} }\sqrt{1+h^2+\dot{h}{}^2}\,.
 \labell{conex6}
 \eea
Now the plan is to make the integral over $h$ finite. To do so, we find
the solution of $\sin(\theta)$ and $\dot{h}$ near the boundary and then
see how the integrand diverges in the limit $h\to 0$. Then, we subtract
the terms up to order $1/h$ in the integrand of \eqref{conex6} to make
it finite. At this step, we will find that it is convenient to write the asymptotic solution for $\sin(\theta)$ for each $d \geq
4$ separately. Hence, now we work case by case.

First, we consider $d=4$ and find $\dot{h}$ and $\sin(\theta)$ in terms
of $h$ in the asymptotic limit, where $h\to 0$. To do so, we make a
change of variable $y=\sin\theta$ and invert \eqref{conex4} into
equation of motion for $y=y(h)$. Using $\dot{h}=\sqrt{1-y^2}/y'(h)$ and
$\ddot{h}=-(y y'{}^2+(1-y^2)y'')/y'{}^3$, we find that
 \bea
0&\,=\,& h \left(1+h^2\right) y \left(1-y^2\right) y'' -y y' \left(3+h^2+\left(3+5 h^2+2 h^4\right) y'{}^2\right)+2 h y{}^2 \left(1+\left(1+h^2\right) y'{}^2\right) \nonumber \\
& & -h \left(1+\left(1+h^2\right) y'{}^2\right)+\left(3+h^2\right)
y{}^3 y' - h y{}^4 \,,
 \labell{conex7}
 \eea
where $y'=dy/dh$ and $y''=d^2y/dh^2$. We can solve this equation of
motion perturbatively near the boundary and find that
 \be
y=\sin(\Omega) - \frac{1}{4} \cos(\Omega) \cot(\Omega) h^2 + \left(
\frac{1}{64}(3-\cos(2 \Omega) ) \cot^2(\Omega) \csc(\Omega)  \log(h) +
y_0 \right) h^4 + \mathcal{O}(h^6)\,,
 \labell{conex8}
 \ee
where we have used $y(\theta=\Omega)=y(h=0)=\sin(\Omega)$. In the above
expression, $y_0$ is a constant and its value is fixed by using the
fact that $y(h)$ has an extrema at $h=h_0$. Further, this solution also
contains $\log(h)$, which will appear in solutions for even $d$. As a
next step, we find $\dot{h}(\theta)$ near the boundary in terms of $h$.
For that, we define  $f(h)=\dot{h}(\theta)$ and write \eqref{conex7} as
 \bea
& 0  \,=\, &h \left(1+h^2\right)  y f f' + h \sqrt{1-y^2} f^3 + \left(3+h^2\right) y f^2  \nonumber \\
&& \qquad \qquad \qquad \qquad \qquad +h \left(1+h^2\right)
\sqrt{1-y^2} f + \left(3+5 h^2+2 h^4\right) y\,.
 \labell{conex9}
 \eea
Now using \eqref{conex8} we can solve this equation near the boundary
and find
 \be
f(h) \,=\, - {2 \tan(\Omega) \ov h} -\frac{1}{2} h (3-\cos(2 \Omega) )
\csc(2 \Omega) \log(h) + f_0 h+\dots \,,
 \labell{conex10}
 \ee
where $f_0$ is a constant. Now using \eqref{conex8}, \eqref{conex10}
and $d=4$ in integrand of \eqref{conex6}, we find that for small $h$
 \be
{ \sin\theta \ov \dot{h} h^3 }\sqrt{1+h^2+\dot{h}{}^2} \sim
-\frac{\sin(\Omega)}{h^3} + \frac{\cos(\Omega) \cot(\Omega)}{8 h}+
\mathcal{O}(h)\,.
 \labell{conex11}
 \ee
This implies that the $h$ integral in the entropy functional has only
divergences coming from first two terms in the limit $h \to
\delta/\rho$. We can separate these divergent terms and for $d=4$,
write the entropy functional \eqref{conex6} in the following form
 \bea
S_4 \big|_{c_1} &\,=\,& {4 \, \pi^2 \,L^3 \ov \lp^3} \left( I_1+I_2 \right)\,.
 \labell{conex12}
 \eea
where
 \bea
I_1 &\,=\,& \int_{\delta/h_0}^H {d \rho \ov \rho} \int_{h_0}^{\delta/\rho} dh \, \left[ { \sin\theta \ov \dot{h} h^3 }\sqrt{1+h^2+\dot{h}{}^2}  + \frac{\sin(\Omega)}{h^3} - \frac{\cos(\Omega) \cot(\Omega)}{8 h}  \right]\,, \notag \\
I_2 &\,=\,& -\int_{\delta/h_0}^H {d \rho \ov \rho}
\int_{h_0}^{\delta/\rho} dh \left( \frac{\sin(\Omega)}{h^3} -
\frac{\cos(\Omega) \cot(\Omega)}{8 h} \right)\,. \notag
 \eea
If we series expand $I_1$ in terms of UV cut-off $\delta$, we find that
leading term is of order $\log(\delta)$. To see that, we use
\eqref{conex11} and find that in $I_1$, integration over $h$ is
actually finite if we set the upper limit $h = 0$. As all the
subleading terms will be of higher order in powers of $\delta$, we find
that
 \bea
I_1 \,=\,  - \log(\delta) \int_{h_0}^{0} dh \, \left[ { \sin\theta \ov
\dot{h} h^3 }\sqrt{1+h^2+\dot{h}{}^2}  + \frac{\sin(\Omega)}{h^3} -
\frac{\cos(\Omega) \cot(\Omega)}{8 h}  \right] +
\mathcal{O}(\delta^0)\,.
 \labell{conex13}
 \eea
Simultaneously, we can evaluate $I_2$ and find that
 \bea
 I_2 &\,=\,& \frac{H^2 \sin(\Omega )	}{4 \delta ^2} -\frac{1}{16} \cos(\Omega) \cot(\Omega)  \log(\delta/H)^2   \labell{conex14} \\
& & \qquad +\left( \frac{1}{8} \cos(\Omega) \cot(\Omega) \log(h_0) +
\frac{\sin(\Omega)}{2 h_0^2}\right) \log(\delta/H) +
\mathcal{O}(\delta^0) \,. \nonumber
 \eea
Now using \eqref{conex13} and \eqref{conex14} in \eqref{conex12}, we
find the complete structure of divergences in the  entanglement entropy
for cone:
 \bea
 S_4 \big|_{c_1} &\,=\,& {4\, \pi^2 \,L^3 \ov \lp^3} \bigg[ \frac{H^2 \sin(\Omega)}{4 \delta ^2}-\frac{1}{16} \cos(\Omega) \cot(\Omega)  \log(\delta/H)^2 + q_4 \log(\delta/H) +
\mathcal{O}(\delta^0)+\dots \bigg] \,,
 \labell{conex15}
 \eea
where
 \bea
q_4 &\,=\,& \frac{1}{8} \cos(\Omega) \cot(\Omega) \log(h_0) + \frac{\sin(\Omega)}{2 h_0^2} \notag \\
&& \qquad \qquad \qquad + \int^{h_0}_{0} dh \, \left[ { \sin\theta \ov \dot{h} h^3
}\sqrt{1+h^2+\dot{h}{}^2}  + \frac{\sin(\Omega)}{h^3} -
\frac{\cos(\Omega) \cot(\Omega)}{8 h}  \right]   \,.
 \labell{conex16}
 \eea

So we find that EE for a cone in $d=4$ CFT has a double logarithmic
term. We can notice from expression of $I_2$ that one of the $\log$
terms comes from integration over $h$ and then second from integration
over $\rho$. Here, first integration over $h$ or $\theta$ actually
brings us close to the cut-off on the smooth part of the entangling
surface. Further, when second integration over $\rho$ is performed, we
approach to the singularity. This idea is consistent with the fact that
in EE for even dimensions, we get a logarithmically divergent term
according to \cite{solodukhin}.

Now, as a next step, we generalize our discussion to cones in higher
dimensions. First, we calculate EE for cone in $d=5$ CFT. In this case
the calculations proceeds similar to the $d=4$ and the complete
expression for the EE is given in the appendix \ref{appa}. However, we
find that that cut-off independent term is the coefficient of $\log
\delta$ divergence and it is given by
 \bea
S_5^{\log} \big|_{c_2} \,=\, {8\, \pi^2 \,L^4 \ov \lp^4}\,
q_5(\Omega)\, \log(\delta/H) \,,
 \labell{conex17x1}
 \eea
where
\bea
q_5 &\,=\,& - \frac{4 \cos^2(\Omega)}{9\, h_0} + \frac{\sin^2(\Omega)}{3\, h_0^3} \notag \\
&& \qquad  + \int^{h_0}_{0} dh \, \left[ { \sin^2(\theta) \ov \dot{h}\,
h^4}\sqrt{1+h^2+\dot{h}{}^2}  + \frac{\sin^2(\Omega)}{h^4} - \frac{4
\cos^2(\Omega)}{9\, h^2}  \right]\,. \labell{q5t}
 \eea
We further draw this universal term in figure \ref{figure1}. There,
$\log|q_d|$ is plotted as a function of $\log(\sin\Omega)$ for $d=3$
and 5. In the limit $\Omega \to 0$, we see that $\log|q_5|$ asymptotes
to a straight line with slope $-1$. This implies that for small
$\Omega$,
 \be
q_5 \propto {1\ov \Omega} \,.
\labell{conex17x1x1}
\ee
Finally for $d=6$, we find that the cut-off independent term is
 \bea
S_6^{\log^2} \big|_{c_3} &\,=\,& {4\, \pi^3 \,L^5 \ov \lp^5} \frac{9
\cos(\Omega) \cot(\Omega) (31-\cos(2 \Omega) ) }{8192}
\log\left(\delta/ H\right)^2 \,.
 \labell{conex17x2}
 \eea
It is straight forward to see from \eqref{conex6} that all the even dimensions will
produce a $\log^2\!\delta$ divergence. For even dimensions, the number
of powers of $h$ in the denominator is odd. When separating the
divergences, similar to \eqref{conex11}, it will produce $1/h$. This term integrated over $h$ and then over $\rho$, similar to
$I_2$, will produce a $\log^2\!\delta$ divergence. Of course these
results will persist in field theories in curved backgrounds and for
dual gravities with higher derivative curvatures. We will discuss these
examples in sections \ref{crease} and \ref{central1}.

%
%\FIGURE[!ht]{
% \begin{tabular}{cc}
%\includegraphics[width=0.5\textwidth]{d4d5_c.pdf}&
%\includegraphics[width=0.5\textwidth]{d6_c.pdf}\\
%(a) & (b)
%\end{tabular}
%\caption{(Colour Online) Panel (a) plots cut-off independent terms for cone in $d=4$ and $d=5$ CFTs. For $d=4$, the cut-off independent term $C_4=-\cos(\Omega)^2/16\sin(\Omega)$ and for $d=5$, it is $C_5=q_5-4\cos(\Omega)^2/9h_0+\sin(\Omega)^2/3h_0^3$. Panel (b) plots the cut-off independent term $C_6=9\cos(\Omega)^2(31-\cos(2\Omega))/8192\sin(\Omega)$ for $d=6$ CFT.}
%\label{figure1}}
%

%
\FIGURE[!ht]{
\includegraphics[width=0.6\textwidth]{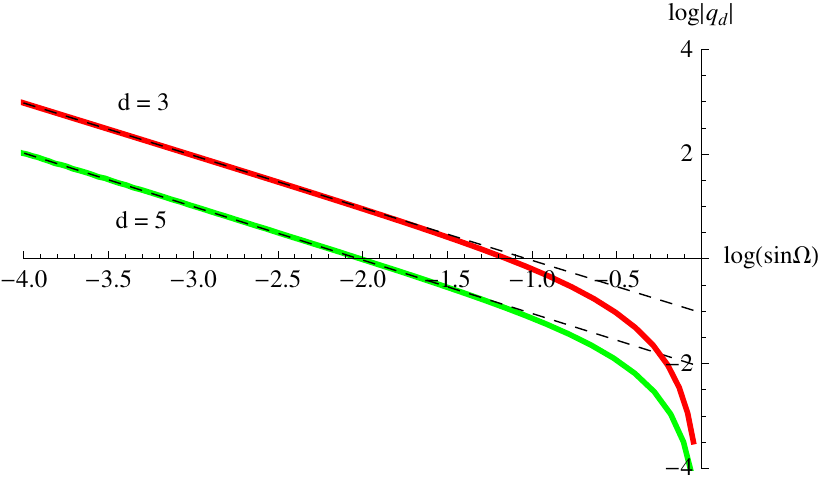}
\caption{(Color online) We have drawn $\log|q_d|$ as a function of $\log(\sin\Omega)$ for $d=3$ and $5$. In the limit $\Omega\to 0$, it is known that $\log|q_3|=-\log(\sin\Omega)+\dots$ \cite{hirata}. For $d=5$, we observe similar behavior and find that for small $\Omega$, the linear fit is $\log|q_5|=-1.0\log(\sin\Omega)-2.1\,$.}
 \label{figure1}}
%

%%%%%%%%%%%%%%%%%%%%%%%%%%%%%%%%%%%%%%
\section{EE for extended singularities}  \label{crease}

In this section, we will calculate the entanglement entropy for various
creases. Obviously the crease can have two types of locus of
singularities: flat or curved. By studying these examples, we will
argue that the contribution from the singularity will be non-zero only
if the locus is even dimensional and curved.

In the first subsection, we will study creases with a flat locus of
singularity. We will see that in this case, we don't find any new
logarithmic divergence coming from the singularity. The creases we
will be considering are $k\times R^m$ with arbitrary $m$ and, $c_n\times R^m$ with $\{n,m\}\in \big\{ \{1,1\},\{1,2\},\{1,3\},\{2,1\},\{2,2\}\big\}$.

In second subsection, we will mainly consider the singular geometries
of the form $k\times S^m$ and $c_n\times S^m$. In these cases, a
generic calculation for arbitrary curvature is tedious. So we will work
in certain limits where curvature of the locus of singularity is very
small and show how the leading order terms behave in these cases. In
the calculations, we will use following approach: If the curvature of
the locus is related to $1/R_1$, then in the limit $R_1 \to \infty$,
the boundary geometry will become flat. Now we make $R_1$ finite but
keep it very large; and calculate the leading order corrections to EE
in this limit. In this process, we will find that leading correction to
EE, in a proper normalization, will be of order $\mathcal{O}(1/R_1^2)$
but not $\mathcal{O}(1/R_1)$. Now, this approach has one more
advantage. As we are going to argue now, for some simple background
geometries like $R^3 \times S^2$, the coefficient of the logarithmic
term at order $1/R_1^2$ will be the complete contribution from the
kink. We will note in our calculations that the only dimensionful
quantities in the problem will be $R_1$, the UV cut-off $\delta$ and
the IR cut-off $H$. As the coefficient of the logarithmic divergence
should be dimensionless and if we series expand it in terms of $1/R_1$,
the numerator can either be $\delta$ or $H$. So the new logarithmic
contribution will be of the form \be S_{new} \,=\, R_1^2 \left({C_0 \ov
R_1^2} + \mathcal{O}\left({\delta^2 \ov R_1^4}\right) +
\mathcal{O}\left( {H^2 \ov R_1^4} \ri)+ \dots  \right) \log(\delta)\,.
\labell{ccx0} \ee In this case, the terms with $\delta$ in the
numerator are not UV cut-off invariant and terms with $H$ in the
numerator are not really the contribution from the kink. That is
because $H$ is the scale of the bulk part of the entangling surface and
a term arising from the singularity should be independent of it. So we
only need to focus on the leading order correction which is precisely
the contribution from the singularity. This issue will again be
discussed when we rigorously calculate EE in section \ref{curvedcrease}.

Let us summarize our strategy for the calculations here: for cases
where boundary theory is on curved background, we use the
Fefferman-Graham expansion to find the dual gravity near the boundary.
For flat boundary, we don't need to go through this step. Then, we
calculate the entanglement entropy for the kinks and use the
Fefferman-Graham expansion to separate the logarithmic divergence from
the singularities. We will see that first few terms of the
Fefferman-Graham expansion will be sufficient to find the leading order
corrections to the logarithmic divergences in the limit $R_1 \to
\infty$. To begin with, we will calculate EE for some singular
geometries with flat locus in the next subsection.

%%%%%%%%%%%%%%%%%%%%%%%%%%%%%%%%%%%%%
\subsection{EE for singularity with a flat locus}  \label{flatcrease}

In this section, we will calculate entanglement entropy for following
two types of geometries\footnote{The EE for wedge $k\times R^1$ is also
calculated in \cite{new}.}: $k\times R^m$ and $c_n\times R^m$. We will
particularly see EE for cases where $n=1,2$ and $m=1,2,3$.

\subsubsection{Crease $k\times R^m$}  \label{cuspflat}

In this section, we will work in the dual Einstein's gravity with bulk
metric \eqref{metric1} and $n=0$. The crease $k\times R^m$ geometry is
given by $\theta \in [-\Omega,\Omega]$, $\rh\in[0,\infty]$ and
$x^i\in[-\infty,\infty]$. We pick $(\rh,\theta,x^i)$ as the induced
coordinates over the minimal area surface and the radial coordinate is
$z=z(\rh,\theta)$. Now the induced metric becomes
 \be
h \,=\, \left[
 \begin {array} {c c c c c}
{L^2 \ov z^2}\left( 1 + z'{}^2 \right) & {L^2 \ov z^2} z' \dot{z}  &  &  & \\
{L^2 \ov z^2} z' \dot{z}  & {L^2 \ov z^2}\left( \rho^2 + \dot{z}^2 \right) & &  & \\
&  & {L^2 \ov z^2} &  & \\
& &   & \ddots &  \\
&  & &   & {L^2 \ov z^2}
\end {array}
\right]\,,
 \labell{csflatx2}
 \ee
and the EE is given by
 \be
S_{d}\big|_{k\times R^{d-3}}\,=\, {2 \pi \,L^{d-1} \tilde{H}^{d-3} \ov \lp^{d-1}} \int d\rho
\, d\theta \, {1 \ov z^{d-1}} \sqrt{\left( \dot{z}^2 +\rho^2 (1+z'{}^2)
\right)} \,,
 \labell{csflatx3}
 \ee
where $\dot{z}=\partial_\theta z$, $z'=\partial_\rho z$ and we have
substituted $m=d-3$. We have also imposed an IR cut-off such that
$x^i\in[-\tilde{H}/2, \tilde{H}/2]$. From this functional, we can find
the equation of motion for $z(\rho,\theta)$ and it turns out to be
 \bea
0&\,=\,&\rho  z \left(\rho ^2+\dot{z}^2\right) z'' + \rho  z \left(1+z'{}^2\right) \ddot{z} -2 \rho  z \dot{z} z' \dot{z}' \nonumber \\
&& \qquad \qquad \qquad \qquad + \dot{z}^2 ((d-1) \rho +2 z z')+\rho ^2
((d-1) \rho +z z') \left(1+z'{}^2\right)
 \labell{csflatx4}
 \eea
Now, in the limit $h \to \infty$, $\rh$ is the only scale in the
problem. Hence, using scaling symmetry of AdS, we can argue that
 \be
z\,=\,\rh \, h(\theta) \,.
 \labell{csflatx5}
 \ee
Using this in \eqref{csflatx3}, we find that
 \be
S_d \big|_{k\times R^{d-3}} \,=\, {4 \pi \,L^{d-1} \tilde{H}^{d-3} \ov \lp^{d-1}}
\int_{\delta/h_0}^{H} {d\rho \ov \rh^{d-2} } \int_{h_0}^{\delta/\rh} d
h {1\ov \dot{h} h^{d-1}} \sqrt{\dot{h}{}^2 + h^2 +1}\,,
 \labell{csflatx6}
 \ee
where we have put the UV cut-off at $z=\delta$ and defined $h_0=h(0)$.
Further, we have used \eqref{csflatx5} to find the limits of the
integrations and the extra factor of two comes from changing the range
of the integration from $\theta \in [-\Omega, \Omega]$ to $\theta
\in[0, \Omega]$. Using \eqref{csflatx5} in \eqref{csflatx4}, we find
the equation of motion for $h$
 \be
h(1+h^2) \ddot{h} + (d-1) \dot{h}{}^2 + h^4	+d h^2 +(d-1) \,=\, 0\,.
 \labell{csflatx6x1}
 \ee
In this case, although it is not explicit, we can find a constant along
the $\theta$ translation from \eqref{csflatx6x1}. It is straight
forward to see that
 \be
K_d \,=\,{(1+h^2)^{(d-1)/2} \ov h^{(d-1)} \sqrt{\dot{h}{}^2+h^2+1} }
\,=\, {(1+h_0^2)^{(d-2)/2}   \ov h_0^{(d-1)}}
 \labell{csflatx6x2}
 \ee
is a conserved quantity and it satisfies the equation of motion. Now we
can use \eqref{csflatx6x2} to separate the divergences in the EE
\eqref{csflatx6}. Using the fact that $h$ decreases near the boundary
and hence $\dot{h}$ should be negative, we find
 \be
\dot{h}\,=\,-{ \sqrt{1+h^2} \sqrt{ (1+h^2)^{d-2} - K_d^2 \, h^{2(d-1)}
} \ov K_d \, h^{d-1} }\,.
 \labell{csflatx6x3}
 \ee
Using this in integrand of \eqref{csflatx6}, in the limit $h \to 0$, we
find that
 \be
{\sqrt{\dot{h}{}^2 + h^2 +1} \ov \dot{h} h^{d-1}}   \sim
-\frac{1}{h^{d-1}}-\frac{1}{2} K_d^2 \, h^{d-1} +
\mathcal{O}(h^{d+1})\,.
 \labell{csflatx7}
 \ee
Now in \eqref{csflatx6}, we can add and subtract $1/h^{d-1}$ in the
integrand and write
 \bea
S_d \big|_{k\times R^{d-3}} &\,=\,& {4 \pi \,L^{d-1} \tilde{H}^{d-3} \ov \lp^{d-1}}
\bigg[\frac{H}{(d-2) \delta^{d-2}}-\frac{1}{(d-3)h_0 \delta^{d-3}
}+\mathcal{O}(\delta^0) + I_1 \bigg]\,,
 \labell{csflatx8}
 \eea
where
 \be
I_1 \,=\, \int_{\delta/h_0}^{H} {d\rho \ov \rh^{d-2} }
\int_{h_0}^{\delta/\rh} d h \left( {\sqrt{\dot{h}{}^2 + h^2 +1} \ov
\dot{h} h^{d-1}} +\frac{1}{h^{d-1}}\right)\,. \notag
 \ee
In \eqref{csflatx8}, first few term are coming from $h$ and $\rh$
integration of $1/h^{d-1}$. To find the divergences in $I_1$, first we
represent the integrand by $J(h)$ and from \eqref{csflatx7}, $J(h)\sim
h^{d-1}$ as $h \to 0$. We perform the integration by parts and write
$I_1$ as
 \bea
I_1 &\,=\,& -{1\ov (d-3) H^{d-3}}\int_{h_0}^{\delta/H} d h \, J(h) - {\delta \ov (d-3)} \int_{\delta/h_0}^{H} {d\rho \ov \rh^{d-1} } J(h)\big|_{h=\delta/\rh} \nonumber \\
&\,=\,& -{1\ov (d-3) H^{d-3}}\int_{h_0}^{\delta/H} d h \, J(h) -
{\delta \ov (d-3)} \, I_2 \,.
 \labell{csflatx9}
 \eea
Now to separate the divergences in $I_2$, we make a change of variable
from $\rh$ to $q=\delta/\rh$ and then Taylor expand the terms around
$\delta=0$:
 \bea
I_2 &\,=\,& -{1 \ov \delta^{d-2}} \int_{h_0}^{\delta/H} dq \, q^{d-3} J(q) \nonumber \\
&\,=\,& -{1 \ov \delta^{d-2}} \left[ \int_{h_0}^{0} dq \, q^{d-3} J(q) + {\delta \ov H} \left( q^{d-3} J(q)\right)_{q=\delta/H}+\dots \right] \nonumber \\
&\,=\,&  -{1 \ov \delta^{d-2}} \int_{h_0}^{0} dq \, q^{d-3} J(q) +
\mathcal(O)(\delta^3)\,.
 \labell{csflatx10}
 \eea
In the above expression, the integral over $q$ is finite because from
\eqref{csflatx7}, $q^{d-3}J(q)\sim q^{2d-4}$ for small $q$. Combining
\eqref{csflatx8}-\eqref{csflatx10}, we can write
 \be
S_d \big|_{k\times R^{d-3}} \,=\, {4 \pi \,L^{d-1} \tilde{H}^{d-3} \ov \lp^{d-1}}
\bigg[\frac{H}{(d-2) \delta^{d-2}} + \left(\int_{h_0}^{0}dq\,q^{d-3}
J(q)-\frac{1}{h_0} \right) {1\ov (d-3)\delta^{d-3}}
+\mathcal{O}(\delta^0)\bigg] \,.
 \labell{csflatx11}
 \ee
Remarkably, we find that as soon as we add a flat locus to the kink,
the $\log$ divergence disappears. However, there is a new divergent term of order $1/\delta^{d-3}$ which does not arise in smooth entangling surfaces. This clearly shows that logarithmic contribution from the crease depends on the curvature of the locus of
the singularity. Further, there can be a logarithmic term from the smooth part of the entangling surface in even $d$. However, the entangling surface in $k\times R^m$ is flat everywhere apart from the singularity, and hence such contributions vanish. This example clearly shows that when the locus is flat, there is no contribution from the singularity. However, we will see ahead that as soon as we turn on the curvature of the locus, the logarithmic divergence will appear only for the cases where locus is even dimensional.

%%%%%%%%%%%%%%%%%%%%%%%%%%%%%%%%%%%%%%%%%%
\subsubsection{Crease $c_n\times R^m$}  \label{coneflat}

In this section, we will calculate EE for the geometries $c_n\times
R^m$. Although such a calculation can be extended to arbitrary $n$ and
$m$, we will particularly focus on $n=1,2$ and $m=1,2$. To begin
with, we consider the metric \eqref{metric1} for the dual gravity with
arbitrary $m$. The conically singular geometry $c_n\times R^m$ is given
by $\rh\in[0,\infty]$ and $\theta \in [0,\Omega]$ in gravitational dual
\eqref{metric1}. We assume that the induced coordinates over the
minimal area surface are $(\rh,\theta,\xi^i,x^j)$, where $\xi^i$'s are
coordinates over unit sphere $S^n$ and $x^j$'s are on $R^m$. As we have rotational symmetry over
the sphere $S^{n}$, the radial coordinate for the minimal area surface
will be given by $z=z(\rh,\theta)$. Now, the induced metric over the
surface is
 \be
h \,=\, \left[
 \begin {array} {c c c c c c c c}
{L^2 \ov z^2}\left( 1 + z'{}^2 \right) & {L^2 \ov z^2} z' \dot{z}  &  &  & & &\\
{L^2 \ov z^2} z' \dot{z}  & {L^2 \ov z^2}\left( \rho^2 + \dot{z}^2 \right) &  & & & & \\
&  & {L^2 \rh^2 \sin^2(\theta) \ov z^2} g_{ab}(S^n) &  & & &  \\
%&  &  & \ddots  & & & & \\
  &  & & {L^2 \ov z^2} &  &  \\
 &   & & & \ddots  & \\
 &  & & & & {L^2 \ov z^2}
\end {array}
\right]\,,
 \labell{cnflatx2}
 \ee
where $g_{ab}(S^n)$ represents the metric over the unit sphere $S^n$ and the EE is
given by
\be
S_{d} \big|_{c_n\times R^m} \,=\, {2 \pi \,L^{d-1} \tilde{H}^{m} \Omega_{n} \ov \lp^{d-1}}
\int d\rho \, d\theta \, {\rh^{n} \sin^{n}(\theta) \ov z^{d-1}}
\sqrt{\left( \dot{z}^2 +\rho^2 (1+z'{}^2) \right)} \,,
\labell{cnflatx3}
\ee
where $\Omega_{n}$ is the area of the unit $n$-sphere,
$\dot{z}=\partial_{\theta} z$ and $z'=\partial_\rho z$. Note that we
have integrated over the $x^i$'s and used the IR cut-off
$x^i\in[-\tilde{H}/2,\tilde{H}/2]$. Now for this case, the equation of
motion becomes \bea 0&\,=\,&\rho ^2 z \left(\rho ^2+\dot{z}^2\right)
z'' + \rho ^2 z \left(1+z'{}^2\right) \ddot{z} - 2 \rho ^2 z \dot{z} z'
\dot{z}' +(d-1) \rho^2 \left(\dot{z}^2+\rho^2
\left(1+z'{}^2\right)\right)
\labell{cnflatx4} \\
&& \qquad +z \left(d' \cot(\theta) \dot{z}^3 + (d'+2) \rho  \dot{z}^2
z'+d' \rho ^2 \cot(\theta) \dot{z} \left(1+z'{}^2\right)+(d'+1) \rho^3
z' \left(1+z'^2\right)\right)\,. \nonumber
 \eea
Once again, here we can use the scaling arguments and say that
 \be
z(\rh,\theta)\,=\,\rh\, h(\theta)\,.
 \labell{cnflatx5}
 \ee
Then, EE reduces to
 \be
S_{d} \big|_{c_n\times R^m} \,=\, {2 \pi \,L^{d-1} \tilde{H}^{m} \Omega_{n} \ov \lp^{d-1}}
\int_{\delta/h_0}^{H} {d\rho \ov 	\rh^{d-n-2}}
\int_{h_0}^{\delta/\rh} dh \, {\sin^{n}(\theta)\sqrt{\dot{h}{}^2+h^2+1}
\ov \dot{h}\, h^{d-1}} \,,
 \labell{cnflatx6}
 \ee
and equation of motion for $h$ becomes
 \bea
0 &\,=\,& h(1+h^2) \ddot{h} +n \cot(\theta)h \dot{h}{}^3 + (d+n h^2-1) \dot{h}{}^2 + n \cot(\theta) h (1+h^2) \dot{h} \nonumber \\
&&\qquad \qquad \qquad \qquad \qquad +(n+1) h^4 +(d+n) h^2 +d-1 \,.
\labell{cnflatx7} \eea In \eqref{cnflatx6}, we have changed the
integration from $\theta$ to over $h$. We have also introduced the UV
cut-off $z=\delta$ and defined $h_0=h(0)$.

Now we set $n=1$ and $d=5$ (that also mean that $m=1$), and calculate
EE for the singular surface $c_1\times R^1$. First, we
need to find $y=\sin(\theta)$ near the boundary in terms of $h$. For
that, we invert the equation of motion \eqref{cnflatx4} and get
 \bea
0&\,=\,& h \left(1+h^2\right) y \left(1-y^2\right) y'' -2 \left(2+3 h^2+h^4\right) y y'{}^3   \labell{cnflatx8} \\
&& \qquad \qquad \qquad \qquad -h \left(1+h^2\right) \left(1-2
y^2\right) y'{}^2 - \left(4+h^2\right) y \left(1-y^2\right) y' - h
\left(-1+y^2\right)^2\,. \nonumber
 \eea
Now solving this equation perturbatively near the boundary, we get the
solution
 \be
y\,=\, \sin(\Omega) -\frac{1}{6} h^2 \cos(\Omega) \cot(\Omega) -
\frac{1}{432} h^4 (19-5 \cos(2\Omega)) \cot^2(\Omega) \csc(\Omega) +
\mathcal{O}(h^5)\,,
 \labell{cnflatx9}
 \ee
where we have used that at $h=0$, $y(0)=\sin(\Omega)$. Further, we can
define $\dot{h}(\theta)=f(h)$ and write the equation \eqref{cnflatx7}
in the form
 \bea
0&\,=\,&h \left(1+h^2\right) y f f' + h \sqrt{1-y^2} f^3+\left(4+h^2\right) y f^2\nonumber \\
&& \qquad \qquad \qquad \qquad +h \left(1+h^2\right) \sqrt{1-y^2} f+2
\left(2+3 h^2+h^4\right) y \,,
 \labell{cnflatx10}
 \eea
where $y'=dy/dh$. Using $y$ from \eqref{cnflatx9} and solving this
equation near the asymptotic boundary, we find
 \bea
f &\,=\,& -\frac{3 \tan(\Omega)}{h} + \frac{1}{3} h (8-\cos(2 \Omega)) \csc(2 \Omega) + f_0 h^2 \nonumber \\
&&\qquad - \frac{1}{216} h^3 (435-404 \cos(2 \Omega) + 52 \cos(4
\Omega)) \csc^3(\Omega) \sec(\Omega) + \mathcal{O}(h^4)\,,
 \labell{cnflatx11}
 \eea
where $f_0$ is a constant which is fixed by the condition $f(h_0)=0$.
Now it is straight forward to see that near the boundary
 \be
\frac{ \sin(\theta) \sqrt{1+h^2+\dot{h}{}^2}}{\dot{h} h^4} \sim
-\frac{\sin(\Omega)}{h^4}+\frac{\cos(\Omega) \cot(\Omega)}{9
h^2}-\frac{1}{18} \cos(\Omega) \cot(\Omega) + \dots \,,
 \labell{cnflatx12}
 \ee
and we can use it to make $h$ integral in entropy functional finite. So
we write EE as
 \bea
S_5\big|_{c_1\times R^1} &\,=\,&  {4\, \pi^2 \,L^{4} \tilde{H} \ov \lp^{4}} \bigg[ \frac{H^2 \sin(\Omega)}{6 \delta^3} + \frac{\cos(\Omega) \cot(\Omega) }{9 \delta } \log(\delta/H) \nonumber \\
&& + \frac{2 h_0^2 \cos(\Omega) \cot(\Omega) \left(1-\log(h_0)\right) - 9 \sin(\Omega)}{18 h_0^2 \delta } \nonumber \\
& & +  \int_{\delta/h_0}^{H} {d\rho \ov 	\rh^{2}}
\int_{h_0}^{\delta/\rh} dh \left(
{\sin^{d'}(\theta)\sqrt{\dot{h}{}^2+h^2+1}  \ov \dot{h} h^{d-1}} +
\frac{\sin(\Omega)}{h^4} - \frac{\cos(\Omega) \cot(\Omega)}{9 h^2}
\right)  +\mathcal{O}(h^0) \bigg]\,,
 \labell{cnflatx13}
 \eea
where we have already performed the integrations over some terms. Now
let us call the term with integration $I_1$ and the integrand $J_5(h)$.
Then, near the boundary $J_5(h)\sim \mathcal{O}(h^0)$. Now using
integration by parts, we can write
 \bea
I_1&\,=\,& \int_{\delta/h_0}^{H} {d\rho \ov \rh^{2}} \int_{h_0}^{\delta/\rh} dh J_5(h) \nonumber \\
&\,=\,& -{1\ov H}  \int_{h_0}^{\delta/H} dh J_5(h)  - \delta
\int_{\delta/h_0}^{H} {d\rho \ov \rh^{3}} J_5(h)|_{h=\delta/\rh}
\nonumber
 \labell{cnflatx14}
 \eea
We further make the coordinate transformation $\rh=\delta/q$ and Taylor
expand the second term in $\delta$:
 \bea
I_1&\,=\,&-{1\ov H}  \int_{h_0}^{\delta/H} dh J_5(h)  + {1\ov \delta}  \int_{h_0}^{\delta/H} dq \, q J_5(q) \nonumber \\
&\,=\,& -{1\ov H}  \int_{h_0}^{0} dh J_5(h)  + {1\ov \delta}
\int_{h_0}^{0} dq \, q^2 J_5(q) - {\delta \ov 18 H^3} \cos(\Omega)
\cot(\Omega) +\mathcal{O}(\delta)\,,
 \labell{cnflatx15}
 \eea
where we have used \eqref{cnflatx12} to get the final expression.
Combining \eqref{cnflatx13} and \eqref{cnflatx15}, we can write
 \bea
S_5 \big|_{c_1\times R^1} &\,=\,&  {4\, \pi^2 \,L^{4} \tilde{H} \ov \lp^{4}} \bigg[ \frac{H^2 \sin(\Omega)}{6 \delta^3} + {1\ov \delta}  \int_{h_0}^{0} dq \, q J_5(q) + \frac{\cos(\Omega) \cot(\Omega) }{9  } {\log(\delta/H)\ov \delta}    \nonumber \\
& &  + \frac{2 h_0^2 \cos(\Omega) \cot(\Omega) \left(1-\log(h_0)\right)
- 9 \sin(\Omega)}{18 h_0^2 \delta }  + \mathcal{O}(\delta^0) \bigg]\,,
 \labell{cnflatx16}
 \eea
Note that in the above expression, we have a new divergence of the form
$\log(\delta/H)/\delta$ which does not arise in EE for smooth
entangling surfaces. This term should be a contribution from the
singularity. Further, we note that as soon as we add a one-dimensional
locus to the conical singularity, both double $\log$ and logarithmic
divergences disappear. Recall that in the previous case for cone $c_1$
in $d=4$, \ie eqn.~\eqref{conex15}, we got a double $\log$ and $\log$
terms.

We can easily generalize the above calculations to the crease
$c_1\times R^2$. For this case, the integrand near the boundary and
complete expression for EE are given by \eqref{cnflatx17} and
\eqref{cnflatx18}. Once again we find a divergent term of the order
$\log(\delta/H)/\delta$ in EE. Also, now the cut-off independent term
is
 \bea
S_6^{\log} \big|_{c_1\times R^2} &\,=\,& {4\, \pi^2 \,L^{5} \tilde{H}^2 \ov \lp^{5}}
\frac{3 (13-19 \cos(2 \Omega) ) \cot^2(\Omega) \csc(\Omega) }{8192
H^2}\log(\delta/H)\,.
 \labell{cnflatx18x1}
 \eea
Note that for singular geometry $c_1\times R^2$ in $d=6$, we only have
$\log$ divergence instead of $\log^2\!\delta$ as compared to the cone
$c_1$ in equations \eqref{conex15}. Also, recall that in
\eqref{conex17x2}, we saw that a cone without a locus, $c_3$ in $d=6$
gives a $\log^2\!\delta$ divergence. In \eqref{cnflatx18x1}, this
$\log^2\!\delta$ term disappears because of flat locus we have added.
We will see in section \ref{conecurved} that as soon as the curvature
of the locus turned on, the $\log^2\!\delta$ divergence will reappear
in \eqref{cc6x20} in $d=6$. In \eqref{cnflatx18x1}, the logarithmic
contribution can be attributed to the fact that in even dimensions, the
entangling surface has a logarithmic divergence. We can explicitly
verify that logarithmic term in \eqref{cnflatx18x1} comes from
integration over $1/h$ in expansion \eqref{cnflatx17}. As $\rho$ need
not to be near the singularity, this contribution is from smooth part
of the entangling surface. Further, it is straight forward to find that
for cone $c_1\times R^3$ in $d=7$ dimensional CFT, the EE has no double
$\log$ or $\log$ divergence. Which is expected because for entangling
surface in odd $d$, there is no logarithmic divergence and as the locus
is odd dimensional and flat, there should not be any logarithmic
contribution from the singularity either.

We can further extend these calculations for the case $n=2$ and
$m=1,2$. For crease $c_2\times R^1$, we find that the universal term in
EE is a logarithmic divergence and it is given by
 \be
S_6^{\log} \big|_{c^2\times R^1} \,=\, {8\, \pi^2\, L^5 \tilde{H} \ov
\lp^5}{(7-9 \cos(2 \Omega)) \cot^2(\Omega) \ov 256\, H} \log({\delta
}/{H})\,.
 \labell{nnx1}
 \ee
We also find a new contribution from the singularity at the order
$1/\delta$. In $c^2\times R^1$, the locus of the singularity is odd
dimensional and hence singularity doesn't contribute in the universal
term. However, smooth part of the entangling surface contributes through
a $\log$ in $d=6$ and this is the contribution \eqref{nnx1}. It can be
verified that this log term arise from integration over $h$ away from
the singularity. Further, we can also calculate EE for crease
$c_2\times R^2$ and find that there is a new divergence from
singularity of order $1/\delta^2$. However, there is neither a $\log$
nor a $\log^2$ term in EE. Note that for crease $c_2\times R^2$, the
locus of the singularity is even dimensional but flat. So singularity
does not contribute through a $\log$ term. However, as we will see in
\eqref{cc6x24} that for a curved locus, we get a $\log$ contribution.

%%%%%%%%%%%%%%%%%%%%%%%%%%%%%%%%%%%%%%%%%%%%%%%%%
\subsection{EE for singularity with curved locus}  \label{curvedcrease}

In this subsection, we will consider several singular embeddings which
have a curved locus. As soon as the curvature of the locus is turned
on, the double $\log$ and $\log$ terms from singularity will make
appearance. In this section, we will consider the creases $k\times
\Sigma$ and $c_n\times \Sigma$, where locus $\Sigma$ will take the form
$S^m$ or $S^{m-p}\times R^p$. These cases will be slightly more
involved and hence we will do two things: first, we will always work in
the limit where curvature of the locus is very small and we will do the
calculations perturbatively. In certain cases, we will see that these
perturbative calculations are sufficient to pick the complete
contributions from the singularity. Second, for calculations in this
section, we will foliate our minimal area surface in different way. In
all the previous cases, the induced coordinates on the minimal area
surface were $(\rho,\theta, \dots)$ and we assumed that the radial
coordinate $z=z(\rh,\theta)$. However, we will find that it is more
convenient to do the calculations in a coordinate system where bulk
radial coordinate $z$ is one of the induced coordinates on the minimal
area surface and we have $\rh=\rh(z,\theta)$. We have shown in figure
\ref{foliage}, how these different induced coordinates foliate the
minimal area surface in the bulk. Note that the new set of induced
coordinate $(z,\theta,\dots)$ are not the well-defined coordinates on
the boundary as $\rh=\rh(z,\theta)$ will be multivalued for $z=0$ and
$\theta=\pm \Omega$. However, until unless we put the UV cut-off, we
find that $\rh=\rh(z,\theta)$ is a well-behaved function and we can
work with it. Now, in the next subsection we will begin with the
geometries $k\times \Sigma$ and then we will move on to conical
singularities $c_n\times \Sigma$.

\FIGURE[!ht]{
\begin{tabular}{cc}
\includegraphics[width=0.5\textwidth]{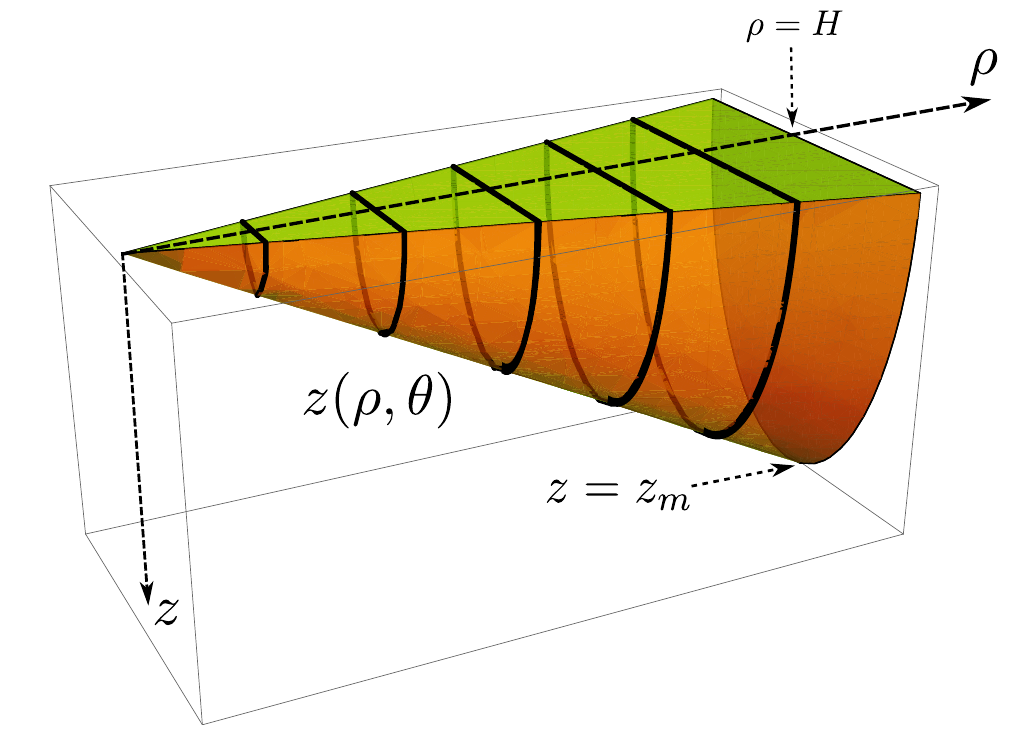}&
\includegraphics[width=0.5\textwidth]{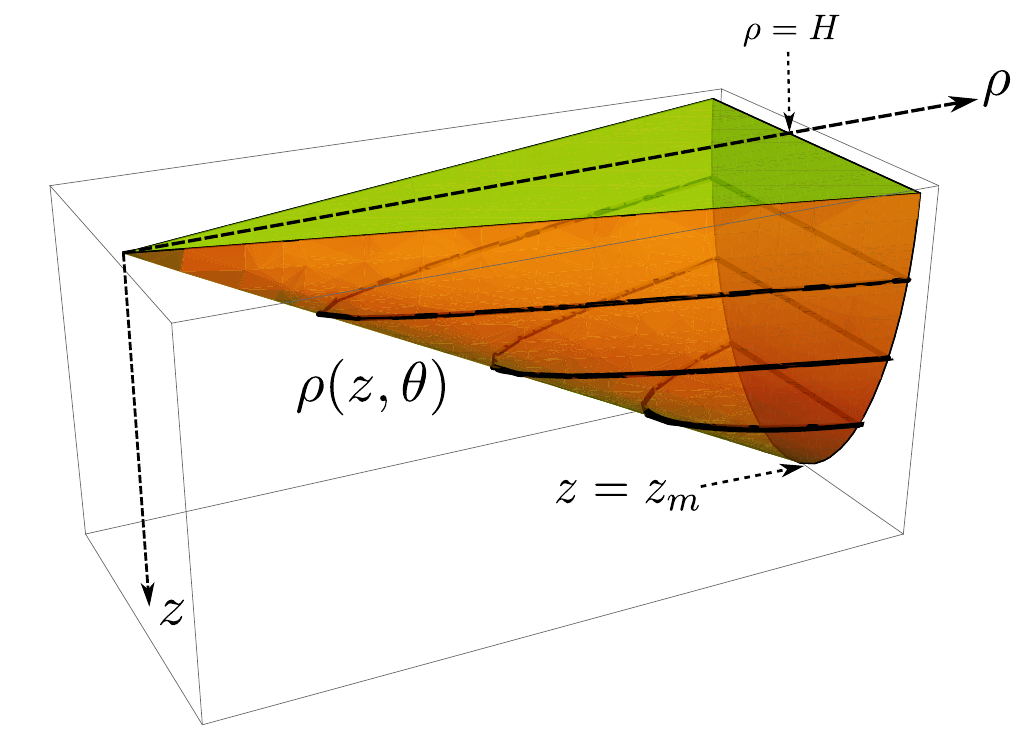}\\
(a) & (b)
\end{tabular}
\caption{Panel (a) shows how minimal area surface is foliated when we
have the induced coordinates $(\rho,\theta,\dots)$ and minimal area
surface is given by $z=z(\rho,\theta)$. Panel (b) shows the foliation
of the minimal area surface when induced coordinates are $(z,\theta,
\dots)$ and we have $\rho=\rho(z,\theta)$. Note that $z=z_m$ is the
maximum value of $z$ on the surface such that $\rho(0,z_m)=H$.}
\label{foliage}}
%

%%%%%%%%%%%%%%%%%%%%%%%%%%%%%%%%%%%%%%%%%
\subsubsection{Crease $k\times \Sigma$}  \label{cuspcurved}

In this section, we will mainly consider the geometries $k\times S^2$,
$k\times R\times S^2$ and $k\times S^3$. We will see that singularities
with even dimensional locus will contribute through a logarithmic term.

To begin with, let us consider $d=5$ CFT on background $R^3\times S^2$.
Before we construct the singular entangling surface in this geometry
and calculate holographic EE, we need to find the dual gravity. So we
begin with the action for six-dimensional dual Einstein gravity
 \be
I_6 \, =\, {1\ov \lp^4}\int d^6x \sqrt{-g}\left[ {20 \ov L^2} + R
\right]\,,
 \labell{cc2x1}
 \ee
for which, the equation of motion is given by
 \be
R_{\mu\nu} - {10 \ov L^2} g_{\mu\nu} \,=\,0\,.
 \labell{cc2x2}
 \ee
Now we ansatz that the bulk metric, which has the boundary $R^3 \times
S^2$, is of the form
 \be
ds^2 \,=\, {L^2 \ov z^2} \left( dz^2 + f_1(z)\left(dt^2 + d\rho^2 +
\rho^2 d\theta^2 \right) + f_2(z) R_1^2 \, d\Omega_2^2 \right)\,,
 \labell{cc2x3}
 \ee
where $d\Omega_2^2=d\xi_0^2 + \sin^2(\xi_0) d\xi_1^2$ represents the
metric over two-sphere and, $f_1$ and $f_2$ are functions of the radial
coordinate. Here $R_1$ is the radius of the sphere on the boundary and
in the limit $R_1 \to \infty$, we recover the flat boundary. To find
$f_1$ and $f_2$, we use the Fefferman-Graham expansion near the
boundary $z=0$. The idea is to insert the ansatz \eqref{cc2x3} in the
equation of motion \eqref{cc2x2} and then solve it near $z=0$ to find
 \bea
& f_1 & \,= \, 1 + {z^2 \ov 12 R_1^2} + {17 z^4 \ov 576 R_1^4}+ \dots \quad \textrm{and} \nonumber \\
& f_2 & \,= \, 1 - {z^2 \ov 4 R_1^2} - {5 z^4 \ov 192 R_1^4} + \dots
\,.
 \labell{cc2x4}
 \eea
Knowing the dual geometry near the boundary, we can calculate
EE for the geometry $k\times S^2$. On the boundary, this geometry is
defined by $\rho \in [0,H]$ and $\theta \in [-\Omega, \Omega]$, where
$H$ is the IR cut-off. Now to define the minimal area surface, we use
slightly different coordinates compared to the previous sections. We
choose the induced coordinates to be $(z, \theta, \xi_0,\xi_1)$ and
because of the rotational symmetry on the sphere,
$\rho=\rho(z,\theta)$. Although, $\rho(z,\theta)$ is not a well defined
function at the boundary $z=0$ but we can definitely work with it until
we impose the UV cut-off and assume that the boundary is at some finite
$z=\delta$. With this coordinate choice, the induced metric becomes
 \be
h\,=\,\left[
 \begin{array}{c c c c}
 {L^2 \ov z^2}\left( f_1 \rho'{}^2 + 1 \right) & {L^2 f_1\ov z^2} \drh \rh'  & & \\
 {L^2 f_1 \ov z^2} \drh \rh'  & {L^2 f_1 \ov z^2} \left( \drh^2 + \rh^2  \right) & & \\
  &  & {L^2 f_2 \, R_1^2 \ov z^2} & \\
  &  &  & {L^2 f_2 \,R_1^2 \ov z^2} \sin^2(\xi_0)
\end{array}
\right]\,,
 \labell{cc2x5}
 \ee
where $\drh=\partial_\theta \rh$, $\rh'=\partial_z \rh$ and hence, the
entanglement entropy is given by
 \be
S_5 \big|_{k\times S^2} \,=\, {8\, \pi^2 L^4 R_1^2 \ov \lp^4} \int dz \, d\theta \,
{\sqrt{f_1}f_2 \ov z^4} \sqrt{ \drh^2 + \rh^2(1 + f_1 \rh'{}^2)} \,.
 \labell{cc2x6}
 \ee
We can easily find the equation of motion of $\rho(z,\theta)$ to be
 \bea
&0 \;=\; & 2 z f_1 f_2 \rh \left(\rh^2+\drh^2\right) \rh''+ 2 z f_2 \rh \left(1+f_1 \rh'{}^2 \right) \ddot{\rh} - 4 z f_1 f_2 \rh \drh \rh' \drh' +2 z f_1  f_2' \rh \rh' \left(\drh^2 + \rh^2 \left(1+f_1 \rh'{}^2 \right)\right)  \labell{cc2x7} \\
& \ & + f_2 \left(-4 z \drh^2 + \rh \left(-8 f_1 + 3 z f_1' \right)
\drh^2 \rh' - 2 z \rh^2 \left(1 + f_1 \rh'{}^2 \right) + \rh^3 \rh'
\left(3 z f_1'- 8 f_1^2 \rh'{}^2 + 2 f_1 \left(-4 + z f_1' \rh'{}^2
\right)\right)\right)\,, \nonumber
 \eea
where $\ddot{\rh}=\partial_\theta^2 \rho$, $\rh''=\partial_z^2 \rh$ and
$\dot{\rh}'=\partial_z \partial_\theta \rh$.

As discussed earlier, we will be working in the approximation $R_1 \to
\infty$. In this approximation, we will find that leading order
correction to the EE will be of the order $\mathcal{O}(1/R_1^2)$. To
show this, first we need to find the form of the solution
$\rho(z,\theta)$ in this approximation. For that, we make following
ansatz:
 \be
\rho(z,\theta) \,=\, {z \ov h(\theta)} + {z^2 \ov R_1 } g_2(\theta) + {z^3 \ov R_1^2} g_3(\theta) + \dots\,. \\
 \labell{cc2x8}
 \ee
This ansatz is made on the basis of the available dimensionful
quantities in the problem. Now we are going to argue that in the
following expansion, the $\theta$ dependent functions $g_{2n}$ are
zero. To see that, we insert the ansatz \eqref{cc2x8} in the equation
of motion for $\rho(z,\theta)$ and find following equations of motion
for $h$, $g_2$ and $g_3$:
 \bea
h \left(1 + h^2\right) \ddot{h} + 4 \dot{h}{}^2 + \left( 1 + h^2 \right) \left( 4 + h^2 \right) & \,=\,&0 \,,  \labell{cc2x9} \\
h^2 \left(1 + h^2 \right)^2 g_2'' + 2 h \left(7 + 2 h^2 \right) \left(1 + h^2 \right) \dot{h} g_2' \qquad \qquad \qquad \qquad \qquad & & \nonumber \\
+ g_2 \left( 14 \dot{h}{}^2 - 2 h^4 (8-\dot{h}^2) -h^6 - 31 h^2 -16  \right) & \,=\, &0 \,,  \labell{cc2x10} \\
h^2 \left(1 + h^2 \right)^2 \ddot{g}_3 + 4 h \left(4 + h^2\right) \left(1 + h^2\right) \dot{h} \dot{g}_3 \qquad \qquad \qquad \qquad \qquad & & \nonumber \\
- g_3 \left( h^6 + h^2 \left(35 - 2 \dot{h}{}^2\right) + 4 \left(5 -6
\dot{h}{}^2\right) + 2 h^4 \left(8 - \dot{h}{}^2\right)\right) &\,=\,&
\mathcal{S}_1\,,
 \labell{cc2x11}
 \eea
where $\mathcal{S}_1$ is given by
 \bea
\mathcal{S}_1& \,=\, & \frac{1}{12 h} \left(8 + 7 h^4 + 3 \dot{h}{}^2 + h^2 \left(15 + 7 \dot{h}{}^2 \right)\right) \nonumber \\
& & + \frac{2 h}{ \left(1 + h^2\right)} \bigg(h^2 \left(1+h^2\right){}^2 \left(4+h^2\right) g_2'{}^2 + 2 h \left(8+7 h^2+h^6\right) \dot{h} g_2 g_2' \nonumber \\
& \ & + \left(h^2 \left(32-15 \dot{h}{}^2\right) + h^6 \left(8 +
\dot{h}{}^2\right) + 2 h^4 \left(15 + \dot{h}{}^2 \right) + 2 \left(5 +
8 \dot{h}{}^2\right) \right) g_2^2  \bigg)\,.
 \labell{cc2x12}
 \eea
Note that here we have arranged the equations of motion such that on
the left hand side, we have the homogeneous part of the equation and on
the right hand side, we have the source terms. So $\mathcal{S}_1$ is
source for $g_3$ and in \eqref{cc2x12}, terms in first line come from
the corrections to $f_1$ and $f_2$, and terms in last two lines are
second order in $g_2$. We also note that in \eqref{cc2x10} and
\eqref{cc2x11}, we have used equations of motion to eliminate the
second derivatives of $h$ and $g_2$. Now we notice following points:
first, the corrections to $f_1$ and $f_2$ are of even powers of $1/R_1$
and hence they only source $g_{2n+1}$'s. Second, $g_{2n}$ are sourced
only by $g_{2i}$'s, where $i <n$. Third, in flat boundary, precisely in
the limit $R_1 \to \infty$, we have $h \neq 0$ and $g_2 =0$. So above
arguments conclude that as we make $R_1$ finite, only
$g_{2n+1}(\theta)$'s will be turned on and all the $g_{2n}=0$, which
clearly are solution of these equations of motion.

Now to separate the logarithmic divergence in entanglement entropy,
first let us write $\rh=\rh_0(z,\theta) + \rh_1(z,\theta)/R_1^2$, where
$\rho_0 = z/h(\theta)$ and $\rh_1$ is higher order corrections for
large $R_1$. Using this and \eqref{cc2x4} in \eqref{cc2x6}, and keeping
only the leading order terms in $R_1$, we find that
 \bea
S_5 \big|_{k\times S^2} & \,=\, &{8\, \pi^2 L^4 R_1^2 \ov \lp^4} \int_{z_m}^\delta dz \int_{-\Omega+\epsilon}^{\Omega-\epsilon} d\theta \bigg[ \frac{ \sqrt{\drh_0^2 + \rh_ 0^2 \left(1 + \rho_0'{}^2\right)}}{z^4} \nonumber \\
& & \qquad \qquad -\frac{ \left(\dot{\rho}_0 \left(5 z^2 \dot{\rho }_0
- 24 \dot{\rho }_1\right) - 24 \rho_0 \rh_1 \left( 1 + \rho
_0'{}^2\right) + \rh_0^2 \left( 5 z^2 + 4 z^2 \rho _0'{}^2 - 24 \rho
_0' \rho _1'\right)\right)}{24 z^4 R_1^2 \sqrt{\dot{\rho }_0^2 +
\rh_0^2 \left(1 + \left(\rho _0'\right){}^2\right)}} \bigg] \,,
 \labell{cc2x13}
 \eea
where $\delta$ is the UV cut-off and $\epsilon=\epsilon(z)$ is defined
such that $\rho(z,\Omega-\epsilon)=H$ and further, $z_m$ is defined
such that $\rho(z_m,0)=H$ or $\epsilon(z_m)=\Omega$. We can also insert
this ansatz in \eqref{cc2x7} and find the equation of motion for
$\rh_0$ and $\rh_1$ by series expanding it in terms of $R_1$. In the
second term of \eqref{cc2x13}, we can convert $\rh_1'$ and $\drh_1$
into $\rh_1$ using the integration by parts and then use equation of
motion for $\rh_0$ to simplify the coefficient of $\rh_1$ to zero. This
process will leave us with some boundary terms and we find that
 \bea
S_5 \big|_{k\times S^2} & \,=\, &{4\, \pi^2 L^4 R_1^2 \ov \lp^4} \int_{z_m}^{\delta} dz \int_{-\Omega+\epsilon}^{\Omega-\epsilon} d\theta  \Bigg[ \frac{ \sqrt{\dot{\rho }_0^2 + \rh_0^2 \left(1+\left(\rho _0'\right){}^2\right)}}{z^4} - \frac{ \left(5 \dot{\rho }_0^2 + \rh_0^2 \left(5 + 4 \rho _0'{}^2\right) \right)}{24 \, R_1^2 z^2 \sqrt{\dot{\rho }_0^2 + \rh_0^2 \left(1 + \left(\rho _0'\right){}^2\right)}} \nonumber \\
& & \qquad + {\partial \ov \partial \theta}\left(\frac{\dot{\rho }_0
\rh_1}{R_1^2 z^4 \sqrt{\dot{\rho }_0^2+\rh_0^2 \left(1+\left(\rho
_0'\right){}^2\right)}}  \right) + {\partial \ov \partial z} \left(
\frac{ \rh_0^2 \rho _0' \rh_1}{R_1^2 z^4 \sqrt{\dot{\rho }_0^2 +
\rh_0^2 \left(1 + \rho_0'{}^2 \right)}}  \right) \Bigg]\,.
 \labell{cc2x13x1}
 \eea
Now we can insert the ansatz $\rh_0=z/h(\theta)$ and $\rh_1=z^3
g_3(\theta)$ and simplify the functional to
 \bea
S_5 \big|_{k\times S^2} & \,=\, &{16\, \pi^2 L^4 R_1^2 \ov \lp^4} \int_{z_m}^{\delta} dz \int_{h_0}^{h_{1c}} {d h \ov \dot{h}}  \bigg[ \frac{\sqrt{1 + h^2 + \dot{h}^2}}{z^3 h^2} -\frac{ 4 + 5 h^2 + 5 \dot{h}{}^2 }{ 24 \, R_1^2 z h^2 \sqrt{1+h^2 + \dot{h}{}^2}} \bigg] \nonumber \\
&& \qquad \qquad \qquad - {14\, \pi^2 L^4 \ov \lp^4}
\int_{z_m}^\delta {dz \ov z} \frac{g_3 \dot{h}}{ \sqrt{1 + h^2 +
\dot{h}{}^2}} \bigg|_{\theta=\Omega-\epsilon}\,,
 \labell{cc2x14}
 \eea
where we have changed the integration limits from $(-\Omega,\Omega)$ to
$(0,\Omega)$ and also changed the integration variable to $h(\theta)$.
Note that the boundary term with derivative with respect to $z$ turns
out to zero up to leading order. We have also defined
$h_0=h(0)$ and $h_{1c}(\rho)=h(\Omega-\epsilon)$ and used $h'_1(0)=0$
in getting the boundary terms.

Now to separate the logarithmic divergences, we consider all the
contributing factors one by one. We will see that the first term in the
first line of \eqref{cc2x14} will not contain any logarithmic
divergence. The second term and the boundary terms will contain
logarithmic divergence. Note that we also need to be careful and consider the divergences
coming from the limits of the integrals. To begin with, we first study
the behavior of $h$ and $g_3$ near the asymptotic boundary. The
equations of motion for $h$ and $g_3$ are given by \eqref{cc2x9} and
\eqref{cc2x11} with $g_2=0$. Now similar to \eqref{csflatx6x2}, we find
that equation of motion for $h$ can be integrated once to get
 \be
K_5 \,=\,{(1+h^2)^2 \ov h^4\sqrt{1+h^2 + \dot{h}{}^2}}\,,
 \labell{cc2x15}
 \ee
where $K_5$ is a constant and it can be further related to $h(0)$ using
$\dot{h}(0)=0$.

To extract the logarithmic divergence, we will only need the asymptotic
behavior of $h$ and $g_3$. Hence we solve $g_3$ in terms of $h$ in
asymptotic limit, where $h$ is small. To do so, we use \eqref{cc2x11}
and change the variable from $\theta$ to $h$ by expressing
$\ddot{g}_3(\theta)=d^2g_3/d\theta^2$ and
$\dot{g}_3(\theta)=dg_3/d\theta$ in terms of
$\ddot{g}_3(h)=d^2g_3/dh^2$ and $\dot{g}_3(h1)=dg_3/dh$. We further use
\eqref{cc2x15} to express $h1'(\theta)$ in terms of $K_5$ and $h$.
Finally, the equation of motion for $g_3(h)$ becomes
 \bea
0 & \,=\,& 12 h^3 \left(1+h^2\right){}^2 \left(h^8 K_5^2-\left(1+h^2\right){}^3\right) \ddot{g}_3 + 12 h^2 \left(h^8 \left(1+h^2\right) \left(16+5 h^2\right) K_5^2 - 4 \left(1+h^2\right){}^4 \left(3+h^2\right)\right) \dot{g}_3 \nonumber \\
& &- 12 h \left(2 \left(1+h^2\right){}^3 \left(12+h^2+h^4\right) - h^8 \left(44+17 h^2+3 h^4\right) K_5^2 \right)g_3 \nonumber \\
& \ & + \left(1+h^2\right){}^3 \left(3+7 h^2\right)+5 h^8 K_5^2 \,.
 \labell{cc2x16}
 \eea
Now this equation has two solutions when we solve it perturbatively in
the limit $h\to 0$. The leading terms of these solutions go like
$1/h^3$ and $1/h^8$. However, $g_3$ must be such that $\rho$ is finite
in the limit $h \to 0$ and $\delta \to 0$. As $g_3$ appears at order
$\delta^3$ in \eqref{cc2x4}, it can only go like $1/h^3$. Hence, one of
the constants, which is the coefficient of $1/h^8$, is fixed to zero.
As a result, the asymptotic solution turns out to be
 \be
g_3\,=\, \frac{b_3}{h^3} + \frac{1+88 b_3}{56 h} + \frac{4+72 b_3}{189}
h - \frac{4 + 72 b_3 }{693} h^3 + \dots \,,
 \labell{cc2x19}
 \ee
where $b_3$ is a constant such that $g_3(\theta)$ has an extrema at
$\theta=0$.
%So the solutions of second order equation of motion for $g_3(\theta)$ has two independent constants. One of these is fixed by the fact that $g_3$ keeps $\rho$ finite at the boundary and second coefficient is $b_3$ which is actually related to the fact that $\dot{g}_3(0)=0$.

Before we begin discussing divergences of various terms, we find the
series expansion of $h_{1c}$ in terms of the UV cut-off $\delta$. As
$h_{1c}=h(\Omega-\epsilon)$ and $z=\delta$ at the cut-off, we can use
\eqref{cc2x8} and perturbative solution \eqref{cc2x19} to find
following series expansion for $h_{1c}$
 \be
h_{1c}(\delta) \,=\, \left(\frac{1}{H}+\frac{b_3 H}{R_1^2}\right)
\delta +\frac{\left(1+88 b_3\right) \delta ^3}{56 H
R_1^2}+\frac{\left(4+72 b_3\right) \delta ^5}{189 H^3 R_1^2}+
\mathcal{O}(\delta^6)\,.
 \labell{cc2x19x1}
 \ee
Note that we have kept only leading corrections in $R_1$ at any order
in $\delta$.

Now we return to \eqref{cc2x14} and analyze the divergences for each
term. First, we use \eqref{cc2x15} in the integrand of first two terms
of \eqref{cc2x14} and find that in the asymptotic limit
 \bea
\frac{\sqrt{1 + h^2 + \dot{h}^2}}{\dot{h} h^2} &\,\sim \,& -\frac{1}{h^2}-\frac{1}{2} K_5^2 h^6 + \mathcal{O}(h^8)  \labell{cc2x20} \\
\frac{5 \dot{h}{}^2+ 5 h^2 +4}{24 \dot{h} \,  h^2 \sqrt{1 + h^2 +
\dot{h}{}^2}} &\,\sim \,& -\frac{5}{24 h^2}-\frac{K_5^2 h^6}{16}
+\mathcal{O}(h^8) \,.
 \labell{cc2x21}
 \eea
So we can make the integrands finite by organizing the terms in
following form
 \bea
I_1 &\,=\,& \int_{z_m}^{\delta} {dz \ov z^3} \int_{h_0}^{h_{1c}} d h  \frac{\sqrt{1 + h^2 + \dot{h}^2}}{ \dot{h} h^2} \nonumber \\
&\,=\,& \int_{z_m}^{\delta} {dz \ov z^3} \int_{h_0}^{h_{1c}} d h  \left[\frac{\sqrt{1 + h^2 + \dot{h}^2}}{ \dot{h} h^2}+{1\ov h^2}\right] + \int_{z_m}^{\delta} {dz \ov z^3} \left({1 \ov h_{1c}} - {1 \ov h_0} \right)   \labell{cc2x22}  \\
& \,=\, & I_1'+I_2' \,,  \labell{cc2x22x1}
 \eea
and
 \bea
I_2 &\,=\,& \int_{z_m}^{\delta} {dz \ov z} \int_{h_0}^{h_{1c}} d h  \frac{5 \dot{h}{}^2+ 5 h^2 +4}{24 \dot{h}  h^2 \sqrt{1 + h^2 + \dot{h}{}^2}} \nonumber \\
&\,=\,& \int_{z_m}^{\delta} {dz \ov z} \int_{h_0}^{h_{1c}} d h  \left[ \frac{5 \dot{h}{}^2+ 5 h^2 +4}{24 \dot{h}  h^2 \sqrt{1 + h^2 + \dot{h}{}^2}} + {5 \ov 24 h^2}\right] +\int_{z_m}^{\delta} dz {5 \ov 24 z} \left({1\ov h_{1c}}-{1\ov h_0} \right)  \labell{cc2x23}  \\
& \,=\, & I_3' +I_4' \,.  \labell{cc2x23x1}
 \eea
In \eqref{cc2x22x1}, $I_1'$ and $I_2'$ represent the first and second
integrals in \eqref{cc2x22}. Similarly in \eqref{cc2x23x1}, $I_3'$ and
$I_4'$ are the first and second integrals in \eqref{cc2x23}. Now first
we consider $I_1'$. We differentiate it with respect to the UV cut-off
$\delta$ and look for $1/\delta$ divergent terms. After taking the
derivative, we find
 \bea
{d I_1' \ov d \delta} &\,=\,&{1 \ov \delta^3} \int_{h_0}^{h_{1c}(\delta)} d h  \left[\frac{\sqrt{1 + h^2 + \dot{h}^2}}{ \dot{h} h^2}+{1\ov h^2}\right] \nonumber \\
&\,=\,& {1 \ov \delta^3} \int_{h_0}^{0} d h  \left[\frac{\sqrt{1 + h^2 + \dot{h}^2}}{ \dot{h} h^2} + {1\ov h^2}\right] + {1 \ov \delta^2} {d h_{1c}(\delta) \ov d \delta} \left[\frac{\sqrt{1 + h^2 + \dot{h}^2}}{ \dot{h} h^2}+{1\ov h^2}\right]_{h=h_{1c}(\delta)}+\dots \nonumber \\
&\,=\,& {1 \ov \delta^3} \int_{h_0}^{0} d h  \left[\frac{\sqrt{1 + h^2
+ \dot{h}^2}}{ \dot{h} h^2}+ {1\ov h^2}\right] -{K_5^2 \ov 2 H^5}
\left(\frac{1}{ H^2}+\frac{7 b_3 }{ R_1^2}\right) \delta ^4
+\mathcal{O}(\delta^6)\,,
 \labell{cc2x24}
 \eea
where we have Taylor expanded the integrand in the second line and in
the third line, we have used \eqref{cc2x15} and \eqref{cc2x19x1} to
find the leading order divergence. Here we don't have any term
of order $1/\delta$ and of $\log(\delta)/\delta$, which come
consecutively from divergences of order $\log(\delta)$ and
$\log(\delta)^2$ in $I_1'$.

Similar to $I_1'$, we can take a derivative of $I_2'$ with respect to
$\delta$ and use \eqref{cc2x19x1} to find
 \be
{d I_2' \ov d \delta} \,=\, \left(1-\frac{b_3 H^2}{R_1^2}\right){H \ov
\delta ^4}-\frac{1}{h_0 \delta ^3}-\frac{1+88 b_3 }{56 R_1^2} {H \ov
\delta^2} -\frac{4+72 b_3}{189 \, R_1^2 H} + \mathcal{O}(\delta)\,.
 \labell{cc2x25}
 \ee
So we find that $I_1$ doesn't have any logarithmic divergence. We can
use similar steps to find
 \bea
{d I_3' \ov d \delta} &\,=\,& {1 \ov \delta} \int_{h_0}^{0} d h  \left[
\frac{5 \dot{h}{}^2+ 5 h^2 +4}{24 \dot{h}  h^2 \sqrt{1 + h^2 +
\dot{h}{}^2}} + {5 \ov 24 h^2}\right] -  {K_5^2 \ov
H^5}\left(\frac{1}{16 H^2} + \frac{7 b_3}{16 R_1^2}\right) \delta ^5 +
\mathcal{O}(\delta^6)
 \labell{cc2x26} \\
{d I_4' \ov d \delta} &\,=\,& \frac{5 H}{24 \delta ^2}
\left(1-\frac{b_3 H^2}{R_1^2}\right)-\frac{5}{24 h_0} {1 \ov \delta} +
\mathcal{O}(\delta^0)
 \labell{cc2x27}
 \eea
We can also use the same procedure on the boundary term and find that
 \bea
{d \ov d \delta} \int_{z_m}^\delta {dz \ov z} \frac{g_3 \dot{h}}{ \sqrt{1 + h^2 + \dot{h}{}^2}} \bigg|_{\theta=\Omega-\epsilon} & \,=\, & \frac{b_3 H^3}{\delta ^4} \left(1-\frac{3 b_3 H^2}{R_1^2}\right) + \frac{\left(1+88 b_3\right) H \left(-4 b_3 H^2+R_1^2\right)}{56 R_1^2 \delta ^2} \nonumber \\
& \ & +\left(\frac{4+72 b_3}{189 H}-\frac{\left(27+16 b_3
\left(521+17100 b_3\right)\right) H}{84672
R_1^2}\right)+\mathcal{O}(\delta)\,.
 \labell{cc2x28}
 \eea
So from \eqref{cc2x24} - \eqref{cc2x28}, we can find the logarithmic
divergence in the EE for $k\times S^2$ geometry:
 \be
S^{log}_5 \big|_{k\times S^2} \,=\,{16\, \pi^2 L^4 \ov \lp^4}\left(- \int_{h_0}^{0} d h
\left[ \frac{5 \dot{h}{}^2+ 5 h^2 +4}{24 \dot{h}  h^2 \sqrt{1 + h^2 +
\dot{h}{}^2}} + {5 \ov 24 h^2}\right] +\frac{5}{24 h_0} \right)
\log(\delta) \,.
\labell{nnx4}
 \ee
From \eqref{cc2x24} - \eqref{cc2x28}, we also notice a divergence of order $1/\delta^2$ in EE which does not appear in EE for smooth entangling surfaces. Note that such a term also appeared in EE for $k\times R^m$ in \eqref{csflatx11}. We further notice that logarithmic contribution in \eqref{nnx4} is of next to the leading order for large $R_1$. As we are
working in odd dimensional spacetime, there is no logarithmic
contribution from the surface itself.

Now we turn to our next example. Having seen the logarithmic
contribution from a even dimensional curved locus, now we consider odd
dimensional locus. We will calculate the entanglement entropy for
$k\times S^3$ geometry in CFT on $R^3\times S^3$. For this case, the
locus is $S^3$ and we will see that there will be no log contribution
from the singularity. However, as the CFT is in even dimensional
spacetime, we should be getting a logarithmic contribution coming from
the entangling surface. In this case, the metric for the dual geometry
is given by
 \be
ds^2 \,=\, {L^2 \ov z^2} \left( dz^2 + f_1(z)\left(dt^2 + d\rho^2 +
\rho^2 d\theta^2 \right) + f_2(z) R_1^2 \,d\Omega_3^2 \right)\,,
 \labell{cc3x1}
 \ee
where $d\Omega_3^2=d\xi_0^2+\sin^2(\xi_0) d\xi_1^2+\sin^2(\xi_0)
\sin^2(\xi_1) d\xi_2^2$ is the unit three-sphere and $f_1$ and $f_2$
are following
 \bea
& f_1 & \,= \, 1 + {3 z^2 \ov 20 R_1^2} + {69 z^4 \ov 1600 R_1^4}+ \dots \,, \nonumber \\
& f_2 & \,= \, 1 - {7 z^2 \ov 20 R_1^2} - {11 z^4 \ov 1600 R_1^4} +
\dots \,.
 \labell{cc3x2}
 \eea
Once again we choose the induced coordinates on the minimal area
surface to be $(z,\theta,\xi_0,\xi_1,\xi_2)$ and assume
$\rho=\rho(z,\theta)$. Then the induced metric over the minimal area
surface will be given by
 \be
h\,=\,\left[
 \begin{array}{c c c c c}
 {L^2 \ov z^2}\left( f_1 \rho'{}^2 + 1 \right) & {L^2 f_1\ov z^2} \drh \rh'  & \\
 {L^2 f_1 \ov z^2} \drh \rh'  & {L^2 f_1 \ov z^2} \left( \drh^2 + \rh^2  \right) & \\
  &  & {L^2 f_2 \, R_1^2 \ov z^2} g_{ab}(S^3)  %\\
%  &  &  & {L^2 f_2 \,R_1^2 \ov z^2} \sin^2(\xi_0) & \\
%  &  &  &  & {L^2 f_2 \,R_1^2 \ov z^2} \sin^2(\xi_0) \sin^2(\xi_1)
\end{array}
\right]\,,
 \labell{cc3x3}
 \ee
where $\drh=\partial_\theta \rh$, $\rh'=\partial_z \rh$ and now the
entanglement entropy is given by
 \be
S_6 \big|_{k\times S^3} \,=\, {4\, \pi^3 L^5 R_1^3 \ov \lp^5} \int dz \, d\theta \,
{\sqrt{f_1 f_2^3} \ov z^5} \sqrt{ \drh^2 + \rh^2(1 + f_1 \rh'{}^2)} \,.
 \labell{cc3x4}
 \ee
We can easily find the equation of motion for $\rho(z,\theta)$ to be
 \bea
&0 \;=\; & 2 z f_1 f_2 \rh \left(\rho^2+\dot{\rho }^2\right) \rh''+ 2 f_2 \rh \left(z + z f_1 \rho'{}^2\right) \ddot{\rh} -4 z f_1 f_2 \rh \drh \rho ' \drh'  \labell{cc3x5} \\
& \ & + 3 z f_1 \rh f_2' \rh' \left(\dot{\rho}^2+\rh^2 \left(1+f_1 \rh'{}^2 \right)\right) - f_2 \Big(4 z \dot{\rho}^2 + \rh \left(10 f_1 - 3 z f_1' \right) \dot{\rho}^2 \rho' + 2 \rh^2 \left(z + z f_1 \rh'{}^2 \right) \nonumber \\
& & - \rh^3 \rh' \left(3 z f_1'- 10 f_1^2 \rh'{}^2 + 2 f_1 \left(-5 + z
f_1' \rh'{}^2 \right)\right)\Big)\,, \nonumber
 \eea
where $\ddot{\rh}=\partial_\theta^2 \rho$, $\rh''=\partial_z^2 \rh$ and
$\drh'=\partial_z \partial_\theta \rho$. Once again, we can write
$\rho=\rho_0+\rho_1/R_1^2$ and keep only the leading order correction
to the entanglement entropy. We can further series expand the equation
of motion \eqref{cc3x5} to get equation of motion for $\rho_0$ and
$\rho_1$. We find that, using these equations of motion and integration
by parts, the entropy functional can be simplified to
 \bea
S_6 \big|_{k\times S^3} & \,=\, &{4\, \pi^3 L^5 R_1^3 \ov \lp^5} \int_{z_m}^{\delta} dz \int_{-\Omega+\epsilon}^{\Omega-\epsilon} d\theta  \Bigg[ \frac{\sqrt{\dot{\rho }_0^2+ \rho_0^2 \left(1 + \rh_0'{}^2\right)}}{z^5} - \frac{3 \left(6 \dot{\rho}_0^2 + \rho_0^2 \left(6 + 5 \rho_0'{}^2 \right)\right)}{40 R_1^2 z^3 \sqrt{\dot{\rho }_0^2 + \rho_0^2 \left(1 + \rho_0'{}^2\right)}} \nonumber \\
& & \qquad + {\partial \ov \partial \theta}\left( \frac{\drh_0
\rh_1}{R_1^2 z^5 \sqrt{\dot{\rho }_0^2 + \rho_0^2 \left(1+ \rho_0'{}^2
\right)}} \right) + {\partial \ov \partial z} \left( \frac{\rho_0^2
\rho _0' \rh_1}{R_1^2 z^5 \sqrt{\dot{\rho }_0^2 + \rh_0^2 \left(1 +
\rho _0'{}^2\right)}}  \right) \Bigg]\,.
 \labell{cc3x6}
 \eea
Further, we can substitute $\rh_0 = z/h(\theta)$ and $\rh_1=z^3
g_3(\theta)$. The equations of motion for $h$ and $g_3$ from
\eqref{cc3x5} becomes
 \bea
h \left(1 + h^2\right) \ddot{h} + 5 \dot{h}{}^2 + \left( 1 + h^2 \right) \left( 5 + h^2 \right) & \,=\,&0 \,,  \labell{cc3x7} \\
h^2 \left(1+h^2\right){}^2 \ddot{g}_3 + 2 h \left(9+11 h^2+2 h^4\right) \dot{h} \dot{g}_3 \qquad \qquad \qquad \qquad \qquad & & \nonumber \\
- g_3 \left(25+45 h^2+21 h^4+h^6 - \left(27 - h^2 + 2 h^4\right)
\dot{h}{}^2\right) &\,=\,& \mathcal{S}_1\,,
 \labell{cc3x8}
 \eea
where source terms for $g_3$ are
 \be
\mathcal{S}_1 \,=\, {3 \left(10+19 h^2+9 h^4+\left(4+9 h^2\right)
\dot{h}^2\right)  \ov 20 h} \,.
 \labell{cc3x9}
 \ee
Here, as in \eqref{csflatx6x2}, we can integrate the equation of motion
for $h$ once and write it as
 \be
K_6  \,=\, \frac{\left(1+h^2\right)^{5/2}}{
h^{5}\sqrt{1+h^2+\dot{h}{}^2}}\,,
 \labell{cc3x10}
 \ee
where $K_6$ is a constant and we can relate it to $h_0$ by using
$\dot{h}(0)=0$. We can further simplify \eqref{cc3x6} to
 \bea
S_6 \big|_{k\times S^3} &\,=\,& {8\, \pi^3 L^5 R_1^3 \ov \lp^5} \left(I_1 +{I_2 \ov
R_1^2} + {I_3 \ov R_1^2} + {I_4 \ov R_1^2} \right)\,,  \labell{cc3x11}
 \eea
where
 \bea
I_1 &\,=\,& \int_{z_m}^{\delta} dz \int_{h_0}^{h_{1c}} {d h \ov \dot{h}}  \frac{\sqrt{1 + h^2 + \dot{h}^2}}{z^4 h^2} \,, \\
I_2 &\,=\,& -\int_{z_m}^{\delta} dz \int_{h_0}^{h_{1c}} {d h \ov \dot{h}} \frac{3 \left(5+6 h^2+6 \dot{h}{}^2\right)}{40 R_1^2 z^2 h^2 \sqrt{1+h^2+\dot{h}{}^2}} \,, \\
I_3 &\,=\,& -\int_{z_m}^\delta {dz \ov z^2} \frac{g_3 \dot{h}}{ \sqrt{1 + h^2 + \dot{h}{}^2}} \bigg|_{\Omega-\epsilon} \,, \\
I_4 &\,=\,& -\int_{z_m}^{\delta} {dz \ov z^2} \int_{h_0}^{h_{1c}} {d h
\ov \dot{h}} \frac{g_3}{h \sqrt{1+h^2 + \dot{h}{}^2}}\,.
 \eea
Note that the last term $I_4$ is not being integrated over $z$. That's
because $h_{1c}(z)$ appears in the limit of $\theta$ integral. Now, we
can use \eqref{cc3x10} in \eqref{cc3x8} to find following equation of
motion for $g_3$ in terms of $h$:
 \bea
0&\,=\,& 20 h^3\left(1+h^2\right){}^2 \left( h^{10}  K_6^2 - \left(1+h^2\right){}^4\right)\ddot{g}_3 + 20 h^2 (1+h^2)\Big( h^{10} \left(18+5 h^2\right) K_6^2 \nonumber \\
&& - \left(1+h^2\right){}^5 \left(13+4 h^2\right)\Big) \dot{g}_3 + 20 h \left(h^{10} \left(52+19 h^2+3 h^4\right) K_6^2 - \left(1+h^2\right){}^4 \left(27-h^2+2 h^4\right) \right) g_3 \nonumber \\
&&+ 3 \left(1+h^2\right){}^4 \left(4+9 h^2\right)+18 h^{10} K_6^2\,.
 \labell{cc3x12}
 \eea
This equation can be solved perturbatively for small $h$ near the
asymptotic boundary. We find that this second order equation of motion
has two different solutions which go like $1/h^3$ and $1/h^9$. As
$\rho$ is finite at the boundary, the solution can go only with power
$1/h^3$. So the solution near the boundary becomes
 \be
g_3(h)\,=\, \frac{b_3}{h^3} + \frac{3+140 b_3}{80 h} + \frac{3+36
b_3}{64}  h - \frac{1+12 b_3}{96} h^3 + \frac{5+60b_3
}{1536}h^5+\dots\,,
 \labell{cc3x13}
 \ee
where constant $b_3$ is fixed by the condition that $\dot{g}_3(h_0)=0$.
We use this in $\rho=z/h +z^3 g_3/R_1^2$ and evaluate it at $z=\delta$, $\rho=H$ and $h=h_{1c}(\delta)$. By inverting the relation
and keeping only leading order terms in $R_1$, we find
 \be
h_{1c}(\delta) \,=\, \left(\frac{1}{H}+\frac{b_3 H}{R_1^2}\right)
\delta +\frac{\left(3+140 b_3\right) \delta ^3}{80 H
R_1^2}+\frac{\left(3+36 b_3\right) \delta ^5}{64 H^3
R_1^2}+\mathcal{O}(\delta^6) \,.
 \labell{cc3x14}
 \ee
Now we use this to separate logarithmic divergence. Using \eqref{cc3x10}, we find that the integrands in $I_1$, $I_2$,
$I_3$ and $I_4$ are of following form for small $h$:
 \bea
\frac{\sqrt{1 + h^2 + \dot{h}^2}}{\dot{h} h^2} &\,\sim\,& -\frac{1}{h^2}-\frac{1}{2} K_6^2 h^8+\mathcal{O}(h^9)\,, \nonumber \\
-\frac{3 \left(5+6 h^2+6 \dot{h}{}^2\right)}{40 \dot{h} h^2 \sqrt{1+h^2+\dot{h}{}^2}}   &\,\sim\,& \frac{9}{20 h^2}+\frac{3}{20} K_6^2 h^8+ \mathcal{O}(h^9) \nonumber \\
-\frac{g_3 \dot{h}}{\sqrt{1+h^2+\left(\dot{h}\right){}^2}} &\,\sim \,& \frac{b_3}{h^3}+\frac{3+ 140 b_3}{80 h}+\frac{3}{64} \left(1+12 b_3\right) h - \frac{1}{96} \left(1+12 b_3\right) h^3 + \mathcal{O}(h^5) \nonumber \\
\frac{g_3}{\dot{h} h \sqrt{1+h^2 + \dot{h}{}^2}} &\,\sim\,& b_3 K_6^2
h^6 +\mathcal{O}(h^7)\,.
 \labell{cc3x15}
 \eea
Using these relations, now we can make the integral over $h$ finite and
organize the terms in following form
 \bea
I_1 &\,=\,& \int_{z_m}^{\delta} {dz \ov z^4} \int_{h_0}^{h_{1c}} d h  \left( \frac{\sqrt{1 + h^2 + \dot{h}^2}}{ \dot{h} h^2} +{1 \ov h^2}\right)+\int_{z_m}^{\delta} {dz \ov z^4} \left({1\ov h_{1c}(z)} - {1 \ov h_0} \right) \nonumber \\
&\,=\,& I_1' +I_2' \,,   \labell{cc3x16} \\
I_2 &\,=\,& - \int_{z_m}^{\delta} {dz \ov z^2} \int_{h_0}^{h_{1c}} d h  \left( \frac{3 \left(5+6 h^2+6 \dot{h}{}^2\right)}{40 h^2 \sqrt{1+h^2+\dot{h}{}^2}} + {9 \ov 20 h^2}\right) - {9 \ov 20} \int_{z_m}^{\delta} {dz \ov z^2} \left({1\ov h_{1c}(z)} - {1 \ov h_0} \right) \nonumber \\
& \,=\, & I_3' +I_4' \,.
 \labell{cc3x17}
 \eea
In \eqref{cc3x16}, $I_1'$ and $I_2'$ represent the first and second
integral in the expression of $I_1$. Similarly in \eqref{cc3x17}, $I_3'$
and $I_4'$ are the first and second integral in expression for $I_2$.
We can further take derivative of $I'$'s with respect to $\delta$ and
then Taylor expand the integrals to get
 \bea
{d I_1' \ov d \delta} &\,=\,& {1 \ov \delta^4} \int_{h_0}^{0} d h \frac{\sqrt{1 + h^2 + \dot{h}^2}}{ \dot{h} h^2}-\frac{K_6^2 \left(9\, b_3 H^2+R_1^2\right) }{2\, H^9 R_1^2} \delta ^5 + \mathcal{O}(\delta^6)\,,   \labell{cc3x18}  \\
{d I_2' \ov d \delta} &\,=\,& {H \ov \delta^5} \left( 1 -\frac{b_3\, H^2}{R_1^2} \right) - \frac{1}{h_0 \delta^4} - \frac{\left(3+140\, b_3\right) H}{80\, R_1^2 \delta ^3} - \frac{3+36\, b_3}{64\, H\, R_1^2 \delta } + \mathcal{O}(\delta^0) \,,  \labell{cc3x19} \\
{d I_3' \ov d \delta} &\,=\,& - {1 \ov \delta^2} \int_{h_0}^{0} d h  \left( \frac{3 \left(5+6\, h^2+6\, \dot{h}{}^2\right)}{40\, h^2 \sqrt{1+h^2+\dot{h}^2}} \right) + \frac{3\, K_6^2 \left(9\, b_3 H^2 + R_1^2\right) \delta^7}{20\, H^9 R_1^2} + \mathcal{O}(\delta^{10}) \,,  \labell{cc3x20} \\
{d I_4' \ov d \delta} &\,=\,& -\frac{9\, H}{20\, \delta^3}
\left(1-\frac{b_3 H^2}{R_1^2} \right) + \frac{9}{20\, h_0 \delta ^2} +
\frac{9 \left(3+140\, b_3\right) H}{1600\, R_1^2 \delta } +
\mathcal{O}(\delta) \,.  \labell{cc3x21}
 \eea
Also, we can simplify the boundary terms in a similar fashion and find
that
 \bea
{dI_3 \ov d\delta} &\,=\,&\left(1-\frac{3\, b_3 H^2}{R_1^2}\right) \frac{b_3 H^3}{\delta^5}  +\left(7\, b_3+\frac{3}{20}-\frac{b_3 \left(3+140\, b_3\right) H^2}{5\, R_1^2}\right) \frac{H}{4\,\delta ^3} \nonumber \\
&  & \qquad \qquad +\left( \frac{3+36\, b_3}{64\, H}-\frac{\left(9+1440\, b_3+26800\, b_3^2\right) H}{6400\, R_1^2} \right){1 \ov \delta} +\mathcal{O}(\delta^0)\,,  \labell{cc3x22} \\
{dI_4 \ov d\delta} &\,=\,& -{1 \ov \delta^2} \int_{h_0}^{0} d h
\frac{g_3}{h\, \dot{h} \sqrt{1+h^2 + \dot{h}^2}} + \mathcal{O}(\delta^5)\,.  \labell{cc3x23}
 \eea
Now using \eqref{cc3x11} and \eqref{cc3x16}-\eqref{cc3x23}, first we notice that there are new divergent terms of order $1/\delta^3$ and $1/\delta$ which does not arise in EE for smooth entangling surfaces in $d=6$. We also find
that the logarithmic divergence of the entanglement entropy are
  \be
S_6^{log} \big|_{k\times S^3} \,=\, {8\, \pi^3 L^5 R_1 \ov \lp^5}
\left(0+\mathcal{O}\left({1 \ov R_1^4}\ri)\ri) \log(\delta)\,.
  \labell{cc3x23a}
  \ee
Remarkably, the contribution to the logarithmic divergence at the
leading order is zero. At higher order, \ie at order
$\mathcal{O}\left({1 / R_1^4}\ri)$, there should be logarithmic
contribution coming from the bulk part of the entangling surface.
According to \cite{ent1}, the logarithmic contribution from
four-dimensional entangling surface will have the coefficient of the
form $\int dx^4 C$, where $C$ is a combination of various curvature
squared terms. These terms will contribute with $1/R_1^4$ and so does
the smooth part of the entangling surface. So up to leading order,
there is no logarithmic contribution from the singularity because the
locus is odd dimensional. We might ask if there can be any contribution
from the singularity alone at higher order in $1/R_1$. However, we
discard such a possibility. It is easy to see that all the higher order
terms will be of even powers of $1/R_1$ and to make the coefficient
dimensionless, the only other available scale will be $H$. As $H$ is
related to the size of the surface, all such contributions actually
result from the contribution from the bulk surface and not the
singularity alone. Hence, we expect that there is no contribution from
the singularity in this case.

Next, we see one more example where odd dimensional locus doesn't
contribute to logarithmic dimensions, in spite of non-zero curvature.
We are going to consider $k\times R^1\times S^2$ geometry. We consider
the background geometry for CFT to be $R^4\times S^2$. Then the bulk
metric for boundary $R^4 \times S^2$ is given by
 \be
ds^2 \,=\, {L^2 \ov z^2} \left( dz^2 + f_1(z)\left(dt^2 + d\rho^2 +
\rho^2 d\theta^2 + dx^2\right) + f_2(z) R_1^2 \, d\Omega_2^2 \right)\,,
 \labell{cc4x1}
 \ee
where $d\Omega_2$ is line element over $S^2$ and now $f_1$, $f_2$ become
 \bea
& f_1 & \,= \, 1 + {z^2 \ov 20 R_1^2} + {z^4 \ov 100 R_1^4}+ \dots \,, \nonumber \\
& f_2 & \,= \, 1 - { z^2 \ov 5 R_1^2} - {7 z^4 \ov 800 R_1^4} + \dots
\,.
 \labell{cc4x2}
 \eea
The $k\times R^1\times S^2$ geometry is given by $\theta \in [-\Omega,
\Omega]$, $x \in [-\infty, \infty]$ and $\rho \in [0,\infty]$. We put
IR cut-offs on $x$ and $\rho$ directions such that
$x\in[-\tilde{H}/2,\tilde{H}/2]$ and $\rho \in [\rho_m, H]$, where
$\rho_m$ is related to the UV cut-off $\delta$. We choose
$(z,\theta,x,\xi_0,\xi_1)$ as induced coordinates on the minimal area
surface with $\rho=\rho(z,\theta)$. Then, the induced metric over the
minimal area surface will become
 \be
h\,=\,\left[
 \begin{array}{c c c c c}
 {L^2 \ov z^2}\left( f_1 \rho'{}^2 + 1 \right) & {L^2 f_1\ov z^2} \drh \rh'  & & &\\
 {L^2 f_1 \ov z^2} \drh \rh'  & {L^2 f_1 \ov z^2} \left( \drh^2 + \rh^2  \right) & & & \\
  &  & {L^2 f_1 \ov z^2}  & & \\
  &  & &{L^2 f_2 \, R_1^2 \ov z^2} & \\
  &  &  & & {L^2 f_2 \,R_1^2 \ov z^2} \sin^2(\xi_0)
\end{array}
\right]\,,
 \labell{cc4x3}
 \ee
where $\drh=\partial_\theta \rh$, $\rh'=\partial_z \rh$ and now the entanglement entropy is given by
 \be
S_6 \big|_{k\times R^1 \times S^2} \,=\, {8\, \pi^2 L^5 R_1^2 \tilde{H} \ov \lp^5} \int dz \,
d\theta \, {f_1 f_2 \ov z^5} \sqrt{ \drh^2 + \rh^2(1 + f_1 \rh'{}^2)}
\,.
 \labell{cc4x4}
 \ee
Note that here we have already performed integration over $x$. In this
case, the equation of motion for $\rho(z,\theta)$ becomes
 \bea
&0 \;=\; & 2 z f_1 f_2 \rh \left(\rho^2+\dot{\rho }^2\right) \rh''+ 2 z f_2 \rh \left(1 + f_1 \rho'{}^2\right) \ddot{\rh} -2 z f_2 \left(\rho^2 + 2 \dot{\rho }^2\right) \drh'  \labell{cc4x5} \\
& \ & + 2 \rh \left(2 z f_2 f_1' + f_1 \left(-5 f_2 + z
f_2'\right)\right) \left(\rh^2 + \dot{\rho }^2\right) \rho' - 2 z f_1
f_2 \rh^2 \rh'{}^2 + f_1 \rho^3 \left(3 z f_2 f_1' + 2 f_1 \left(- 5
f_2 + z f_2'\right)\right) \rh'{}^3\,. \nonumber
 \eea
where $\ddot{\rh}=\partial_\theta^2 \rho$, $\rh''=\partial_z^2 \rh$ and
$\drh'=\partial_z \partial_\theta \rho$. Once again, we can write
$\rho=\rho_0+\rho_1/R_1^2$ and keep only the leading order correction
to the entanglement entropy. We can further series expand the equation
of motion \eqref{cc4x5} to get equation of motion for $\rho_0$ and
$\rho_1$. We find that, using these equations of motion and integration
by parts, the entropy functional can be simplified to
 \bea
S_6 \big|_{k\times R^1 \times S^2} & \,=\, &{8\, \pi^2 L^5 R_1^2 \tilde{H} \ov \lp^5} \int_{z_m}^{\delta} dz \int_{-\Omega+\epsilon}^{\Omega-\epsilon} d\theta  \Bigg[ \frac{\sqrt{\dot{\rho}_0^2 + \rho_0^2 \left(1 + \rh_0'{}^2\right)}}{z^5} - \frac{3 \left(6 \dot{\rho}_0^2 + \rho_0^2 \left(6 + 5 \rho_0'{}^2 \right)\right)}{40 R_1^2 z^3 \sqrt{\dot{\rho }_0^2 + \rho_0^2 \left(1 + \rho_0'{}^2\right)}} \nonumber \\
& & \qquad + {\partial \ov \partial \theta}\left( \frac{\drh_0
\rh_1}{R_1^2 z^5 \sqrt{\dot{\rho }_0^2 + \rho_0^2 \left(1+ \rho_0'{}^2
\right)}} \right) + {\partial \ov \partial z} \left( \frac{\rho_0^2
\rho _0' \rh_1}{R_1^2 z^5 \sqrt{\dot{\rho }_0^2 + \rh_0^2 \left(1 +
\rho _0'{}^2\right)}}  \right) \Bigg]\,.
 \labell{cc4x6}
 \eea
Further, we can substitute $\rh_0 = z/h(\theta)$ and $\rh_1=z^3
g_3(\theta)$. The equations of motion for $h$ and $g_3$ from
\eqref{cc4x5} becomes
 \bea
h \left(1 + h^2\right) \ddot{h} + 5 \dot{h}{}^2 + \left( 1 + h^2 \right) \left( 5 + h^2 \right) & \,=\,&0 \,,  \labell{cc4x7} \\
h^2 \left(1+h^2\right){}^2 \ddot{g}_3 + 2 h \left(9+11 h^2+2 h^4\right) \dot{h} \dot{g}_3 \qquad \qquad \qquad \qquad \qquad & & \nonumber \\
- g_3 \left(25+45 h^2+21 h^4+h^6 - \left(27 - h^2 + 2 h^4\right)
\dot{h}{}^2\right) &\,=\,& \mathcal{S}_1\,,
 \labell{cc4x8}
 \eea
where source terms for $g_3$ are
 \be
\mathcal{S}_1 \,=\, { \left(10+19 h^2+9 h^4+\left(4+9 h^2\right)
\left(\dot{h}\right){}^2\right)  \ov 20 h} \,.
 \labell{cc4x9}
 \ee
Note that homogeneous part of the equations of motion for $h$ and $g_3$
are same as the case of kink in $R^3\times S^3$. Similar to
\eqref{csflatx6x2}, we can further integrate the equation of motion for
$h$ to get a conserved quantity
 \be
K_6 \,=\, \frac{\left(1+h^2\right)^{5/2}}{
h^{5}\sqrt{1+h^2+\dot{h}{}^2}}\,.
 \labell{cc4x10}
 \ee
The simplified expression for the entanglement entropy then becomes
 \bea
S_6\big|_{k\times R^1 \times S^2} &\,=\,& {16\, \pi^2 L^5 \tilde{H} R_1^2 \ov \lp^5} \left(I_1
+{I_2 \ov R_1^2} + {I_3 \ov R_1^2} + {I_4 \ov R_1^2} \right)\,,
\labell{cc4x11}
 \eea
where
 \bea
I_1 &\,=\,& \int_{z_m}^{\delta} dz \int_{h_0}^{h_{1c}} {d h \ov \dot{h}} \frac{\sqrt{1 + h^2 + \dot{h}^2}}{z^4 h^2}\,, \\
I_2 &\,=\,& -\int_{z_m}^{\delta} dz \int_{h_0}^{h_{1c}} {d h \ov \dot{h}} \frac{ \left(5+6 h^2+6 \dot{h}{}^2\right)}{40 R_1^2 z^2 h^2 \sqrt{1+h^2+\dot{h}{}^2}} \,, \\
I_3 &\,=\,& - \int_{z_m}^\delta {dz \ov z^2} \frac{g_3 \dot{h}}{ \sqrt{1 + h^2 + \dot{h}{}^2}} \bigg|_{\Omega-\epsilon} \,, \\
I_4 &\,=\,& - \int_{z_m}^{\delta} {dz \ov z^2} \int_{h_0}^{h_{1c}} {d h
\ov \dot{h}} \frac{g_3}{h \sqrt{1+h^2 + \dot{h}{}^2}}\,.
 \eea
Now we can convert the equation of motion for $g_3(\theta)$ into
equation of motion for $g_3(h)$ and solve it perturbatively near the
boundary in the limit $h\to0$. We find that
 \be
g_3 \,=\, \frac{b_3}{h^3} + \frac{1+140 b_3}{80 h} + \frac{1}{64}
\left(1 + 36 b_3 \right) h - \frac{1}{288} \left(1 + 36 b_3\right) h^3
+ \frac{\left(5+180 b_3\right) h^5}{4608} + \mathcal{O}(h^5)\,,
 \labell{cc4x12}
 \ee
where constant $b_3$ is fixed by the condition $\dot{g}_3(h_0)=0$.
Using this in $\rho=z/h +z^3 g_3/R_1^2$ and evaluating at $z=\delta$,
$\rho=H$ and $h=h_{1c}(\delta)$, we find
 \be
h_{1c}(\delta) \,=\, \left(\frac{1}{H}+\frac{b_3 H}{R_1^2}\right)
\delta +\frac{\left(1+140 b_3\right) \delta^3}{80 H
R_1^2}+\frac{\left(1+36 b_3\right) \delta ^5}{64 H^3 R_1^2}
+\mathcal{O}(\delta^6) \,.
 \labell{cc4x13}
 \ee
Note that we have only kept the terms of order $1/R_1^2$ in the above
expansion. Now for small $h$, the integrands in $I_i$'s behave as
 \bea
\frac{\sqrt{1 + h^2 + \dot{h}^2}}{\dot{h} h^2} &\,\sim\,& -\frac{1}{h^2}-\frac{1}{2} K_6^2 h^8+\mathcal{O}(h^9)\,, \nonumber \\
-\frac{ \left(5+6 h^2+6 \dot{h}{}^2\right)}{40 \dot{h} h^2 \sqrt{1+h^2+\dot{h}{}^2}}   &\,\sim\,& \frac{3}{20 h^2}+\frac{1}{20} K_6^2 h^8+ \mathcal{O}(h^9) \nonumber \\
-\frac{g_3 \dot{h}}{\sqrt{1+h^2+\left(\dot{h}\right){}^2}} &\,\sim \,& \frac{b_3}{h^3}+\frac{1+ 140 b_3}{80 h}+\frac{ \left(1+36 b_3\right)}{64} h - \frac{ \left(1+36 b_3\right)}{288} h^3 + \mathcal{O}(h^5) \nonumber \\
\frac{g_3}{\dot{h} h \sqrt{1+h^2 + \dot{h}{}^2}} &\,\sim\,& b_3 K_6^2
h^6 +\mathcal{O}(h^7)\,.
 \labell{cc4x14}
 \eea
Using these, we can make $\theta$ integrals in $I_1$ and $I_2$ finite
by separating the divergences:
 \bea
I_1 &\,=\,& \int_{z_m}^{\delta} {dz \ov z^4} \int_{h_0}^{h_{1c}} d h  \left( \frac{\sqrt{1 + h^2 + \dot{h}^2}}{ \dot{h} h^2} +{1 \ov h^2}\right)+\int_{z_m}^{\delta} {dz \ov z^4} \left({1\ov h_{1c}(z)} - {1 \ov h_0} \right) \nonumber \\
&\,=\,& I_1' +I_2' \,,   \labell{cc4x15} \\
I_2 &\,=\,& - \int_{z_m}^{\delta} {dz \ov z^2} \int_{h_0}^{h_{1c}} d h  \left( \frac{ \left(5+6 h^2+6 \dot{h}{}^2\right)}{40 h^2 \sqrt{1+h^2+\dot{h}{}^2}} + {3 \ov 20 h^2}\right) - {3 \ov 20} \int_{z_m}^{\delta} {dz \ov z^2} \left({1\ov h_{1c}(z)} - {1 \ov h_0} \right) \nonumber \\
& \,=\, & I_3' +I_4' \,.
 \labell{cc4x16}
 \eea
In \eqref{cc4x15}, $I_1'$ and $I_2'$ are first and second integrals in
$I_1$ and similarly in \eqref{cc4x16}, $I_3'$ and $I_4'$ represent
first and second integrals in $I_2$. Now we can further differentiate
$I_i$'s with respect to $\delta$ and Taylor expand the expressions to
separate the divergences
 \bea
{d I_1' \ov d \delta} &\,=\,& {1 \ov \delta^4} \int_{h_0}^{0} d h \frac{\sqrt{1 + h^2 + \dot{h}^2}}{ \dot{h} h^2}-\frac{K_6^2 \left(9 b_3 H^2+R_1^2\right) }{2 H^9 R_1^2} \delta ^5 + \mathcal{O}(\delta^6)\,,   \labell{cc4x17}  \\
{d I_2' \ov d \delta} &\,=\,& {H \ov \delta^5} \left( 1 -\frac{b_3 H^2}{R_1^2} \right) - \frac{1}{h_0 \delta^4} - \frac{\left(1+140 b_3\right) H}{80 R_1^2 \delta ^3} - \frac{1 + 36 b_3}{64 H R_1^2 \delta } + \mathcal{O}(\delta^0) \,,  \labell{cc4x18} \\
{d I_3' \ov d \delta} &\,=\,& - {1 \ov \delta^2} \int_{h_0}^{0} d h  \left( \frac{ \left(5+6 h^2+6 \dot{h}{}^2\right)}{40 h^2 \sqrt{1+h^2+\dot{h}{}^2}} \right) + \frac{ K_6^2 \left(9 b_3 H^2 + R_1^2\right) \delta^7}{20 H^9 R_1^2} + \mathcal{O}(\delta^{10}) \,,  \labell{cc4x19} \\
{d I_4' \ov d \delta} &\,=\,& -\frac{3 H}{20 \delta^3}
\left(1-\frac{b_3 H^2}{R_1^2} \right) + \frac{3}{20 h_0 \delta ^2} +
\frac{3 \left(3+140 b_3\right) H}{1600 R_1^2 \delta } +
\mathcal{O}(\delta) \,.  \labell{cc4x20}
 \eea
Also, we can simplify the boundary terms in a similar fashion and find
that
 \bea
{dI_3 \ov d\delta} &\,=\,&\left(1-\frac{3 b_3 H^2}{R_1^2}\right) \frac{b_3 H^3}{\delta ^5}  + \left(1-\frac{4 b_3 H^2}{ R_1^2}\right) \frac{H(1+140 b_3)}{80\delta ^3} \nonumber \\
&  & \qquad \qquad +\left( \frac{1+36 b_3}{64 H}-\frac{\left(1+ 480 b_3+26800 b_3^2\right) H}{6400 R_1^2} \right){1 \ov \delta} +\mathcal{O}(\delta^0)\,,  \labell{cc4x21} \\
{dI_4 \ov d\delta} &\,=\,& -{1 \ov \delta^2} \int_{h_0}^{0} d h
\frac{g_3}{h \dot{h} \sqrt{1+h^2 + \dot{h}{}^2}} +
\mathcal{O}(\delta^5)\,.  \labell{cc4x22}
 \eea
Using \eqref{cc4x11} and \eqref{cc4x15}-\eqref{cc4x22}, we find new divergence of order $1/\delta^3$ and $1/\delta$, and
the universal term in the entanglement entropy
 \be
S_6{}^{log} \big|_{k\times R^1 \times S^2} \,=\, {16\, \pi^2 L^5 \tilde{H} \ov \lp^5} \left(0
+\mathcal{O}\left({1 \ov R_1^2}\ri) \right) \log(\delta)\,.
 \labell{cc4x23}
 \ee
Once again, we find that leading contribution in logarithmic divergence
disappears, which is consistent with the fact that there is no
contribution from the singularity with odd dimensional locus. In the
next subsection, we will consider some more geometries with conical
singularities to push our hypothesis.

%%%%%%%%%%%%%%%%%%%%%%%%%%%%%%%%%%%%%%%%%%%%%%%%%%%%%%%%%%%%%%%%%%%%%%%%%%%%%
\subsubsection{Crease $c_n\times \Sigma$}   \label{conecurved}

In this section, we will calculate EE for conical singularities of the
form $c_n\times S^m$. We will mainly consider following singular
geometries: $c_1\times S^1$, $c_1\times S^2$, $c_1\times S^3$, $c_2\times S^1$ and
$c_2\times S^2$. The case with $\{n,m\}=\{1,1\}$ will turn out to be
trivial and it will be straightforward to see that there is no new
contribution to the $\log^2\!\delta$ or $\log$ divergence. For
$\{n,m\}=\{1,2\}$, we will find that there is a $\log^2\!\delta$
contribution in EE which had disappeared when the locus was taken to be
flat in \eqref{cnflatx18x1}. For $\{n,m\}=\{1,3\}$, we will see that
there is no $\log^2\!\delta$ contribution from the singularity as the
locus is odd dimensional. Finally for $\{n,m\}=\{2,1\}$ and $\{2,2\}$, we will find
that entanglement entropy contains a logarithmic divergence. In $\{n,m\}=\{2,1\}$, this contribution is actually coming from the smooth part of the surface as $d=6$ and logarithmic contribution from the trace anomaly is non-zero. In $\{n,m\}=\{2,2\}$, the logarithmic contribution comes from the singularity and this is
consistent with the idea that for even dimensional curved locus, the
singularity in odd $d$ will contribute through a logarithmic
divergence.

To begin with, we consider the simplest case with $m=1$. In this case,
the background geometry for the CFT is $R^4\times S^1$. Then, the dual
bulk geometry is given by
 \be
ds^2 \,=\, {L^2 \ov z^2} \left( dz^2 + f_1(z)\left(dt^2 + d\rho^2 +
\rho^2 d\theta^2 +\rho^2 \sin^2(\theta) d\phi^2 \right) + f_2(z) R_1^2
\, d\xi^0 \right)\,,
 \labell{cc5x1}
 \ee
where $f_1=1 + \mathcal{O}( 1/R_1^6)$ and $f_2=1 + \mathcal{O}(
1/R_1^6)$. For this bulk, we consider the singular surface given by
$\theta \in [0,\Omega]$, $\phi \in [0,2 \pi]$, $\rho \in [0,H]$ and $\xi^0\in[0,2\pi]$. For
the minimal area surface which gives us the entanglement entropy, we
assume that $\rho=\rho(z,\theta)$. Then, EE is given by
 \be
S_5\big|_{c_1\times S^1} \,=\, {4\, \pi^2 L^4 R_1 \ov \lp^4} \int dz \, d\theta \,
{\sin(\theta)  \rho  f_1 \sqrt{f_2} \ov z^4} \sqrt{ \drh^2 + \rh^2(1 +
f_1 \rh'{}^2)} \,,
 \labell{cc5x3}
 \ee
where $\drh=\partial_\theta \rh$ and $\rh'=\partial_z \rh$. Using this,
we can find the equation of motion for $\rho(z,\theta)$ to be
 \bea
&0\,=\,& 2 z f_1 f_2 \sin(\theta) \rh^2 \left(\rh^2 + \dot{\rho }^2\right) \rh'' + 2 z f_2 \sin(\theta) \rho^2 \left(1 + f_1 \rho'{}^2\right) \ddot{\rho} -4 z f_1 f_2 \sin(\theta) \rh^2 \dot{\rho} \rh' \drh'  \nonumber  \\
& & + 2 z f_1 f_2 \rho^2 \left(-2 \sin(\theta) \rho +\cos(\theta) \dot{\rho} \right) \rho'{}^2 + f_1 \sin(\theta) \rho^4 \left(3 z f_2 f_1' + f_1 \left(-8 f_2 + z f_2'\right)\right) \rho'{}^3 \nonumber \\
& & + 2 z f_2 \left(\cos(\theta) \dot{\rho } \left(\rh^2 + \dot{\rho}^2 \right)-\sin(\theta) \rh \left(2 \rh^2 + 3 \dot{\rho}^2 \right)\right) \nonumber \\
& & +\sin(\theta)\rho^2 \left(4 z f_2 f_1' + f_1 \left(-8 f_2 + z
f_2'\right)\right) \left(\rho^2 + \dot{\rho}^2 \right) \rho'\,. \labell{cc5x4}
 \eea
As the conical singularity has a one dimensional locus, we expect that
there will be no logarithmic contribution from the singularity. Now we
substitute $\rho = \rh_0 + \rh_1/R_1^2$ in \eqref{cc5x3} and find the
leading order correction to the entanglement entropy. However, when we
use the equation of motion for $\rho_0$ and simplify the entropy
functional, the contribution depending on the coefficients of leading
order terms of $f_1$ and $f_2$ will vanish because $f_1=1 +
\mathcal{O}( 1/R_1^6)$ and $f_2=1 + \mathcal{O}( 1/R_1^6)$. So we find
that simplified expression has following form
 \bea
S_5\big|_{c_1\times S^1} &\,=\,& {4\, \pi^2 L^4 R_1 \ov \lp^4} \int dz \, d\theta \bigg[ {\sin(\theta)  \rh_0  \ov z^4} \sqrt{ \drh_0^2 + \rh_0^2(1 + \rh_0'{}^2)} + {1 \ov R_1^2} {\partial \ov \partial \theta} \left( \frac{ \rho_0 \dot{\rho}_0 \rh_1 }{z^4 \sqrt{\drh_0^2 + \rho_0^2 \left(1 + \rh_0'{}^2\right)}} \right) \nonumber \\
& & \qquad + {1 \ov R_1^2} {\partial \ov \partial z} \left(
\frac{\rh_0^3 \rh_0' \rh_1}{z^4 \sqrt{\dot{\rho}_0^2+ \rho_0^2 \left(1
+ \rho_0'{}^2 \right)}} \right) \bigg]\,.
 \labell{cc5x5}
 \eea
Now we can insert the ansatz $\rh_0=z/h(\theta)$ and $\rh_1=z^3
g_3(\theta)$ in the functional and find that
  \bea
S_5\big|_{c_1\times S^1} &\,=\,& {4\, \pi^2 L^4 R_1 \ov \lp^4} \bigg[ \int_{z_m}^\delta {dz \ov z^2} \int_{h_0}^{h_{1c}(z)} {d h \ov \dot{h}} \frac{\sin(\theta) \sqrt{1+h^2 + \dot{h}{}^2 }}{h^3}  - {1 \ov R_1^2} \int_{z_m}^{\delta} dz \frac{g_3 \sin(\theta) \dot{h}}{h \sqrt{1 + h^2 + \dot{h}{}^2}} \bigg|_{h_{1c}(z)}  \nonumber \\
&  & \qquad + {1 \ov R_1^2} \int_{z_m}^\delta dz \int_{h_0}^{h_{1c}(z)}
{d h \ov \dot{h}} \frac{ g_3 \sin(\theta) }{h^2 \sqrt{1 + h^2 +
\dot{h}{}^2}} \bigg] \,.
  \labell{cc5x5a}
  \eea
Also from \eqref{cc5x4}, the equations of motion for $h$ and $g_3$
reduces to
 \bea
0 &\,=\,& \ddot{h} h \left(1+h^2\right) \sin(\theta)+\cos(\theta) h \dot{h}{}^3 +\left(4+h^2\right) \sin(\theta) \dot{h}{}^2 \qquad \qquad \qquad  \labell{cc5x6} \\
& & \qquad \qquad \qquad  \qquad \qquad  +\cos(\theta) \left(1+h^2\right) h \dot{h} + 2 \left(2 + 3 h^2 + h^4\right) \sin(\theta) \nonumber \\
0 &\,=\,& h^2 \left(1+h^2\right){}^2 \sin(\theta) \ddot{g}_3 + h \left(1+h^2\right) \left(2 \left(8+3 h^2\right) \sin(\theta) \dot{h}+\cos(\theta) h \left(1+h^2+ 3 \dot{h}{}^2\right)\right) \dot{g}_3  \labell{cc5x7} \\
&  & - \left( 2 \left(1+h^2\right) \left(10 + 8 h^2 + h^4\right)
\sin(\theta) + 3 \left(8+3 h^2+h^4\right) \sin(\theta) \dot{h}{}^2 + 2
\cos(\theta) h \left( 4 + h^2\right) \dot{h}{}^3 \right) g_3\,.
\nonumber
 \eea
Now, we find that $g_3$ is not sourced by $h$ and hence, a homogeneous
solution $g_3=0$ will be the exact solution for this case. This implies
that excluding the first term in the above expression, all the other
terms will vanish. As $g_3$ is zero, there will not be any new
contribution to the limits of the integrations either. This result is consistent with the idea that singularity will contribute in EE only if the locus is curved and even dimensional.

As a next example, we consider the singular geometry $c_1\times S^2$ in
CFT background $R^4 \times S^2$. For this case, the bulk metric is
given by
 \be
ds^2 \,=\, {L^2 \ov z^2} \left( dz^2 + f_1(z)\left(dt^2 + d\rho^2 +
\rho^2 d\theta^2 + \rho^2 \sin^2(\theta) d\phi^2 \right) + f_2(z) R_1^2
\, d\Omega_2^2 \right)\,,
 \labell{cc6x1}
 \ee
where $d\Omega_2^2=d\xi_0^2 + \sin^2(\xi_0) d\xi_0^2$ is metric over unit two-sphere and
 \bea
f_1 &\,=\,& 1 + {z^2 \ov 20 R_1^2} + {z^4 \ov 100 R_1^4}+\dots \,, \nonumber \\
f_2 &\,=\,& 1 - {z^2 \ov 5 R_1^2} - {7 z^4 \ov 800 R_1^4} + \dots\,.
 \labell{cc6x2}
 \eea
Now the entanglement entropy is given by
 \be
S_6 \big|_{c_1\times S^2} \,=\, {16\, \pi^3 L^5 R_1^2 \ov \lp^5}\int dz d\theta
\frac{ f_1 f_2 \sin(\theta) \rh \sqrt{\drh^2+\rho^2 \left(1+f_1
\rho'{}^2\right)}}{z^5}\,,
 \labell{cc6x4}
 \ee
and equation of motion for $\rho(z,\theta)$ becomes
 \bea
0 &\,=\,& 2 z f_1 f_2 \sin(\theta) \rho^2 \left(\rho^2+\dot{\rho }^2\right) \rh'' + 2 z f_2 \sin(\theta) \rho^2 \left(1+f_1 \rho'{}^2\right) \ddot{\rh} - 4\, z\, f_1 f_2 \sin(\theta) \rho^2 \dot{\rho }\, \rho'\,\dot{\rho}'  \nonumber \\
& & + 2 z \cos(\theta) f_2 \dot{\rho} \left(\dot{\rho}^2+\rho^2 \left(1+f_1 \rho'{}^2\right)\right) +\sin(\theta) \rho \Big( -2\, z f_2 (2 \,\rho^2+3 \dot{\rho}^2) \nonumber \\
&& +2\, \rho \left(2\, z\, f_2 f_1'+f_1 (-5\, f_2+z\, f_2')\right) (\rho^2+\dot{\rho }^2) \rho'- 4\, z\, f_1 f_2 \rho^2 \rho'{}^2 \nonumber \\
&& + f_1 \rho^3 \left(3\, z\, f_2 f_1'+2\, f_1 \left(-5\, f_2+z\, f_2'\right)\right) \rho'{}^3 \Big)\,.
 \labell{cc6x5}
 \eea
Now we can plug in the ansatz $\rho = \rho_0 + \rho_1/R_1^2$ in the
entropy functional and equation of motion. We can Taylor expand the
entropy functional in $R_1$ and keep the leading correction. In this
leading correction, we can use the integration by parts and the
equation of motion for $\rho_0$ to simplify the entropy functional to
 \bea
S_6\big|_{c_1\times S^2} &\,=\,& {16\, \pi^3 L^5 R_1^2 \ov \lp^5} \int_{z_m}^{\delta} dz \int_{0}^{\Omega-\epsilon} d\theta \bigg[ {\sin(\theta) \rho_0 \ov z^5}\sqrt{\dot{\rho}_0^2 + \rho_0^2 \left(1 + \rho_0'{}^2\right)} -\frac{\sin(\theta) \rho_0 \left(6 \dot{\rho }_0^2+\rho_0^2 \left(6+5 \rho _0'{}^2\right)\right)}{40 R_1^2 z^3 \sqrt{\dot{\rho }_0^2+\rho_0^2 \left(1+ \rho _0'{}^2\right)}} \nonumber \\
& & \qquad - {\partial \ov \partial z} \left(
\frac{\sin(\theta)\rho_0^3 \rho_0' \rh_1 }{z^5 R_1^2 \sqrt{\dot{\rho
}_0^2 + \rho_0^2 \left(1+\rho_0'{}^2\right)}} \right) \bigg] + {16\, \pi^3
L^5 \ov \lp^5} \int_{z_m}^{\delta} dz \frac{
\sin(\theta) \rho_0 \dot{\rho}_0 \rh_1}{z^5 \sqrt{\dot{\rho
}_0^2+\rho_0^2 \left(1+
\rho_0'{}^2\right)}}\bigg|_{\theta=\Omega-\epsilon}\,.
 \labell{cc6x6}
 \eea
We can further use $\rh_0=z/h(\theta)$ and $\rh_1=z^3 g_3(\theta)$ and
find that
 \bea
S_6\big|_{c_1\times S^2} &\,=\,& {16\, \pi^3 L^5 R_1^2  \ov \lp^5} \left[I_1 +
{I_2\ov R_1^2} + {I_3 \ov R_1^2} \right]\,,
 \labell{cc6x7}
 \eea
where
 \bea
I_1 &\,=\,& \int_{z_m}^{\delta}  {dz \ov z^3 } \int_{h_0}^{h_{1c}(z)} d h {\sin(\theta) \sqrt{1 + h^2 + \dot{h}{}^2}  \ov \dot{h}\, h^3}\,, \\
I_2 &\,=\,& - \int_{z_m}^{\delta} {dz \ov z} \int_{h_0}^{h_{1c}(z)} d h \frac{\sin(\theta) \left(5+6 h^2+6 \dot{h}{}^2\right)}{40 \, \dot{h}\, h^3 \sqrt{1 + h^2 + \dot{h}{}^2}} \,, \\
I_3 &\,=\,& - \int_{z_m}^{\delta} {dz \ov z} \frac{ g_3 \sin(\theta)
\dot{h}}{ h \sqrt{1+h^2+\dot{h}{}^2}} \bigg|_{h=h_{1c}(z)}\,.
 \eea
Note that term with derivative with respect to $z$ in \eqref{cc6x6} vanishes. Now using the above ansatz in \eqref{cc6x5}, we can also find the equations of motion for $h$ and $g_3$. For these equations, we can make a change of
variable from $\theta$ to $y=\sin(\theta)$ and find
 \bea
y(1-y^2)h (1+h^2) \ddot{h} +(1-y^2)^2 h \dot{h}{}^3 + y (1-y^2) (5+h^2) \dot{h}{}^2 \qquad \qquad \qquad \qquad & & \labell{cc6x8} \\
 + (1-2 y^2) h(1+h^2) \dot{h}+y (1+h^2) (5+2 h^2) &\,=\,& 0 \nonumber  \\
y (1-y^2) h^2 (1+h^2)^2 \ddot{g}_3 + h \Big(-\left(-1+2 y^2\right) h \left(1+h^2\right)^2 - 6 y \left(-1+y^2\right) \left(3+4 h^2+h^4\right) \dot{h}  & &\nonumber \\
+ 3 \left(-1+y^2\right)^2 h \left(1+h^2\right) \dot{h}^2\Big) \dot{g}_3 + \Big(-y \left(1+h^2\right) \left(25+21 h^2+2 h^4\right) \qquad & &  \labell{cc6x9} \\
-3 y \left(-1+y^2\right) \left(9+2 h^2+h^4\right) \dot{h}{}^2 + 2
\left(-1+y^2\right)^2 h \left(4+h^2\right) \dot{h}{}^3\Big) \dot{g}_3 &
\,=\,& \mathcal{S}_1 \,, \nonumber
 \eea
where $\mathcal{S}_1$ is the source terms and it is given by
 \be
\mathcal{S}_1 \,=\, \frac{y \left(1+h^2\right) \left(10+9 h^2\right)+ 4
y \left(1-y^2\right) \left(1+2 h^2\right) \dot{h}{}^2 +
\left(1-y^2\right)^2 h \dot{h}{}^3}{20 h}\,.
 \ee
Now to separate the logarithmic divergence, we want to find the asymptotic behavior of integrand in terms of $h$, where $h\to 0$. So
we invert equation \eqref{cc6x8} using
$\ddot{h}(y)=-(\left(1-y^2\right) y''+y y'^2)/y'{}^3$ and
$\dot{h}(y)=\sqrt{1-y^2}/y'$, where on the right hand side we have
$y=y(h)$ and $y'=dy/dh$. We can also change the independent variable in
\eqref{cc3x9} from $y$ to $h$. Apart from previous two relations, we
also use
 \bea
\dot{g}_3(\theta) &\,=\,& \dot{g}_3(h){\sqrt{1-y^2} \ov y'} \nonumber \\
\ddot{g}_3(\theta) &\,=\,& {\sqrt{1-y^2} \ov y'} {d \ov d h}\left(
{\dot{g}_3(h) \sqrt{1-y^2} \ov y'} \right) \,.
 \labell{cc6x10}
 \eea
Finally, we can rewrite \eqref{cc6x8} and \eqref{cc6x9} as
 \bea
h \left(1+h^2\right) y \left(1-y^2\right) y'' -\left(5+7 h^2+2 h^4\right) y y'{}^3 \qquad \qquad \qquad \qquad \qquad \qquad &&  \labell{cc6x11} \\
- h \left(1+h^2\right) \left(1-2 y^2\right) y'{}^2 - \left(5+h^2\right) y \left(1- y^2\right) y' - h \left(1-y^2\right)^2 &\,=\,&0 \nonumber \\
h^2 \left(1+h^2\right)^2 y \left(1-y^2\right) y' \ddot{g}_3 + h \left(1+h^2\right) \Big(2 h \left(1-y^2\right)^2 + \left(13+5 h^2\right) y \left(1-y^2\right) y' \qquad & &\nonumber \\
- \left(1+h^2\right) \left(5+2 h^2\right) y y'^3 \Big)\dot{g}_3 + \Big(2 h \left(4+h^2\right) \left(1-y^2\right)^2 + 3 \left(9+2 h^2+h^4\right) y \left(1-y^2\right) y' &&  \labell{cc6x12} \\
-\left(1+h^2\right) \left(25+21 h^2+2 h^4\right) y y'^3 \Big) g_3
&\,=\,& \mathcal{S}_2\,, \nonumber
 \eea
where now $\dot{g}_3=dg_3/dh$ and
 \be
\mathcal{S}_2 \,=\, \frac{h \left(1-y^2\right)^2 - 4 \left(1+2
h^2\right) y \left(1-y^2\right) y' - \left(1+h^2\right) \left(10+9
h^2\right) y y'{}^3}{20 h}\,.
 \ee
Now we can try to solve these equations perturbatively in terms of $h$
near the boundary, where $h$ is small. As $y=\sin(\Omega)$ at $h=0$, we
find the solution
 \be
y\,=\,\sin(\Omega) - \frac{\cos(\Omega) \cot(\Omega)}{8} h^2 +
\frac{1}{512} \left(2 \csc(\Omega) -7 \csc^3(\Omega) + 5 \sin(\Omega)
\right) h^4 + \dots\,.
 \labell{cc6x13}
 \ee
Using this in \eqref{cc6x12} and solving it perturbatively, we find
that
 \bea
g_3\,=\, -\frac{1}{20 h}+\frac{1+\csc^2(\Omega)}{96}
h\, \log\left(h\right)+b_3 h + \dots\,,
 \labell{cc6x14}
 \eea
where $b_3$ is a constant and it is fixed by the condition
$\dot{g}_3(0)=0$. Now, we evaluate the expression $\rh = z/h + z^3
g_3/R_1^2$ at $z=\delta$, $\rh=H$ and $h(z)=h_{1c}$. In this relation,
we use \eqref{cc6x14} and invert it to find
 \be
h_{1c}\,=\, \frac{\delta }{H}-\frac{\delta ^3}{20 H
R_1^2}+\frac{\left(96 b_3 +(1+\csc^2(\Omega))\log(\delta/H)  \right)
\delta ^5}{96 H^3 R_1^2}+\mathcal{O}(\delta^6)\,.
 \labell{cc6x15}
 \ee
Now, we can use \eqref{cc6x13} and \eqref{cc6x14} in integrands of
$I_1$ and $I_2$ to find their behavior near the boundary:
 \bea
{\sin(\theta) \sqrt{1 + h^2 + \dot{h}{}^2}  \ov \dot{h} h^3} &\,\sim \,& -\frac{\sin(\Omega)}{h^3}+\frac{3 \cos(\Omega) \cot(\Omega)}{32 h} - \frac{3 h (13-19 \cos(2 \Omega)) \cot^2(\Omega) \csc(\Omega)}{4096}   + \mathcal{O}(h^3) \nonumber \\
\frac{\sin(\theta) \left(5+6 h^2+6 \dot{h}{}^2\right)}{40  \dot{h} h^3
\sqrt{1 + h^2 + \dot{h}{}^2}}  &\,\sim \,& -\frac{3 \sin(\Omega)}{20
h^3}+\frac{\cos(\Omega) \cot(\Omega)}{64 h}-\frac{h (67-157 \cos(2
\Omega)) \cot^2(\Omega) \csc(\Omega)}{81920} + \mathcal{O}(h^3)\,.
 \labell{cc6x16}
 \eea
Using these, we can make the integrations in $I_1$ and $I_2$ finite and
write in the form
 \bea
I_1 &\,=\,& \int_{z_m}^{\delta} {dz \ov z^3} \int_{h_0}^{h_{1c}(z)} d h \bigg[ {\sin(\theta) \sqrt{1 + h^2 + \dot{h}{}^2}  \ov \dot{h} h^3} + \frac{\sin(\Omega)}{h^3} - \frac{3 \cos(\Omega) \cot(\Omega)}{32 h} \nonumber \\
& & \qquad \qquad \qquad \qquad \qquad \qquad + \frac{3 h (13-19 \cos(2 \Omega)) \cot^2(\Omega) \csc(\Omega)}{4096} \bigg] \nonumber \\
&& + \int_{z_m}^\delta {dz \ov z^3} \bigg( \frac{ \sin(\Omega)}{2} \left(-\frac{1}{h_{1c}^2}+\frac{1}{h_0^2}\right) + \frac{3}{32} \cos(\Omega) \cot(\Omega)\log(h_{1c}/h_0) \nonumber \\
&& \qquad \qquad \qquad \qquad \qquad + \frac{3\left(h_0^2-h_{1c}^2\right) (13 - 19 \cos(2 \Omega)) \cot^2(\Omega) \csc(\Omega)}{8192}\bigg) \nonumber \\
&\,=\,& I_1' +I_2'  \labell{cc6x17} \\
I_2 &\,=\,& - \int_{z_m}^{\delta} {dz \ov z} \int_{h_0}^{h_{1c}(z)} d h \bigg[\frac{\sin(\theta) \left(5+6 h^2+6 \dot{h}{}^2\right)}{40 R_1^2 \dot{h} h^3 \sqrt{1 + h^2 + \dot{h}{}^2}} + \frac{3 \sin(\Omega)}{20 h^3} - \frac{\cos(\Omega) \cot(\Omega)}{64 h} \nonumber \\
&& \qquad \qquad \qquad \qquad \qquad \qquad + \frac{(67-157 \cos(2 \Omega)) \cot^2(\Omega) \csc(\Omega) h}{81920}  \bigg] \nonumber \\
& & + \int_{z_m}^\delta {dz \ov z} \bigg( \frac{3\sin(\Omega)}{40} \left(\frac{1}{h_0^2}-\frac{1}{h_{1c}^2}\right) + \frac{\cos(\Omega) \cot(\Omega)}{64}  \log(h_0/h_{1c}) \nonumber \\
&& \qquad \qquad \qquad \qquad -\frac{\left(h_0^2 + h_{1c}^2 \right) (67-157 \cos(2 \Omega)) \cot^2(\Omega) \csc(\Omega)}{163840} \bigg) \nonumber \\
&\,=\,& I_3' +I_4'  \labell{cc6x18}\,.
 \eea
In \eqref{cc6x17} and \eqref{cc6x18}, $I_1'$ and $I_2'$ are first and
second integrals in $I_1$ and $I_3'$ and $I_4'$ are first and second
integrals in $I_2$. Now we can take derivatives of $I_i'$'s with respect to $\delta$, and then Taylor expand the terms to find:
 \bea
{d I_1'\ov d \delta} &\,=\,& {1 \ov \delta^3}  \int_{h_0}^{0} d h \bigg[ {\sin(\theta) \sqrt{1 + h^2 + \dot{h}{}^2}  \ov \dot{h} h^3} + \frac{\sin(\Omega)}{h^3} - \frac{3 \cos(\Omega) \cot(\Omega)}{32 h}+ \frac{3 h (13-19 \cos(2 \Omega)) \cot^2(\Omega) \csc(\Omega)}{4096} \bigg] \nonumber \\
&& \qquad + \frac{3 \left(\csc(\Omega) +11 \csc^3(\Omega) + 3 \csc^5(\Omega) - 15 \sin(\Omega) \right)}{8192 H^4} \delta \log(\delta/H)+\dots\,,  \nonumber \\
{d I_2' \ov d \delta} &\,=\,& \frac{H^2 \sin(\Omega)}{2 \delta ^5}+\dots - \frac{\csc(\Omega)+\sin(\Omega)}{96 R_1^2} {\log(\delta/H) \ov \delta}+\mathcal{O}(1/\delta)\,, \nonumber \\
{d I_3' \ov d \delta} &\,=\,& -{1\ov \delta} \int_{h_0}^{0} d h \bigg[\frac{\sin(\theta) \left(5+6 h^2+6 \dot{h}{}^2\right)}{40 R_1^2 \dot{h} h^3 \sqrt{1 + h^2 + \dot{h}{}^2}} + \frac{3 \sin(\Omega)}{20 h^3} - \frac{\cos(\Omega) \cot(\Omega)}{64 h} \nonumber \\
&& \qquad \qquad \qquad \qquad \qquad + \frac{(67-157 \cos(2 \Omega)) \cot^2(\Omega) \csc(\Omega) h}{81920}  \bigg] +\mathcal{O}(\delta^3)\,, \nonumber \\
{d I_4' \ov d \delta} &\,=\,& -\frac{3 H^2 \sin(\Omega)}{40 \delta^3}
-\frac{\cos(\Omega) \cot(\Omega) }{64}{\log(\delta) \ov \delta} +
\mathcal{O}(1/\delta) \,, \nonumber \\
{d I_3 \ov d \delta} &\,=\,& -\frac{H^2 \sin(\Omega)}{20 \delta ^3}+\frac{(3 - \cos(2 \Omega)) \csc(\Omega)}{192} {\log(\delta) \ov \delta} + \mathcal{O}(1/\delta)\,.
 \labell{cc6x19}
 \eea
Now using \eqref{cc6x17}-\eqref{cc6x19} in \eqref{cc6x7}, we find that
the double $\log$ contribution in the EE becomes
 \be
S_6^{\log^2} \big|_{c_1\times S^2} \,=\, - { \pi^3 L^5 \cos(\Omega)
\cot(\Omega) \ov 8\, \lp^5}  \log(\delta)^2\,.
 \labell{cc6x20}
 \ee
Note that there are other contribution to the double $\log$ term but
they are of higher order in $R_1$. As the only other dimensionful
quantity in the problem is the IR cut-off $H$, these terms will be of
the form $\mathcal{O}(H^2/R_1^2)$. Interestingly, all such terms, which
scale with $H$ are contributions from the smooth part of the entangling
surface. Hence \eqref{cc6x20} is the complete contribution from the
singularity alone. In section \ref{coneflat}, we saw that there was no
such double logarithmic term when the locus of the singularity was
either flat or it is odd dimensional. Hence, \eqref{cc6x20} is also
consistent with the idea that similar to \eqref{suggest}, generically the contribution from the
singularity should be of the following form
\be
S_{univ}\,\sim \,  \int_\sigma d^{2m}y\,\sqrt{h}\, [\mathcal{R}^m]\, \log(\delta)^2 \,.
\labell{cc6x21}
\ee
Here $\sigma$ is the $2m$-dimensional locus of the singularity and $[\mathcal{R}^m]$ is the curvature invariants with $m$ powers of the curvatures.

Having seen the appearance of the $\log^2\!\delta$ divergence for even
dimensional curved locus, now we turn towards odd dimensional locus. So we consider EE for geometry $c_1\times S^3$. For this
case, the calculations proceed in a similar fashion and we find that,
near the boundary $y=\sin(\theta)$ and $g_3$ in terms of $h$ are given
by
 \bea
y(h) &\,=\,& \sin(\Omega) - \frac{1}{10} \cos(\Omega) \cot(\Omega) h^2 - \frac{(63 - 17 \cos(2 \Omega) ) \cot^2(\Omega) \csc(\Omega) h^4}{6000} \nonumber \\
& & \qquad \qquad \qquad +\frac{\left(373 \csc(\Omega) - 1853 \csc^3(\Omega) - 889 \csc^5(\Omega) + 2369 \sin(\Omega) \right) h^6}{450000} + \mathcal{O}(h^7)\,, \nonumber \\
g_3(h) &\,=\,& -\frac{1}{12 h} - \frac{h (39 - 11 \cos(2 \Omega) )
\csc^2(\Omega) }{2250} + b_3 \, h^2 + \mathcal{O}(h^3)\,,
 \labell{cc6x22}
 \eea
where $b_3$ is a constant and it is fixed by the condition that $g_3$
has a minimum at $\theta=0$. However, we will see that this constant
is zero. If we calculate EE, we find that there is no $\log^2\!\delta$
term but there is a $\log$ term in EE at the order next to the leading
order in $R_1$:
 \bea
S_7^{\log} \big|_{c_1\times S^3} &\,=\,& - {8\,\pi^4 L^6 R_1 \ov H \lp^6} b_3
\, \sin(\Omega)  \log( \delta)\,.
 \labell{cc6x23}
 \eea
Now, we are going to argue that this logarithmic term is not coming from the
singularity. We can see that, if we set $\Omega=\pi/2$, the conical
singularity in the entangling surface disappears. However, the
logarithmic term in \eqref{cc6x23} is still non-zero. Hence, this
contribution is actually coming from the non-singular part of the
entangling surface. Further, as we are considering the EE in odd
dimensional CFTs, there should not be any $\log$ term from the smooth part of the entangling surface. Hence, the
contribution \eqref{cc6x23} should be zero and this is only if $b_3=0$.
This result further assures that a singularity with odd dimensional locus does not contribute through a $\log$ or $\log^2\!\delta$ term.

We can further calculate EE for singularity $c_2\times S^1$ in $d=6$. In this case, the singularity has an odd dimensional, curved locus which does not contribute in the universal term. However, as $d$ is even, we find that the smooth part of the surface contribute through a log term and it is given by
 \be
S^{\log}_6\big|_{c_2\times S^1} \,=\, {\pi^3 L^5 R_1 \ov \lp^5} \frac{(7-9 \cos(2 \Omega)) \cot^2(\Omega)}{16\, H } \log(\delta/H)\,.
\labell{nnx2}
 \ee
Further, there is a new divergent term of the order $1/\delta$ in this
case. Note that this term matches with the logarithmic term in EE for
$c_2\times R^1$ in \eqref{nnx1}. Finally, we give results for the case
where a conical singularity in odd dimensions have a curved, even
dimensional locus. In background $R^5\times S^2$, we can consider the
singular geometry $c_2\times S^2$ which contains a new divergence of
order $1/\delta^2$. Further, in odd $d$ the smooth part of the
entangling geometry does not contribute through a logarithmic term.
However, a singularity will contribute through a $\log(\delta)$ term if
the locus of the singularity is even dimensional and curved. Precisely
this is what we see in EE for $c_2\times S^2$ and find following
universal term in EE
 \bea
S_7^{\log}\big|_{c_2\times S^2} &\,=\,& {32\,\pi^3 L^6\, \ov
\lp^6}\log(\delta) \bigg[-\int_0^{h_0} dh\left({\mathcal{L}_0\ov
\dot{h}}-{7\sin^2\Omega \ov 60\,h^4} +{\cos^2 \Omega \ov 25\, h^2}
\right)
 \nonumber \\
&& \qquad \qquad \quad \qquad \qquad  +{7\sin^2\Omega \ov
180\,h_0^3}-{\cos^2 \Omega \ov 25\, h_0} \bigg]\,,
 \labell{cc6x24}
 \eea
where $\dot{h}=dh/d\theta$ and
 \be
\mathcal{L}_0\,=\, - {\sin^2(\theta) \left(7\,\dot{h}^2 + 7\, h^2 +6
\right) \ov 60 \, h^4 \sqrt{\dot{h}^2 + h^2 +1}}\,.
 \labell{cc6x25}
 \ee
Note that similar to previous cases, $h$ is defined such that
$\rho=z/h(\theta)+z^3\,g_3(\theta)/R_1^2$ and it is the solution of
following equation of motion
 \be
h(1+h^2)\ddot{h} + 2\cot^2(\theta)\,h\,\dot{h}^3 + 2 \left(h^2+3\right)
\dot{h}^2 + 2 \cot(\theta) h_1 \left(1+h_1^2\right) \dot{h} + 3
\left(2+3 h^2+h^4\right) \,=\,0 \,,
 \ee
with $h_0=h(0)$. The $\log$ divergence \eqref{cc6x24} is a contribution
from the singularity and it is non-zero because the locus of the
singularity is even dimensional and curved. So the examples in this
section reaffirm that an extended singularity contributes in the
cut-off independent terms through $\log$ or $\log^2\!\delta$ terms only
if its locus is even dimensional and curved.

%%%%%%%%%%%%%%%%%%%%%%%%%%%%%%%%%%%%%%%%
%%%%%%%%%%%%%%%%%%%%%%%%%%%%%%%%%%%%%%%%
\section{Universal terms and the central charges}  \label{central1}

In previous sections, we calculated the EE for various surfaces and
found that singularity produces new $\log\delta$ and $\log^2\!\delta$
terms in the EE. As it has been seen that the regulator independent
coefficients contain central charges when considering smooth surfaces,
it is natural to ponder what function of central charges appears in
these new universal contributions arising from the singularities. In
the calculations above, we have been working with CFTs which are dual
to the Einstein gravity. For these CFTs, all the central charges are
equal and there is no way to distinguish these in the universal term of
the EE. It has long been known that to construct a holographic model
where the various central charges are distinct from one another, the
gravity action must include higher curvature interactions \cite{highc}.
In part, this motivated the recent holographic studies of Gauss-Bonnet
gravity \cite{lovel} --- for example, see \cite{renyi,EtasGB}. Hence we
apply this approach here as a first step towards determining the
dependence of the new universal contributions on the central charges of
the dual CFT. In particular, we will calculate EE for some singular
geometries in Gauss-Bonnet gravities below.

%%%%%%%%%%%%%%%%%%%%%%%%%%%%%%%%%%%%%%
\subsection{Singular embedding} \label{GB1}

In this section, we will discuss EE for cone geometry in $d=4,5,6$
dimensional CFTs. We will first calculate the EE for $d=4$
systematically and discuss the results for other cases.

For the Gauss-Bonnet gravity in $d+1$ dimensions, the action is given
by
 \be
I_d\,=\,\frac{1}{2\lp^{d-2}}\int d^{d+1}x \sqrt{-g}\left[ R +
{d(d-1)\ov L^2} + \frac{\lambda L^2}{(d-2)(d-3)} \X_4 \right]\,,
 \labell{gbx1}
 \ee
where
 \be
\X_4=R_{abcd}R^{abcd}-4R_{ab}R^{ab}+R^2\,
 \labell{gbx2}
 \ee
corresponds to the Euler density on a four-dimensional manifold (and
hence we must have $d\ge4$ here). We have introduced $L$ as a canonical
scale in the curvature-squared interaction so that strength of this
term is controlled by $\lambda$, a dimensionless coupling constant. We
may write metric for the AdS vacuum solution as
 \be
ds^2 \,=\, {\tL^2 \ov z^2}\left(dz^2+d\te^2+d\rho^2+\rho^2
\left(d\theta^2 + \rho^2 \sin^2(\theta) \, d\Omega^2_{d-3}
\right)\right)\,.
 \labell{gbx3}
 \ee
Here the AdS curvature scale $\tL$ is  related to canonical scale $L$
by the following relation
 \be
\tL^2=L^2/\fin \quad{\rm where}\ \
\fin=\frac{1-\sqrt{1-4\lambda}}{2\lambda} \,.
 \labell{gbx4}
 \ee
Now in Gauss-Bonnet gravity, the holographic EE is again determined by
a minimization problem over surfaces $m$ in the bulk geometry which
match the entangling surface $\Sigma$ at the asymptotic AdS$_{d+1}$
boundary. However, the functional that is minimized is now given by
\cite{ent1,ent2}
 \be
S_d\,=\, {2\pi \ov \lp^{d-1}} \int d^{d-1}x\sqrt{h} \left[1+ {2\lambda
L^2 \ov (d-2)(d-3)} \mathcal{R} \ri]\,,
 \labell{gbx2x1}
 \ee
where $\mathcal{R}$ is the Ricci scalar for the induced metric $h$.

The bulk theory can now be characterized in terms of two dimensionless
couplings, the ratio $\tL/\lp$ and the new coupling constant $\lambda$.
As a result, there are two distinct central charges which characterize
the boundary CFT dual to Gauss-Bonnet gravity. A convenient choice for
these which applies for any value of $d$ is, \eg see \cite{renyi}:
 \bea
\widetilde{C}_\mt{T} &\,=\,& {\pi^{d/2} \ov \Gamma(d/2)}\left({\tL\ov
\lp}\right)^{d-1}[1-2\lambda f_\infty]\,, \labell{ct}\\
a^*_d &\,=\,&{\pi^{d/2} \ov \Gamma(d/2)}\left({\tL\ov \lp}\right)^{d-1}
\left[1-2{d-1 \ov d-3} \lambda f_\infty \right]\,.
 \labell{adstar}
 \eea
The physical role of these charges in the boundary theory is as
follows: The central charge $\tilde{C}_T$ controls the leading
singularity of the two-point function of the stress tensor and $a^*_d$
appears as the universal coefficient in the entanglement entropy of a
sphere $S^{d-2}$ \cite{cthem1,cthem2}. The latter has also been shown
to satisfy a holographic c-theorem in arbitrary dimensions. Below we
will calculate the EE for various singular surfaces and would like to
see if the universal terms have some simple dependence on these central
charges. If we consider the case of $d=4$, the above expressions
simplify to
 \be
c\,=\,\widetilde{C}_\mt{T}=\pi^2{\tL^3 \ov \lp^3}(1-2\lambda
f_{\infty}) \quad \textrm{and} \quad a\,=\,a^*_4= \pi^2{\tL^3 \ov
\lp^3}(1-6\lambda f_{\infty})\,.
 \labell{central}
 \ee
Here, $c$ and $a$ are the standard central charges that appear in the
trace anomaly or entanglement entropy, \eg eq.~\reef{solo}, of the
four-dimensional boundary CFT.

Now for calculation of the holographic EE, the cone geometry is
defined by $\rh\in[0,H]$, $\theta \in[0,\Omega]$ and $\phi
\in[0,2\pi]$. We consider the induced coordinates to be
$(\rho,\theta,\phi)$ and radial coordinate $z=z(\rho,\theta)$. For this
case, the induced metric becomes
 \be
h \,=\,\left[
 \begin{array}{c c c}
 {\tL^2 \ov z^2}\left( 1+ (\partial_\rho z)^2  \right) & {\tL^2 \ov z^2} \partial_\rho z \partial_\theta z  & 0 \\
 {\tL^2 \ov z^2} \partial_\rho z \partial_\theta z & {\tL^2 \ov z^2}\left( \rho^2 + (\partial_\theta z)^2  \right) & 0 \\
 0 & 0 & {\tL^2 \ov z^2} \rho^2 \sin^2(\theta)
\end{array}
\right]\,.
 \labell{gbx6}
 \ee
For this metric, the expression for Ricci scalar $\mathcal{R}$ in
\eqref{gbx2x1} contains the terms like $\partial_\rho^2 z$ and
$\partial_\theta^2 z$. However, it is straightforward to see that the
equation of motion is still second order. This is because Gauss-Bonnet
term is topological in nature. Further, we impose the UV cutoff at
$z=\delta$ and define $\epsilon(\rho)$ such that at
$\theta=\Omega-\epsilon$, $z(\rh,\Omega-\epsilon)=\delta$. As the
background geometry has scaling symmetry and apart from $\rh$, there
are no dimensionful quantities in the problem, the solution for $z$
should be of the following form
 \be
z\,=\,\rho \, h(\theta)\,.
 \labell{gbx8}
 \ee
Here $h(\theta)$ is a function such that $h(\Omega)=0$ and
$\dot{h}(0)=0$. Also, the maximum value of $h(\theta)$ is $h(0)=h_0$.
By plugging this ansatz in equation of motion for $z(\rh,\theta)$,
which we get by applying the variational principle on entropy
functional \eqref{gbx2x1} with $d=4$, the equation of motion for
$h(\theta)$ turns out to be
 \bea
0 &\,=\,& h \left(1+h^2\right) \left(\left(1+h^2\right) \sin(\theta) \left(1+4 \lambda  f_{\infty }\right) + 6 \lambda  \cos(\theta) h f_{\infty } \dot{h}+ \sin(\theta) \left(1-2 \lambda  f_{\infty }\right) \dot{h}{}^2\right) \ddot{h} \nonumber \\
& & + \cos(\theta) h \left(1-2 \lambda  f_{\infty }\right) \dot{h}{}^5 - \left(3+h^2\right) \sin(\theta) \left(-1+2 \lambda  f_{\infty }\right) \dot{h}{}^4 \nonumber \\
& & + 2 \cos(\theta) h \left(1 + h^2 + \lambda  \left(1-2 h^2\right) f_{\infty }\right) \dot{h}{}^3 - 3 \left(1+h^2\right) \sin(\theta) \left(-2 - h^2 + 2 \lambda  \left(1 + h^2\right) f_{\infty }\right) \dot{h}{}^2 \nonumber \\
&& + \cos(\theta) h \left(1+h^2\right){}^2 \left(1+4 \lambda  f_{\infty
}\right) \dot{h}+ \left(1+h^2\right){}^2 \sin(\theta) \left(3+2 h^2
\left(1+\lambda  f_{\infty }\right)\right) \,.
 \labell{gbx9}
 \eea
Further, we can simplify the entropy functional using this equation of
motion and find that
 \be
S_4 \big|_{c_1} \,=\,{4\, \pi^2 \tL^3 \ov \lp^{3}} \int_{\delta/h_0}^{H}{d\rh
\ov \rh} \int_{h_0}^{\delta/\rh} dh {\sin(\theta) \mathcal{L}_1 \ov
\mathcal{L}_2}  \,,
 \labell{gbx10}
 \ee
where we have changed the integration from $\theta$ to $h$ and
 \bea
\mathcal{L}_1&\,=\,& \left(1+h^2+ \dot{h}{}^2\right) \Big(\sin(\theta) \left(1+h^2+ \dot{h}{}^2\right) -2 \lambda  f_{\infty } \Big( \left(h^2 \cos(\theta) \cot(\theta) +4 \sin(\theta)\right) \dot{h}{}^2 \nonumber \\
& & + 2 h^3 \cos(\theta) \dot{h} + \left(2+h^2\right){}^2 \sin(\theta) \Big) \Big) + 4 \lambda ^2 f_{\infty }^2  \Big( \left(h^2 \cos(\theta) \cot(\theta) + 3 \sin(\theta)\right) \dot{h}{}^4 \nonumber \\
& & - 2 h \left(3 - h^2\right) \cos(\theta) \dot{h}{}^3 + h^2 \left(2+h^2+2 \cos(2 \theta) \right) \csc(\theta) \dot{h}{}^2 + 2 h^3 \left(1+h^2\right) \cos(\theta) \dot{h} \nonumber \\
& & + h^4 \left(1+h^2\right) \sin(\theta) \Big) \,,  \labell{gbx11} \\
\mathcal{L}_2 &\,=\,& h^3 \dot{h} \sqrt{1+h^2+\dot{h}{}^2} \Big(
\sin(\theta) \left(1+h^2 + \dot{h}{}^2\right) + 2 \lambda  f_{\infty}
\left(3 h \cos(\theta) \dot{h} + \sin(\theta) \left(2+2 h^2 -
\dot{h}{}^2\right)\right)  \Big) \,. \nonumber
 \eea
Now we want to make $h$ integrand in \eqref{gbx10} finite. For that, we
define $y=\sin(\theta)$ and find $y$ and $\dot{h}$ in terms of $h$ near
the asymptotic boundary. For that, we convert \eqref{gbx9} into the
equation of motion for $y$ with independent variable $h$. Solving this
equation of motion perturbatively, we find that near the boundary
 \bea
y(h) \,=\, \sin(\Omega) - \frac{1}{4} h^2 \cos(\Omega) \cot(\Omega) +
\frac{1}{64} h^4 \log(h) (3 - \cos(2 \Omega)) \cot^2(\Omega)
\csc(\Omega) + c_0 h^4 + \dots \,,
 \labell{gbx12}
 \eea
where we have used the condition $y(0)=\sin(\Omega)$ and $c_0$ is a
constant which is fixed by the condition that $y'(h_0)=0$. We can now
assume $y=y(\theta)$ and $h=h(\theta)$ in \eqref{gbx12} and then invert
it to find
 \bea
\dot{h}(\theta) &\,=\,& -{2\tan(\theta) \ov h} - \frac{(3 - \cos(2 \Omega) ) \csc(2 \Omega)}{2}h \log(h) \nonumber \\
&& \qquad \qquad \qquad - {\sec(\Omega) \big(256 c_0 \tan^2(\theta) - 5
\cos(2 \Omega) \csc(\Omega) + 7 \big) \ov 16} h \dots \,.
 \labell{gbx12x1}
 \eea
Now using \eqref{gbx12} and \eqref{gbx12x1}, we can find that near the
boundary
 \bea
{\sin(\theta) \mathcal{L}_1 \ov \mathcal{L}_2} &\,\sim \,& -
\frac{\sin(\Omega) \left(1 - 6 \lambda  f_{\infty }\right)}{h^3} +
\frac{\cos(\Omega) \cot(\Omega) \left(1-2 \lambda  f_{\infty
}\right)}{8 h}+\mathcal{O}(h)\,,
 \labell{gbx13}
 \eea
which can be used to write the EE as
 \bea
S_4 \big|_{c_1} &\,=\,& {4\, \pi^2 \tL^3 \ov \lp^{3}} \bigg[ \frac{H^2 \sin(\Omega) \left(1-6 \lambda  f_{\infty }\right)}{4 \delta ^2} -\frac{1}{16} \cos(\Omega) \cot(\Omega) \left(1-2 \lambda  f_{\infty }\right) \log^2(\delta/H)  \labell{gbx14} \\
& & + \left(\frac{\sin(\Omega) \left(1-6 \lambda  f_{\infty }\right)}{2 h_0^2} + \frac{1}{8} \log(h_0) \cos(\Omega) \cot(\Omega) \left(1 - 2 \lambda  f_{\infty }\right) \right) \log(\delta/H) \nonumber \\
&& + \int_{\delta/h_0}^{H}{d\rh \ov \rh} \int_{h_0}^{\delta/\rh} dh
\left( {\sin(\theta) \mathcal{L}_1 \ov \mathcal{L}_2} +
\frac{\sin(\Omega) \left(1 - 6 \lambda  f_{\infty }\right)}{h^3} -
\frac{\cos(\Omega) \cot(\Omega) \left(1-2 \lambda  f_{\infty
}\right)}{8 h} \right) \bigg] \,.\nonumber
 \eea
In the last term, the $h$ integration is finite in the limit $\delta
\to 0$ and hence the leading order divergence is logarithmic. Hence, we
write
 \bea
S_4 \big|_{c_1} &\,=\,& {4\, \pi^2 \tL^3 \ov \lp^{3}} \bigg[ \frac{H^2 \sin(\Omega) \left(1-6 \lambda  f_{\infty }\right)}{4 \delta ^2} -\frac{1}{16} \cos(\Omega) \cot(\Omega) \left(1-2 \lambda  f_{\infty }\right) \log^2(\delta/H)  \labell{gbx15} \\
& & + \left(\frac{\sin(\Omega) \left(1-6 \lambda  f_{\infty }\right)}{2 h_0^2} + \frac{1}{8} \log(h_0) \cos(\Omega) \cot(\Omega) \left(1 - 2 \lambda  f_{\infty }\right) \right) \log(\delta/H) \nonumber \\
&& + \log(\delta/H) \int^{h_0}_{0} dh \left( {\sin(\theta)
\mathcal{L}_1 \ov \mathcal{L}_2} + \frac{\sin(\Omega) \left(1 - 6
\lambda  f_{\infty }\right)}{h^3} - \frac{\cos(\Omega) \cot(\Omega)
\left(1-2 \lambda  f_{\infty }\right)}{8 h} \right)
+\mathcal{O}(\delta^0)\bigg] \,.\nonumber
 \eea
Now we can compare the coefficient of the $\log^2\!\delta$ divergence
with the central charge \eqref{ct} and find that
 \bea
S_4^{\log^2} \big|_{c_1} &\,=\,& - {\tilde{C}_T \ov 4}\cos(\Omega)
\cot(\Omega) \log^2(\delta/H) \labell{d4univ} \\
&\,=\,& - {c \ov 4}\ \frac{\cos^2(\Omega)}{ \sin(\Omega)}\
\log^2(\delta/H)\,,
 \notag
 \eea
where the second equation follow from \eqref{central}. So we find that
for EE of the cone $c_1$ in $d=4$, the new universal term arising from
the singularity is directly proportional to the central charge
$\tilde{C}_T$ and has a relatively simple dependence on the opening
angle $\Omega$. We will return to discussing this result in more detail
in section \ref{dis}.

We can also calculate the EE for five-dimensional cone in Gauss-Bonnet
gravity. For this case, the action for bulk geometry is \eqref{gbx1}
with $d=5$ and canonical scale $L$ is related to $\tL$ by relation
\eqref{gbx4}. Now the complete expression of entanglement entropy is
given by \eqref{gbx16} and the universal term is
 \bea
S_5^{\log} \big|_{c_2} &\,=\,& - {8\,\pi^2 \tL^4 \ov \lp^4} \left(
\frac{\left(2\cos^2(\Omega) \left(2-7 \lambda  f_{\infty }\right) h_0^2
- 3 \sin^2(\Omega) \left(1 - 4 \lambda  f_{\infty }\right) \right) }{9
h_0^3} + \int^{h_0}_0 dh {\mathcal{L}_3 \ov \mathcal{L}_4 }
\right)\log(\delta/H)\,, \labell{houseGB3}
 \eea
where $\mathcal{L}_3$ and $\mathcal{L}_4$ are given by \eqref{gbx17}.
In this case, we can further compare the universal term with the
central charges \eqref{ct} and \eqref{adstar}. However, we observe from
expression of $\mathcal{L}_3$ and $\mathcal{L}_4$ in \eqref{gbx17} that
there are terms of order $\mathcal{O}(\lambda^2)$ in the universal
term. This implies that the expression \eqref{houseGB3} is not a simple
function of the central charges $\tilde{C}_T$ and $a^*_5$ and in
particular, it is not a linear function.

Further, we calculate the EE for cone in $d=6$ and find that the
regulator independent term is $\log^2\!\delta$ and it is given by
 \bea
S_6^{\log^2} \big|_{c_3} &\,=\,& {12\,\pi^3 \tL^5 \cos(\Omega)
\cot(\Omega) \ov \lp^5} \left( {93-190 \lambda f_{\infty} \ov 8192}- {
3-2 \lambda  f_{\infty } \ov 8192} \cos(2 \Omega) \ri) \log^2(\delta/H)
\,.
 \labell{gbx18x1}
 \eea
We have given the complete expression of EE in the appendix \ref{appa}
in eqn.~\eqref{gbx18}. In this expression, we would like to compare the
$\Omega$ independent coefficients with the central charges of the CFT.
So there are two terms, $(93-190 \lambda  f_{\infty})$ and $(3-2
\lambda  f_{\infty })$, which we can express in terms of $\tilde{C}_T$
and $a^*_6$. Using equations \eqref{ct} and \eqref{adstar}, we find
that the universal terms in EE for $c_3$ can be written as
 \bea
S_6^{\log^2} \big|_{c_3} &\,=\,& {3 \cos(\Omega) \cot(\Omega) \ov 1024}
\left[ \left(90\,\tilde{C}_T + 3\, a^*_6\right) - \left(
6\,\tilde{C}_T-3\, a^*_6 \right) \cos(2 \Omega) \ri]
\log^2(\delta/H)\,.
 \labell{d6central}
 \eea
Hence we see here that in $d=6$, the new universal term for $c_3$
depends linearly on both of the central charges, $\tilde{C}_T$ and
$a^*_6$, but it still has a relatively simple dependence on the opening
angle $\Omega$.

%%%%%%%%%%%%%%%%%%%%%%%%%%%%%%%%%%%%%%%%%%%%%%%%%%%%%%%%%%%%%
\subsection{Singularity with a curved locus}

In this section, we will repeat the calculation of EE for geometry
$k\times S^2$ in $d=5$ CFT, which is dual to the Gauss-Bonnet gravity.
For this case, the bulk action and EE are given by \eqref{gbx1} and
\eqref{gbx2x1}. We further consider the metric ansatz
 \be
ds^2\,=\, {\tL^2 \ov z^2} \left(dz^2 +f_1(z) \left(dt^2 + d\rh^2 +
\rh^2 \sin^2(\theta) \right) + R_1^2 f_2(z) d\Omega_2^2 \right)\,,
 \labell{gbx19}
 \ee
where $d\Omega_2$ is line element over the two-sphere. Further, similar to
\eqref{gbx4}, $\tL^2 = L^2/f_\infty$ where
$f_\infty=(1-\sqrt{1-4\lambda})/2\lambda$. Here $f_1$ and $f_2$ are
functions of $z$ and one can use the Fefferman-Grahm expansion to find
their values near the boundary.

Now, to calculate EE, we first use Fefferman-Grahm expansion and find
that
 \bea
f_1(z) & \,=\, & 1+{z^2\ov 12 R_1^2}+ \frac{z^4 (51-58 \lambda f_\infty)}{1728 R_1^4 (1-2 \lambda f_\infty)} +\dots\,, \nonumber \\
f_2(z) &\,=\,& 1-{z^2\ov 4 R_1^2} - \frac{z^4 (15-2 \lambda
f_\infty)}{576 R_1^4 (1 - 2 \lambda f_\infty)}+\dots\,.
 \labell{gbx20}
 \eea
Once again, we choose the parametrization $\rh=\rh(z,\theta)$ and
similar to \eqref{cc2x5}, we can find the induced metric for the
entropy functional. Here, we restrain to give the complete expression
for the entropy functional but the important point to note is that now
there are terms with higher derivatives, like $\rh''$, $\ddot{\rh}$ and
$\drh'$ in the entropy functional. However, in spite of this, the
equation of motion for $\rh(z,\theta)$ is still second order and that
is because of the topological nature of the Gauss-Bonnet terms. Now we
can simplify the entropy function by inserting the ansatz $\rh = \rh_0
+ \rh_1/R_1^2$ and then series expending it for large $R_1$. We find
that up to the leading order, the EE is given by
 \bea
S_5 \big|_{k\times S^2} &\,=\,& {16\,\pi^2 \tL^4 R_1^2 \ov \lp^4} \int_{z_m}^{\delta} dz \int_0^{\Omega-\epsilon} d\theta \bigg(\mathcal{L}_0(\rh_0) + {1\ov R_1^2} \Big(\mathcal{L}_N(\rh_0) + \mathcal{L}_f(\rh_0,\mathbf{c}_1,\mathbf{c}_2) + \rh_1' \mathcal{L}_1(\rh_0) \nonumber \\
&& \qquad \qquad +\dr\dot{h}\mathcal{L}_2(\rh_0) +
\ddot{\rh}_1\mathcal{L}_3(\rh_0) + \rh_1 \mathcal{L}_4(\rh_0) + \drh_1
\mathcal{L}_5(\rh_0) +\rh_1 \mathcal{L}_6(\rh_0) \Big) \bigg)\,,
\labell{gbx21}
 \eea
where $\mathcal{L}_0(\rh_0)$ is the term which comes from the limit
$R_1\to \infty$ in the original lagrangian. The term
$\mathcal{L}_N(\rh_0)$ is a new term which doesn't appear for
$\lambda=0$ case. This term comes from the contribution of sphere in
the Ricci scalar in \eqref{gbx2x1}. The term
$\mathcal{L}_f(\rh_0,\mathbf{c}_1,\mathbf{c}_2)$ is the term which is
independent of $\rh_1$ and comes from the leading order corrections in
$f_1=1+\mathbf{c}_1 z^2/R_1^2$ and $f_2=1+\mathbf{c}_2 z^2/R_1^2$,
where $\mathbf{c}_1=1/12$ and $\mathbf{c}_2=-1/4$. Further, terms with
$\mathcal{L}_1(\rh_0)$-$\mathcal{L}_6(\rh_0)$ are independent of
$\mathbf{c}_1$ and $\mathbf{c}_2$ and are linear in $\rh_1$, as it is
written. Now, we can write
 \bea
\rh'_1 \mathcal{L}_1 &\,=\,& \partial_z(\mathcal{L}_1 \rh_1 - \mathcal{L}_1' \rh_1) + \mathcal{L}_1'' \rh_1 \,, \nonumber \\
\drh'_1 \mathcal{L}_2 &\,=\,& \partial_z(\mathcal{L}_2 \drh_1) - \partial_\theta(\mathcal{L}_2' \rh_1) + \dot{\mathcal{L}_1}' \rh_1 \,, \nonumber \\
\ddot{\rh}_1 \mathcal{L}_3 &\,=\,& \partial_\theta(\mathcal{L}_3 \drh_1) - \partial_\theta(\dot{\mathcal{L}_3} \rh_1) +\ddot{\mathcal{L}_3} \rh_1 \,, \nonumber \\
\rh_1 \mathcal{L}_4 &\,=\,& \partial_z(\mathcal{L}_4 \rh_1) - \mathcal{L}_4' \rh_1 \,, \nonumber \\
\drh_1 \mathcal{L}_5 &\,=\,& \partial_\theta(\mathcal{L}_5 \rh_1) -
\mathcal{L}_5' \rh_1 \,,  \labell{gbx22}
 \eea
where prime and upper dot denote the partial derivative with respect to
$z$ and $\theta$. Using these in \eqref{gbx21}, we write
 \bea
S_5 \big|_{k\times S^2} &\,=\,& {16\,\pi^2 \tL^4 R_1^2 \ov \lp^4} \int_{z_m}^{\delta} dz \int_0^{\Omega-\epsilon} d\theta \bigg(\mathcal{L}_0(\rh_0) + {1\ov R_1^2} \Big( \mathcal{L}_N(\rh_0)+ \mathcal{L}_f(\rh_0,\mathbf{c}_1,\mathbf{c}_2) \nonumber \\
&& \qquad + \rh_1 (\mathcal{L}_1'' + \dot{\mathcal{L}_2}' + \ddot{\mathcal{L}_3} - \mathcal{L}_4' - \dot{\mathcal{L}_5})  + \partial_z(\mathcal{L}_1 \rh_1 - \mathcal{L}_1' \rh_1 + \mathcal{L}_2 \drh_1 + \mathcal{L}_4 \rh_1) \Big) \bigg) \nonumber \\
&& + {16\,\pi^2 \tL^4 \ov \lp^4} \int_{z_m}^{\delta} dz
\left(-\mathcal{L}_2' \rh_1 +\mathcal{L}_3 \drh_1 - \dot{\mathcal{L}_3}
\rh_1 + \mathcal{L}_5 \rh_1  \right)_{\theta=\Omega-\epsilon} \,,
 \labell{gbx23}
 \eea
where the coefficient of the $\rh_1$ vanishes from the equation of
motion of $\rh_0$ and we have performed the integration over $\theta$
in terms in the last line. Note that we can not integrate over $z$ in
any term as $\epsilon=\epsilon(z)$ and both $\theta$ and $z$
integrations don't commute. Now we further insert the ansatz $\rh_0 =
z/h(\theta)$ and $\rh_1=z^3g_3(\theta)$ in the above entropy functional
and in the equations of motion for $\rh_0$ and $\rh_1$ to find the
equations of motion for $h$ and $g_3$. Once again, we restrain to give
the complete expression of the equations of motion as they are not very
illuminating. Similar to \eqref{cc2x14}, at this stage we find that the
terms with the partial derivative with respect to $z$ vanish. Finally,
the entropy functional reduces to
 \bea
S_5 \big|_{k\times S^2} &\,=\,& {16\,\pi^2 \tL^4 R_1^2 \ov \lp^4} \int_{z_m}^{\delta} dz \int_{h_0}^{h_{1c}(z)} {d h \ov \dot{h}} \left({\hat{\mathcal{L}}_0(h) \ov z^3} + {1\ov R_1^2} {\hat{\mathcal{L}}_N(h)+\hat{\mathcal{L}}_f(h,\mathbf{c}_1,\mathbf{c}_2) \ov z} \right) \nonumber \\
&& \qquad \qquad + {16\,\pi^2 \tL^4 \ov \lp^4} \int_{z_m}^{\delta}
dz \, {\hat{\mathcal{L}}_B |_{\theta=\Omega-\epsilon} \ov z} \,,
 \labell{gbx24}
 \eea
where we have defined $\mathcal{L}_0=\hat{\mathcal{L}}_0(h) / z^3$,
$\mathcal{L}_N=\hat{\mathcal{L}}_N(h) / z$,
$\mathcal{L}_f=\hat{\mathcal{L}}_f(h,\mathbf{c}_1,\mathbf{c}_2) / z$
and expressions for $\hat{\mathcal{L}}$'s are given by \eqref{gbx25} in
appendix \ref{appa}. Now we can solve the equations of motion for $h$
and $g_3$ near the asymptotic boundary to find
 \bea
\dot{h} &\,=\,& -\frac{a_1}{h^4}-\frac{2 a_1}{h^2}-a_1 + \frac{h^4 \left(1 - 6 f_{\infty } \lambda \right)}{2 a_1 \left(1 - 2 f_{\infty } \lambda \right)} - \frac{h^6 \left(1 - 10 f_{\infty } \lambda \right)}{2 a_1 \left(1 - 2 f_{\infty } \lambda \right)} + \dots \,,  \labell{gbx26} \\
g_3 &\,=\,&  \frac{b_3}{h^3} + \frac{9-26 f_{\infty } \lambda +792 b_3
\left(1-2 f_{\infty } \lambda \right)}{504 h \left(1-2 f_{\infty }
\lambda \right)} +\frac{4 h \left(3-11 f_{\infty } \lambda +54 b_3
\left(1-2 f_{\infty } \lambda \right)\right)}{567 \left(1-2 f_{\infty }
\lambda \right)}+ \dots \,, \nonumber
 \eea
where $a_1$ and $b_3$ are constants which are fixed by ensuring that
both $h$ and $g_3$ have extrema at $\theta=0$. Note that here $a_1$ is
related to a quantity which is conserved along the $\theta$ translation
similar to \eqref{cc2x15}. Using these solutions near the boundary, we
first find the value of $h_{1c}(z)$ at the UV cut-off $z=\delta$. To
do that, we use the above solutions in the ansatz $\rh=z/h + z^3
g_3/R_1^2$ and inverting the relations iteratively, we find
 \bea
h_{1c}(\delta) &\,=\,& \left(\frac{1}{H}+\frac{b_3 H}{R_1^2}\right) \delta +\frac{\left(9-26 f_{\infty } \lambda +792 b_3 \left(1-2 f_{\infty } \lambda \right)\right) \delta ^3}{504 H R_1^2 \left(1-2 f_{\infty } \lambda \right)}\nonumber \\
&& \qquad \qquad \qquad +\frac{4 \left(3-11 f_{\infty } \lambda +54 b_3
\left(1-2 f_{\infty } \lambda \right)\right) \delta ^5}{567 H^3 R_1^2
\left(1-2 f_{\infty } \lambda \right)}+ \dots \,.
 \labell{gbx27}
 \eea
Note that above relation reduces to \eqref{cc2x19x1} for $\lambda=0$.
Now we use \eqref{gbx26} to study the behavior of integrands in
\eqref{gbx24} near the asymptotic boundary, where we have $h \to 0$ :
 \bea
{\hat{\mathcal{L}}_0 \ov \dot{h}} &\,\sim\,&  -\frac{1-4 f_{\infty } \lambda }{ h^2}+ \mathcal{O}(h^4) \,, \nonumber \\
{\hat{\mathcal{L}}_N+\hat{\mathcal{L}}_f \ov \dot{h}} &\,\sim\,&
\frac{15-68 f_{\infty } \lambda }{72 h^2} + \mathcal{O}(h^2) \,.
 \labell{gbx28}
 \eea
Using these, similar to \eqref{cc2x22} and \eqref{cc2x23}, we can make
the integrands in \eqref{gbx25} finite. We can break the terms in
following components
 \bea
I_1 &\,=\,& \int_{z_m}^{\delta} {dz \ov z^3} \int_{h_0}^{h_{1c}(z)} d h \left( {\hat{\mathcal{L}}_0 \ov \dot{h}} + \frac{1-4 f_{\infty } \lambda }{ h^2} \right) \,,\\
I_2 &\,=\,& - \int_{z_m}^{\delta} {dz \ov z^3} \int_{h_0}^{h_{1c}(z)} dh \, \frac{1-4 f_{\infty } \lambda }{ h^2} \,, \\
I_3 &\,=\,& -\int_{z_m}^{\delta} {dz \ov z} \int_{h_0}^{h_{1c}(z)} d h  \left( {\hat{\mathcal{L}}_N + \hat{\mathcal{L}}_f \ov \dot{h}} - \frac{15-68 f_{\infty } \lambda }{72 h^2} \right) \,, \\
I_4 &\,=\,& -\int_{z_m}^{\delta} {dz \ov z} \int_{h_0}^{h_{1c}(z)} d h
\,\frac{15-68 f_{\infty } \lambda }{72 h^2}\,,
 \eea
and then take a derivative with respect to $\delta$ and find following
series expansion:
 \bea
{d I_1 \ov d \delta} &\,=\,& {1\ov \delta^3} \int_{h_0}^0 dh \left({\hat{\mathcal{L}}_0 \ov \dot{h}} + \frac{1-4 f_{\infty } \lambda }{ h^2}\right) + \mathcal{O}(\delta^4) \nonumber \\
{d I_2 \ov d \delta} &\,=\,& \left(1-{b_3 H^2 \ov R_1^2}\right)\frac{H \left(1-4 f_{\infty } \lambda \right)}{ \delta ^4}-\frac{1-4 f_{\infty } \lambda }{h_0 \delta ^3} \nonumber \\
&& \qquad - \frac{H \left(1-4 f_{\infty } \lambda \right) \left(9-26 f_{\infty } \lambda +792 b_3 \left(1-2 f_{\infty } \lambda \right)\right)}{504  \left(1-2 f_{\infty } \lambda \right) R_1^2\delta ^2} + \mathcal{O}(\delta^0) \nonumber \\
{d I_3 \ov d \delta} &\,=\,& -{1\ov \delta} \int_{h_0}^0 dh \left( {\hat{\mathcal{L}}_N+\hat{\mathcal{L}}_f \ov \dot{h}} - \frac{15-68 f_{\infty } \lambda }{72 h^2}  \right) +\mathcal{O}(\delta^5)  \labell{gbx29} \\
{d I_4 \ov d \delta} &\,=\,& -\left(1 -{b_3 H^2 \ov R_1^2}\right) \frac{H \left(15-68 f_{\infty } \lambda \right)}{72 \delta ^2} +\frac{15-68 f_{\infty } \lambda }{72 h_0 \delta } + \mathcal{O}(\delta^0)  \nonumber \\
{d\ov d\delta}\int_{z_m}^{\delta} dz \, {\hat{\mathcal{L}}_B |_{\theta=\Omega-\epsilon} \ov z} &\,=\,& \left(1 -{3b_3 H^2 \ov R_1^2}\right) \frac{b_3 H^3\left(1-4 f_{\infty } \lambda \right)}{ \delta ^4} \nonumber  \\
&& + \left(1 -{4 b_3 H^2 \ov R_1^2}\right) \frac{H \left(1-4 f_{\infty
} \lambda \right) \left(9-26 f_{\infty } \lambda +792 b_3 \left(1-2
f_{\infty } \lambda \right)\right)}{504 \left(1-2 f_{\infty } \lambda
\right) \delta ^2 } +\mathcal{O}(\delta^0)\,. \nonumber
 \eea
Note that we have done the same with the boundary term in \eqref{gbx24}
too. Using these relations, we can read off the logarithmic term in the
EE, which is given by
 \be
S_5^{\log} \big|_{k\times S^2} \,=\, {16\,\pi^2 \tL^4 \ov \lp^4}  \left(\int_{h_0}^0 dh \left(
{\hat{\mathcal{L}}_N+\hat{\mathcal{L}}_f \ov \dot{h}} - \frac{15-68
f_{\infty } \lambda }{72 h^2}  \right) + \frac{15-68 f_{\infty }
\lambda }{72 h_0} \right) \log(\delta/H)\,,
 \ee
where $\hat{\mathcal{L}}_N$ and $\hat{\mathcal{L}}_f$ are given by
\eqref{gbx25}. Now we can try to compare this cut-off independent term
with the central charges \eqref{ct} and \eqref{adstar}. However, by
looking at the expression of $\hat{\mathcal{L}}_N$ and
$\hat{\mathcal{L}}_f$, we find that there are terms of order
$\mathcal{O}(\lambda^2)$. Hence the universal term is not a simple (and
particularly linear) function of the central charges.

Now, it would have been interesting to further investigate the
universal term for the geometry $c_1\times S^2$. In this case, the
singularity has a even dimensional, curved locus and we will get a
$\log^2\!\delta$ divergence. So it would have been very interesting to
see if this contribution from the locus is a specific central charge.
However, this calculate is tedious and demands more patience and
ingenuous simplifications.

%%%%%%%%%%%%%%%%%%%%%%%%%%%%%%%%%%%%%%%%
\section{Discussion}  \label{dis}

In this paper, we have used holography to study entanglement entropy
for various singular surfaces in higher dimensions and our results are
summarized in table \ref{table2}. In particular, in section \ref{cone},
we considered cones in various dimensions. This analysis suggests that
for CFT's in an odd number of spacetime dimensions, an additional
universal contribution appears:
 \be
 S_\mt{univ}= q_d(\Omega)\,\log(\delta/L)\,,
 \labell{bend2}
 \ee
where  $\Omega$ is the opening angle of the cone, as shown in figure
\ref{sing}. In particular, for $c_2=R^+\times S^2$ in $d=5$,
eq.~\reef{conex17x1} provides an expression for the coefficient
$q_5(\Omega)$ in the holographic CFT dual to Einstein gravity. We
showed that this coefficient satisfies the required properties: (a)
$q_d(\Omega=\pi/2)=0$ since the entangling surface is actually a flat
plane for this angle and (b) $q_d(\Omega) = q_d(\pi-\Omega)$ since the
entanglement entropy of the CFT ground state is identical for the
density matrix describing the degrees of freedom inside or outside of
the cone. Further we found that $q_d(\Omega\to0)\propto 1/\Omega$ as in
eq.~\eqref{conex17x1x1}. Again our expectation is that this behaviour
extends generally to cones for odd dimensional theories. That is, the
entanglement entropy acquires a universal contribution of the form
given in eq.~\reef{bend2} for a cone $c_{d-3}=R^+\times S^{d-3}$ in any
odd $d$.

Here we might recall that in three dimensions, strong subadditivity was
used to derive further constraints on the form of $q_3(\Omega)$
\cite{log3,hirata}. In particular, one shows that this function must
satisfy $q_3(\Omega)\ge 0$ and $q_3'(\Omega)\leq 0$. Of course, our
holographic result \reef{cuspx7} satisfies both of these inequalities.
It is noteworthy then that in five dimensions, our holographic result
satisfies $q_5(\Omega)\leq 0$ and $q_5'(\Omega)\geq 0$. It would be
interesting to understand if these inequalities are again general
properties originating from strong subadditivity.

Further we add that in eq.~\reef{conex17x1}, the coefficient $q_5$ is
proportional to $L^4/\lp^4$ which can be identified with a central
charge in the dual boundary CFT. However, with Einstein gravity in the
bulk, all of the central charges in the five-dimensional boundary
theory are identical and so in section \ref{GB1}, we extended the
calculation to Gauss-Bonnet gravity in the bulk, which allows us to
distinguish at least two such central charges, as described there.
However, given the result for $q_5(\Omega)$ implicit in
eq.~\reef{houseGB3} and \reef{gbx17}, it appears that this coefficient
is a complicated nonlinear function of both central charges. It may in
fact be that this coefficient is not determined by the central charges
alone -- this would then be similar to the results found for Renyi
entropies in \cite{renyi}.

The case of even dimensions was particularly interesting. In section
\ref{cone} with $d=4$ and 6, we found that a cone yields a universal
contribution of the form:
 \be
 S_\mt{univ}= \hat{q}_d(\Omega)\,\log^2(\delta/L)\,,
 \labell{bend3}
 \ee
where  $\Omega$ is again the opening angle, as shown in figure
\ref{sing}. Again we believe this is a generic result for even
dimensional theories. In our holographic examples, the coefficient
functions have a relatively simple form, as shown in
eqs.~\reef{conex15} and \reef{conex17x2}. Hence for these examples, it
is straightforward to verify that $\hat{q}_d(\Omega)$ satisfies the
same properties as described for $q_d(\Omega)$ in the first paragraph
above. From these results, we also see that $\hat{q}_d(\Omega \to0)
\propto 1/\Omega$, which parallels the behaviour seen for $q_3$ and
$q_5$. Further, in section \ref{GB1}, using Gauss-Bonnet gravity, we
were able to verify that this coefficient is proportional to the
central charge \reef{ct} which controls the leading singularity of the
two-point function of the stress tensor for $d=4$. In particular, we
found the simple result \reef{d4univ} which yields
 \be
\hat{q}_4=-\frac{c}{4}\,\frac{\cos^2\Omega}{\sin\Omega}\,.
 \labell{khat}
 \ee
For $d=6$, we also found a simple expression \reef{d6central} for the
new universal term for $c_3$. However, in this case, $\hat{q}_6$
depends linearly on both of the central charges, $\tilde{C}_T$ and
$a^*_6$, and it is still a relatively simple function of the opening
angle $\Omega$. Given the simplicity of these results, particularly in
eq.~\reef{khat}, one might conjecture that the same form may arise for
CFT's in general beyond our holographic calculations.

We note here that with even $d$, there are also contributions
proportional to a single power of $\log\delta/L$, however, these terms
are no longer universal. Rather the corresponding coefficient will vary
if the details of the cut-off (or the macroscopic scale $L$) are
changed because of the presence of the $\log^2\delta/L$ term -- see
further discussion below.

Recall that the trace anomaly in an even dimensional CFT gives rise a
universal contribution in the entanglement entropy with a smooth
entangling surface \cite{ryu2,solodukhin,finn,cthem1}. In particular,
if we consider a four-dimensional CFT in a flat background, this
contribution takes the form given in eq.~\reef{solo}. Given this
explicit expression, it is interesting to compare this contribution for
a conical entangling surface $c_1$ in $d=4$ to the universal term
\reef{d4univ} found in our holographic calculation.

Now in $d=6$, we do not know the precise expression of the contribution
of the trace anomaly to EE, \ie $d=6$ generalization of \eqref{solo}.
Hence, now we can not distinguish the contribution from the trace
anomaly with the contribution from the singularity in
\eqref{d6central}. Hence, we can not say what part of the universal
term comes from the singularity and particularly, we can not confirm if
singularity contributes only through the central charge $\tilde{C}_T$.

\FIGURE[!ht]{
\includegraphics[width=0.45\textwidth]{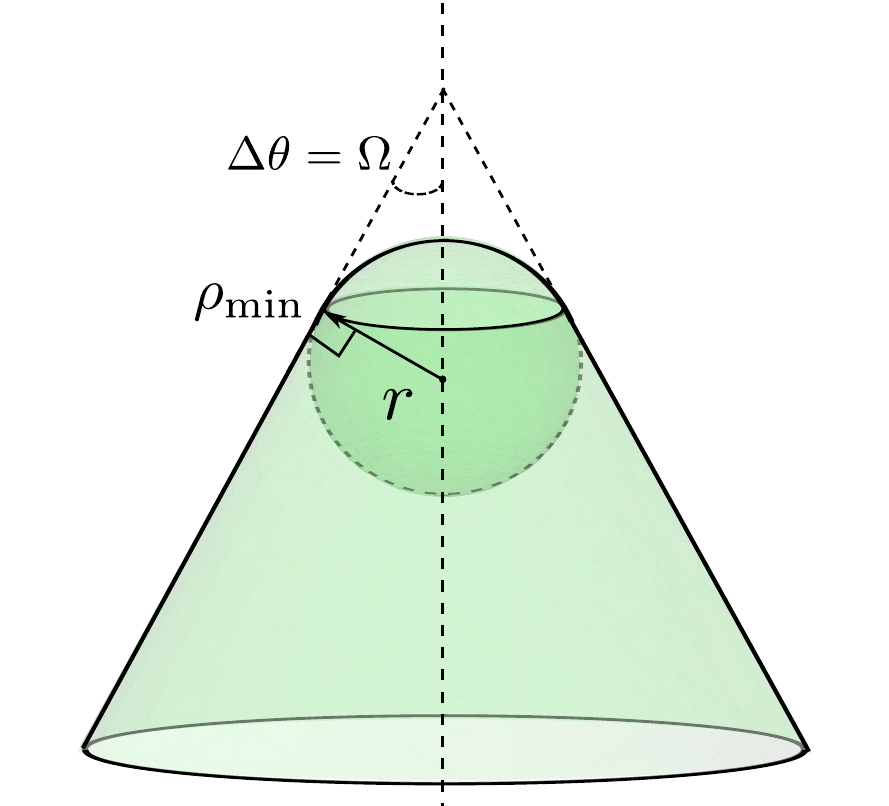}
\caption{(Colour Online) The illustration shows regulated cone $\tilde
c_1$, where the tip is replaced by a spherical cap. }
 \label{regu}}
%
%%%%%%%%%%%%%%%%%%%%%%%%%%%%%%%%%%%%%%%%%%%%%%%%%%%%
%
%\FIGURE[!ht]{
% \begin{tabular}{cc}
%\includegraphics[width=0.5\textwidth]{1cone.pdf}&
%\includegraphics[width=0.5\textwidth]{1wedge.pdf}\\
%(a) & (b)
%\end{tabular}
%\caption{(Colour Online) Panel (a) shows regulated cone $\tilde c_1$,
%where the tip is replaced by a spherical cap. Panel (b) shows a similar
%regulated surface which smoothly models the crease $k\times R^1$.}
% \label{regu}}
%
Of course, eq.~\reef{solo} is only expected to apply for a smooth
entangling surface and so cannot be applied directly to the cone $c_1$.
Hence our approach is to construct a `regulated cone' $\tilde c_1$ by
cutting off the cone at some $\rmin$ and replacing the tip with a
spherical cap of radius $r=\rmin\,\tan\Omega$, as shown in figure
\ref{regu}. Hence $\tilde c_1$ provides a smooth model of the desired
conical entangling surface to which we can apply eq.~\reef{solo} and we
can consider the limit $\rmin\to0$ to recover the result for the
singular surface $c_1$. Hence working first with finite $\rmin$, it is
relatively straightforward to show that eq.~\reef{solo}
yields\footnote{On the conical portion of $\tilde c_1$, ${\cal R}=0$
while the only nontrivial component of the extrinsic curvature is
$K^{\hat \theta}_{\phi\phi}=\rho\,\sin\Omega\,\cos\Omega$. On the
spherical cap, ${\cal R}=2/r^2$ while the combination of extrinsic
curvatures in eq.~\reef{solo} vanishes.}
 \beqa
S_\mt{univ} &=& - \frac{c}{2}\,\frac{\cos^2\Omega}{\sin\Omega}\,
\log(\delta/L)\log(\rmin/L)
 \nonumber\\
&&\quad +\ 2\,a\,\left(1-\sin\Omega\right) \,\log(\delta/L)\,.
 \labell{bang8}
 \eeqa
Unfortunately the first term will diverge if we take the limit
$\rmin\to0$. However, if the underlying CFT has been regulated with the
cut-off $\delta$, it should not be able to resolve any features in
geometry at shorter distances. Hence it is natural to consider the
limit $\rmin\to\delta$ which yields
 \be
S_\mt{univ} = - \frac{c}{2}\,\frac{\cos^2\Omega}{\sin\Omega}\,
\log^2(\delta/L) + 2\,a\,\left(1-\sin\Omega\right) \,\log(\delta/L)\,.
 \labell{bang9}
 \ee
Hence with this construction, the universal contribution \reef{solo}
contains terms proportional to both $\log^2(\delta/L)$ and
$\log(\delta/L)$. The most surprising aspect of this result is that the
coefficient of the leading term is almost identical to the holographic
result in eq.~\reef{d4univ}. However, there is a mismatch by a factor
of two.\footnote{This observation was also made in \cite{new}. We also
note that the same result appears in an alternate calculation of the
holographic entanglement entropy presented in appendix \ref{conformal}.
There the mismatch can be regarded as an anomaly arising from a
singular conformal transformation.} Our interpretation of this result
is that with the conical entangling surface $c_1$, part of the
$\log^2(\delta/L)$ divergence should be associated with correlations
across the smooth part of the entangling surface away from the
singularity. The above calculation then suggests a pile up of short
distance correlations in the vicinity of the tip. However, the full
accumulation of correlations near the tip of the cone does not have
(precisely) the form expected from the `smooth' expression \reef{solo}.
Hence a part of this universal contribution should be thought of as
intrinsic to the singularity at the tip of the cone itself. 

As already noted, if we examine our holographic entanglement entropy
for $c_1$ in eq.~\reef{conex15}, we also find a term proportional to
$\log(\delta/L)$. The angular dependence of the corresponding
coefficient is different from that for the $\log^2(\delta/L)$ term and
so one might envision that certain universal aspects can still be
extracted from this coefficient. For example, in eq.~\reef{bang9}, the
second central charge $a$ appears in the $\log(\delta/L)$ term and so
one might hope to extract this charge by studying the entanglement
entropy for cones with various opening angles $\Omega$. However, we
wish to emphasize that this $\log(\delta/L)$ contribution is simply not
universal in the presence of the $\log^2(\delta/L)$ term. For example,
let us return to our regulated model of cone above where we argued it
was natural to take the limit $\rmin\to\delta$ for the radius of the
spherical cap. The latter was motivated by observing that $\delta$ is a
short distance cut-off in the underlying CFT and so the latter can not
resolve any geometric features involving shorter distance scales.
However, let us note that the radius of the spherical cap was
$r=\rmin\,\tan\Omega$ and so even if we set $\rmin=\delta$, for small
$\Omega$, the cap is effectively much smaller than the short distance
cut-off. Hence one might instead choose
 \be
 \rmin =\left\lbrace
 \begin{matrix}
\delta&{\rm for}\ \ \Omega>\pi/4\,,\\
\frac{\delta}{\tan\Omega} &{\rm for}\ \ \Omega\leq\pi/4\,.
 \end{matrix}\right.
 \labell{confuse}
 \ee
While such a choice leaves the $\log^2(\delta/L)$ term unchanged, the
coefficient in the $\log(\delta/L)$ contribution acquires a complicated
new angular dependence. We present this discussion here simply to
illustrate that all of the details (including the angular dependence)
of the coefficient of the $\log(\delta/L)$ can be expected to depend on
the precise choice of the regulator in the calculation of the
entanglement entropy. While eq.~\reef{confuse} gives an illustrative
example, we might observe that the same regulator in the holographic
calculations is far more subtle. The analog of eq.~\reef{confuse} would
be $\rmin=\delta/h_0$ in section \ref{cone}.

We also extended our holographic analysis to consider creases or
extended singularities in section \ref{crease}. Examining the examples
summarized in table \ref{table2}, we find that the crease of the form
$k\times R^m$ or $c_n\times R^m$ creates no new universal
contributions.\footnote{Let us re-iterate that we are ignoring certain
cases here, \eg $c_1\times R^2$, where a $\log\delta$ term appears but
it can be attributed to the smooth part of the entangling surface. We
also do not consider the possibility that the finite contributions may
exhibit some new universal behaviour.} However, in general, we found
that creases can contribute additional universal terms, but singular
locus must have an even dimension and must be curved. These results
suggest that these new universal contributions to the entanglement
entropy take the form given in eq.~\reef{suggest} with a $\log\delta$
divergence in odd dimensions and $\log^2\!\delta$ in even dimensions.
These results indicate that there is a rich variety of new geometric
contributions to entanglement entropy that can be associated with
singular entangling surfaces. However, our analysis only considered
simple families of singularities and does not suffice to reveal the
full geometric structure of these universal terms. It would, of course,
be interesting to consider more general singularities, \eg a crease of
the form $k\times R^1$ but where the opening angle varies along $R^1$
or where $R^1$ was not entirely straight. Another step towards a
clearer picture of these geometric coefficients would be to carry out
the holographic calculations using the Fefferman-Graham expansion
\cite{fefferman} to compute entanglement entropy along the lines
discussed in \cite{janet}.

In part, our motivation for these studies was the possibility that
these new universal contributions may be used to identify the central
charges of the underlying CFT. Recently there has been a great deal of
interest in using entanglement to identify central charges that obey a
c-theorem, in particular for odd spacetime dimensions, \eg see
\cite{liu,new,cthem1,cthem2,sthem,casini6}.  In this regard, our
results only point towards a clear result for even dimensions, namely,
that the coefficient of the $\log^2\!\delta$ contribution is
proportional to a particular central charge. Unfortunately, this
central charge $\widetilde{C}_\mt{T}$ is not the one expected to
satisfy a c-theorem.\footnote{Setting holography aside, it was shown
that for four-dimensional QFT's, there is no possible (linear)
combination of the two central charges, $c=\widetilde{C}_\mt{T}$ and
$a=\ads$, that can satisfy a c-theorem apart from $a$ alone
\cite{anselmi2}.} However, our holographic result seems particularly
simple and so it may be that there is a general derivation for any CFT,
perhaps connected to the trace anomaly as for the universal terms
identified for smooth surfaces \cite{ryu2,solodukhin,cthem1,finn}. It
is also worthwhile to investigate if other central charges appear if we
consider, \eg cones with a more general cross-section than simply
$S^{d-3}$.

In odd dimensions, our results in section \ref{GB1} indicate that the
universal terms associated with singularities in the entangling surface
will be complicated nonlinear functions of many parameters in the
underlying CFT. Here it must be said that we refer to these
contributions as universal since they will be independent of the
details of the UV regulator and so should characterize properties of
the underlying CFT (or QFT more generally). However, the precise nature
of the information contained in these terms remains to be understood. A
similar result was found for Renyi entropies of spherical entangling
surfaces in \cite{renyi}.

It would also be interesting to study these universal contributions to
entanglement entropy in non-holographic theories. In particular, one
might consider heat kernel methods for free field theories for simple
surfaces, \eg along the lines of \cite{frank,next}. We have been
informed by Brian Swingle that he found similar $\log^2\!\delta$ terms
in the entanglement entropy of surfaces with a conical singularity
using a twist field calculation in the Gaussian approximation, as
discussed in \cite{brian}. It is also of interest to extend our
holographic calculations to consider entangling surfaces with flat
faces and edges, \eg a cube. In particular, one would like to confirm
that the new $\log^2\!\delta$ contributions persist for the `corners'
in such cases.\footnote{For example, in four dimensions, we can regard
such a corner on the surface of a cube as a `cone' with a triangular
cross-section.} This would have practical implications for lattice
calculations in higher dimensions, where it is difficult to avoid such
corners. It would also be interesting to further investigate the Renyi
entropy for singular entangling surfaces \cite{log1}.

\vskip 2cm

\section*{Acknowledgements:} We would like to thanks Horacio Casini,
Alioscia Hamma, Janet Hung, Spiro Karigiannis, Igor Klebanov, Ben
Safdi, Misha Smolkin and Brian Swingle for various useful discussions.
Research at Perimeter Institute is supported by the Government of
Canada through Industry Canada and by the Province of Ontario through
the Ministry of Research \& Innovation. RCM also acknowledges support
from an NSERC Discovery grant and funding from the Canadian Institute
for Advanced Research.

%%%%%%%%%%%%%%%%%%%%%%%%%%%%%%%%%%%%%%%%%%%%

\appendix

%%%%%%%%%%%%%%%%%%%%%%%%%%%%%%%%%%%%%
\section{Conformal transformations and EE for $c_{d-3}$}   \label{conformal}
%rcm

In this section, we consider an alternate approach to calculating the
entanglement entropy associated with a conical singularity. In
particular, we begin by performing a conformal transformation which
takes $R^d$ to $R\times S^{d-1}$. Under such a conformal
transformation, the conical entangling surface $c_{d-3}$, considered in
sections \ref{results} and \ref{singular}, becomes simply a uniform
cylinder $R\times S^{d-3}$ in the latter background. The analogous
mapping was applied to calculating the cusp anomaly for a Wilson with a
sharp corner in N=4 SYM, \eg see \cite{cusp}. In this case, calculating
the cusp anomaly becomes problem of determining the quark-antiquark
potential on a three-sphere. In fact, the holographic calculation for
the leading result corresponds to determining an extremal surface in
the bulk with a `kink' boundary condition and so it precisely matches
the calculation of the holographic EE for a kink in $d=3$
\cite{cusp,cusp2}.

We would like to calculate EE using the cylindrical background geometry
$R\times S^{d-1}$ and in particular, to see if we get the same
$\log\delta$ and $\log^2\delta$ terms that appeared in our previous
calculations for the cones $c_{d-3}$. The key feature in these
calculations is that the conformal mapping takes the conical entangling
surface in $R^d$ to a new surface $R\times S^{d-3}$ in the cylindrical
background. Now this new surface has an infinite length along the $R$
direction and so this length must be regulated to properly account for
the logarithmic divergences, as we will discuss below in section
\ref{cylinder} when we calculate the holographic EE. We first describe
desired conformal transformation in the CFT and then we consider how
this transformation is implemented with a coordinate transformation in
the dual bulk spacetime.

In the flat background \eqref{met0x} with $\{n,m\}=\{d-3,0\}$, we can
make the coordinate transformations $t_E=r\cos\xi$ and $\rho=r\sin\xi$
and find following metric
 \be ds^2\,=\,dr^2 +
r^2(d\xi^2+\sin^2\xi\,(d\theta^2+\sin^2\theta\,d\Omega_{d-3}^2))\,.
\labell{ax00}
 \ee
In these coordinates, the cone geometry discussed in section
\ref{results} translates to the surface: $c_{d-3}= \lbrace r
=[0,\infty), \xi=\pi/2, \theta =\Omega \rbrace$. We can further perform
the coordinate transformation $r=L\, e^{Y/L}$, and make a Weyl
transformation to remove the overall factor $e^{2Y/L}$ from the
resulting metric. After these transformations, we find background
geometry becomes $R\times S^{d-1}$ with the metric
 \be
ds^2\,=\, dY^2 +
L^2(d\xi^2+\sin^2\xi\,(d\theta^2+\sin^2\theta\,d\Omega_{d-3}^2))\,,
 \labell{ax000}
 \ee
where $Y\in[-\infty,\infty]$. Further, the conical entangling surface
above is now mapped to a cylinder of infinite length, \ie
 \be
c_{d-3}= \lbrace Y =(-\infty,\infty), \xi=\pi/2, \theta =\Omega
\rbrace\,.
 \labell{wind7}
 \ee

Next we discuss the coordinate transformation in the bulk geometry
which implements the conformal transformation between the two boundary
metrics in eqs.~\eqref{ax00} and \eqref{ax000}. We begin with the
standard description of (Euclidean) AdS$_{d+1}$ as a hyperbola embedded
in the following $(d+2)$-dimensional Minkowski space
 \be
ds^2\,=\,-dU^2 + dV^2+(dX^1){}^2+\dots+(dX^{d}){}^2 =\,-dU^2 +
dV^2+dR^2+R^2 d\Omega_{d-1}^2\,,
 \labell{ax1}
 \ee
where in the second expression, we introduced polar coordinates on the
space spanned by the $X^i$. Now the AdS geometry is defined by the
hyperbola
 \be
-U^2+V^2+(\vec{X})^2 \,=\, -U^2 + V^2+R^2\,=\,-L^2\,.
 \labell{ax2}
 \ee
We can solve this constraint by writing $U\,=\,\sqrt{R^2+L^2}\,\cosh
(Y/L)$ and $V\,=\,\sqrt{R^2+L^2}\,\sinh (Y/L)$, in which case the
induced metric on this surface becomes
 \be
ds^2\,=\, {1\ov \left( 1+{R^2\ov L^2}\right) }dR^2 + \left(1+{R^2 \ov
L^2}\right) dY^2 + R^2
\left[d\xi^2+\sin^2\xi\,(d\theta^2+\sin^2\theta\,d\Omega_{d-3}^2)\right]\,.
 \labell{ax3}
 \ee
Here we have written $d\Omega_{d-1}^2= d\xi^2+ \sin^2\xi\, (d\theta^2+
\sin^2\theta\, d\Omega_{d-3}^2)$. Of course, we recognize this metric
as (Euclidean) AdS$_{d+1}$ in global coordinates and this geometry is
dual to the boundary CFT with the background metric \eqref{ax000}.
Alternatively, to get the bulk metric dual to eq.~\eqref{ax00}, we
write $R=rL/z$ and $U+V=L^2/z$ and eq.~\eqref{ax2} yields
$U-V=z+r^2/z$. With these coordinates, the induced metric on the
hyperbola becomes
 \be
ds^2\,=\, {L^2 \ov z^2}(dz^2 + dr^2 + r^2 d\Omega_{d-1}^2)\,,
 \labell{ax4}
 \ee
which matches the `Poincar\'e patch' metric \eqref{metric1} with
$r^2=t_E^2+\rho^2$. Clearly, the boundary metric matches the desired
form given in eq.~\reef{ax00}. This bulk metric \reef{ax4} was used in
our calculation of the holographic EE for the cone in section
\ref{singular}. So now we use the metric \eqref{ax3} in calculating the
EE for the cylindrical surface produced by the conformal mapping.

%%%%%%%%%%%%%%%%%%%%%%%%%%%%%%%%%%%%%%%
\subsection{EE for cylinders} \label{cylinder}

In this section, we will calculate the holographic EE for cylindrical
entangling surface $R\times S^{d-3}$ on the background $R\times
S^{d-1}$ for $d=3,\,4$ and $5$. As we will see the calculations here
are closely related to those already presented in section
\ref{singular}. We find that the coefficient of the universal
$\log\delta$ term matches our previous results for $d=3$ and 5. In
$d=4$, the universal term is a $\log^2\delta$ contribution but we find
the coefficient in the following does not give a precise match with
that in eq.~\eqref{d4univ} from our previous calculation. In fact, the
$\log^2\delta$ term here shows the same mismatch by a factor of two
that we found in eq.~\eqref{bang9} from considering the contribution of
the trace anomaly on a regulated conical surface. We will show that
this difference in the holographic results comes from choosing
different UV cut-offs in the different coordinate systems. Although
these cut-offs are natural in the particular coordinate system in which
they are chosen, they produce different coefficients of the
$\log^2\delta$ divergence.

Now we begin with a general discussion for a $d$-dimensional CFT on the
background $R\times S^{d-1}$ and later focus on the specific cases with
$d=3,\,4$ and $5$. The cylindrical entangling surface on the boundary
is given in eq.~\reef{wind7}. However, to produce a finite result, the
length the cylinder must be regulated which we do by restricting
$Y\in[Y_-,Y_+]$, where $Y_-$ and $Y_+$ are cut-offs to be fixed below.
Now the holographic EE is calculated by first finding the minimal area
surface hanging into the bulk geometry with metric \eqref{ax3} and if
we define this surface by coordinates $(Y,\theta,\Omega_{d-3})$, we
will have a radial profile $R(\theta)=L\, g(\theta)$. In particular,
note that $R(\theta)$ is independent of $Y$ because we have
translational symmetry along this direction.\footnote{The corresponding
symmetry for the cone in $R^d$ is invariance under dilatations.} The
only scale in the geometry is $L$, which coincides with the radius of
the sphere $S^{d-1}$ on the boundary. With this ansatz, the holographic
EE is given by
 \be
S_d|_{cylinder}\,=\, {2\,\pi\, L^{d-2}\, \Omega_{d-3} \ov \lp^{d-1}}
\int_{Y_-}^{Y_+} dY \int_0^{\Omega-\epsilon} d\theta\,g^{d-3}
(\sin\theta)^{d-3} \sqrt{\dot{g}^2+g^2+g^4}\,,
 \labell{ax5}
 \ee
where $\dot{g}=dg/d\theta$ and $\Omega_{d-3}$ represents the area of
the unit $(d-3)$-sphere. Further, $\epsilon$ is related to the UV
cut-off in the theory. Note that the natural way to choose this UV
cut-off for cylinder should keep $\epsilon$ independent of $Y$, the
coordinate along the length of the cylinder. So a natural UV cut-off
for the minimal area surface can be $R=L^2/\delta$, which will fix
$\epsilon$ such that $g(\Omega-\epsilon)=L/\delta$.

To compare the above expression \reef{ax5} with our previous results,
we need to relate the quantities for the cylinder with the quantities
for the cone discussed in section \ref{singular}. First we note that
from the coordinate transformation taking us from eq.~\eqref{ax3} to
eq.~\eqref{ax4}, we have $\rho=r$ and $R=r\,L/z=\rho\, L/z$. In our
previous calculation of the holographic EE in eq.~\eqref{cuspx3} or
\reef{conex6}, we used the ansatz $z=\rho\, h(\theta)$. Hence, above
relation implies that $R=L/h(\theta)$. Comparing with our ansatz above,
we see that $g(\theta)=1/h(\theta)$ and $h(\Omega-\epsilon)=\delta/L$.
Now we turn to consider the cut-offs in the coordinate $Y$. First note
that these are independent of our UV cut-off on the radial coordinate,
\ie $R=L^2/\delta$. As the conformal transformation has mapped the cone
in $R^d$ to the present cylinder, $Y_-$ and $Y_+$ should be related to
the cut-offs in $\rho$ appearing in eq.~\eqref{conex6}. We use the
coordinate transformation $Y=L\log(\rho/L)$, as given above
eq.~\eqref{ax000}, and from the range $\rho\in[\delta/h_0,H]$ in
eq.~\eqref{conex6}, we find that $Y_-=L \log(\delta/h_0L)$ and $Y_+=L
\log(H/L)$. Using all these results, we can now rewrite eq.~\eqref{ax5}
as
 \be
S_d|_{cylinder}\,=\, {2\,\pi\, L^{d-1} \Omega_{d-3} \ov \lp^{d-1}}
\int_{\delta/h_0}^{H} {d\rho \ov \rho} \int_{h_0}^{\delta/L} dh\
{\sin^{d-3} (\theta) \ov \dot{h}\,h^{d-1}}\sqrt{\dot{h}^2+h^2+1} \,.
 \labell{ax6}
 \ee
At first sight, this expression is identical to eq.~\eqref{conex6}
found in the previous calculation. However, we should notice that there
is crucial difference. Namely, the upper limit of integration over $h$
here is a fixed constant while in eq.~\reef{conex6}, it is a function
of $\rho$.

Now we consider cylinder in different dimensions one by one. In
eq.~\eqref{ax5}, we first take $d=3$ to consider a cylinder dual to a
kink in $R^3$. Note that in this case $\theta\in[-\Omega,\Omega]$ and
we will also have an extra factor of two coming from the change in the
limits of integration. We can further use the variable
$y=\sqrt{1/h^2-1/h_0^2}$ as around eq.~\eqref{cuspx4x1} and follow the
steps in section \ref{singular}. Finally, we find that for $d=3$,
eq.~\eqref{ax6} will produce
 \be
S_3|_{cylinder}\,=\, {4\pi L^2 \ov \lp^2}\left[ {L\ov \delta}
\log(H/\delta) + {L\ov \delta} \log(h_0) -q_3(\Omega)
\log(H/\delta)+\dots  \right]\,,
 \labell{ax8}
 \ee
where $q_3$ is given by \eqref{cuspx7}. We note that the universal term
in EE for the cylinder here precisely matches with that for the kink.
However, now the leading order divergence is different as it contains
an extra factor of $\log(h_0H/\delta)$. Of course, these divergent
terms are not expected to be universal.

In part, the aim of the present calculations is to show that our
`universal' terms are independent of the details of the choice of
regulator. Here we chose the natural cut-off adapted to the new
coordinates after the conformal transformation in the boundary geometry
(or a coordinate transformation in the bulk geometry). While the entire
structure of the divergent contributions was not unchanged, our
universal logarithmic term matched the previous calculations. Note that
in eq.~\eqref{ax6}, if we would have used $z=\delta$ as the UV cut-off,
using $R=\rho L/z$, we would have had
$g(\Omega-\epsilon)=\rho/\delta=1/h(\Omega-\epsilon)$. In this case,
eq.~\eqref{ax6} (with an extra factor of two) would have become
 \be
\tilde{S}_3|_{cylinder}\,=\, {4\pi L^2 \ov \lp^2} \int_{\delta/h_0}^{H}
{d\rho \ov \rho} \int_{h_0}^{\delta/\rho} dh {\sqrt{\dot{h}^2+h^2+1}\ov
h^2 \dot{h}}\,,
 \ee
precisely matching with eq.~\eqref{cuspx3}. However, choosing the new
radial coordinate $R$ to be a function of $\rho$ (or more clearly $Y$)
would not have been a natural UV cut-off in this case. Although we have
found that for $d=3$, the universal term in eqs.~\eqref{cuspx3} and
\eqref{ax8} match, this will not be the case in $d=4$ which we discuss
next.

For a cylinder in a four-dimensional CFT on $R\times S^3$, the EE is
given by eq.~\eqref{ax6} with $d=4$. Now following the calculations and
steps in section \ref{cone}, we find that
 \bea
S_4|_{cylinder} &\,=\,&  {4\pi^2 L^3 \ov \lp^3} \bigg[{ L^2
\sin^2\Omega \ov 2 \, \delta^2} \log(H/\delta) + {L^2 \sin^2\Omega \ov
2 \, \delta^2} \log(h_0)
\notag \\
&& \qquad \qquad - {\cos\Omega\,\cot\Omega \ov 8} \log^2(\delta/H) +
\mathcal{O}(\log(\delta)) \bigg]\,.
 \labell{ax10}
 \eea
In comparing eq.~\eqref{ax10} with eq.~\eqref{conex16}, we see that the
universal term is off by a factor of two. However, this $\log^2\delta$
contribution precisely matches with the contribution from the trace
anomaly as given in eq.~\eqref{bang9}. Here we can see that the new
universal contribution from the singularity discussed in
eq.~\eqref{d4univ} is not invariant under the conformal transformation
considered here. In eq.~\eqref{ax6}, if we had chosen the UV cut-off on
$R=\rho L/z=L\,g(\theta)$ to be $z=\delta$, the upper limit of
integration would have been $\delta/\rho$, instead of $\delta/L$. In
that case, the results for the holographic EE here would have precisely
matched our previous results for the cone $c_1$.

Here we would note that singularities appear in two places here. Of
course, there is the geometric singularity in the entangling surface
$c_1$. However, the conformal transformation taking us from the flat
metric \reef{ax00} to the cylindrical metric \reef{ax000} is also
singular at precisely the same point, \ie this transformation maps the
origin in $R^d$ to $Y\to-\infty$ in $R\times S^{d-1}$. Hence it seems
this transformation is `anomalous' in that it does not preserve the
coefficient of the universal contribution to the entanglement entropy.
This effect is somewhat reminiscent of the anomaly that arises in
mapping a straight Wilson line in N=4 SYM to a circular Wilson loop,
both in $R^4$. While the former has a vanishing expectation value, the
expectation of the latter yields a nontrivial result
\cite{cusp2,cusp3}.

As a final example, we consider the EE for a cylinder in $R\times S^4$.
In this case, we insert $d=5$ into eq.~\eqref{ax6} and follow the steps
in section \ref{cone} to find that
 \bea
S_5|_{cylinder} &\,=\,&  {8\pi^2 L^4 \ov \lp^4} \bigg[{ L^3
\sin^2\Omega \ov 3 \, \delta^3} \log(H/\delta) + { L^3 \sin^2\Omega \ov
3 \, \delta^3} \log(h_0) - {4\,L\,\cos^2\Omega \ov 9\,\delta}
\log(H/\delta)
 \notag \\
&& \qquad - {4\,L\,\cos^2\Omega \ov 9\,\delta} \log(h_0) + q_5
\log(\delta/H) + \mathcal{O}(\delta^0) \bigg]\,,
 \labell{ax12}
 \eea
where $q_5$ precisely the same as given in eq.~\eqref{q5t}. Hence in
comparing eqs.~\eqref{ax12} and \eqref{conex17x1}, we clearly see that
the universal logarithmic term for a cylinder in $R\times S^4$ matches
with that for cone $c_2$. However, as compared to EE for $c_2$ in
\eqref{conex17}, there are again new non-universal divergences, which
take the form $\log(\delta)/\delta^3$ and $\log(\delta)/\delta$.

We conclude this section by saying that although we have studied
examples in $d=3,4$ and $5$, we expect that the $\log^2\delta$ term for
cone $c_{d-3}$ in even $d$ is not invariant under the conformal mapping
from flat space to a cylinder. However, for odd $d$, the $\log\delta$
term will be invariant under the corresponding conformal
transformation.

%%%%%%%%%%%%%%%%%%%%%%%%%%%%%%%%%
\section{Intermediate quantities for calculation of EE}   \label{appa}

In this section, we mention various intermediatory steps in calculation
of EE.

%%%%%%%%%%%%%%%%%%%%
\subsection{EE for cone $c_2$ in $d=5$}
 \bea
S_5 \big|_{c_2} \,=\, {8\, \pi^2 \,L^4 \ov \lp^4} \bigg[\frac{H^3
\sin^2(\Omega)}{9 \delta^3} -\frac{4 H \cos^2(\Omega)}{9 \delta } +
q_5 \log(\delta/H) + \mathcal{O}(\delta^0) \bigg]\,,
 \labell{conex17}
 \eea
where $q_5$ is given by \eqref{q5t}. The $h(\theta)$ in \eqref{q5t} is solution of following equation of motion
\bea
&& h (1+h^2) \sin(\theta) \ddot{h}+ 2 \,h\, \cos(\theta) \dot{h}^3 + 2\, (2+h^2)\sin(\theta) \dot{h}^2 \notag \\
&& \qquad \qquad \qquad \qquad + 2\,h \, (1+h^2) \cos(\theta) \dot{h}+ (4+7 h^2+3 h^4) \sin (\theta)\,=\,0\,.
\labell{d5h}
\eea

%%%%%%%%%%%%%%%%%%%%%%
\subsection{EE for cone $c_3$ in $d=6$}
 \bea
S_6 \big|_{c_3} &\,=\,& {8\, \pi^3 \,L^5 \ov \lp^5} \bigg[ \frac{H^4 \sin^3(\Omega)}{16 \delta^4} - \frac{27 H^2 \cos^2(\Omega) \sin(\Omega) }{128 \delta^2} + \frac{9 \cos(\Omega) \cot(\Omega) (31-\cos(2 \Omega) ) }{8192}  \log\left(\delta/ H\right)^2   \nonumber \\
& & \,+ \bigg( q_6 + \frac{\sin^3(\Omega)}{4 h_0^4} - \frac{27 \cos^2(\Omega)  \sin(\Omega) }{64 h_0^2}  \labell{conex19} \\
& & \qquad \qquad \qquad \qquad  - \frac{9 \cos(\Omega) \cot(\Omega)
(31 - \cos(2 \Omega) )  \log( h_0)}{4096}  \bigg) \log\left(\delta/
H\right) + \mathcal{O}(\delta^0)    \bigg]\,, \nonumber
 \eea
where $q_6$ is given by
 \bea
q_6 &\,=\,& \int^{h_0}_{0} dh \, \bigg[ { \sin^3(\theta) \ov \dot{h} h^5 }\sqrt{1+h^2+\dot{h}{}^2}  + \frac{\sin^2(\Omega)}{h^5} - \frac{27 \cos^2(\Omega) \sin(\Omega)}{32 h^3}  \nonumber \\
&& \qquad \qquad \qquad \qquad \qquad \qquad \qquad + \frac{9
\cos(\Omega) (31 - \cos(2 \Omega) ) \cot(\Omega) }{4096 h}  \bigg] \,.
 \eea

%%%%%%%%%%%%%%%%%%%%%%
\subsection{EE for conical singularity $c_1\times R^2$ in $d=6$}
For this case, the integrand in EE behaves as
 \bea
&\frac{ \sin(\theta) \sqrt{1+h^2+ \dot{h}{}^2}}{\dot{h} h^5} & \sim -\frac{\sin(\Omega)}{h^5}+\frac{3 \cos(\Omega) \cot(\Omega)}{32 h^3} - \frac{3 (13-19 \cos(2 \Omega)) \cot^2(\Omega) \csc(\Omega)}{4096 h} + \nonumber \\
& & \frac{\cot^2(\Omega) \csc^3(\Omega) (-375+348 \cos(2 \Omega) - 45 \cos(4 \Omega) - 128 b_3 \sin(2 \Omega) + 64 b_3 \sin(4 \Omega) )}{32768}h \log(h) \nonumber \\
& & \qquad + \mathcal{O}(h)\,,
 \labell{cnflatx17}
 \eea
near the boundary. Using this, we find that EE is given by
 \bea
S_6 \big|_{c_1\times R^2} &\,=\,& {4\, \pi^2 \,L^{5} h^2 \ov \lp^{5}} \bigg[ \frac{H^2 \sin(\Omega)}{8 \delta^4} + \frac{3 \cos(\Omega) \cot(\Omega) \log(\delta) }{64 \delta ^2} \nonumber \\
&& +\frac{3 h_0^4 (13 - 19 \cos(2 \Omega) ) \cot^2(\Omega) \csc(\Omega) +384 h_0^2 \cos(\Omega) \cot(\Omega) \left(1 - 2 \log(h_0)\right)-4096 \sin(\Omega)}{16384 h_0^2 \delta ^2} \nonumber \\
& & -{1\ov 2 \delta^2} \int^{h_0}_{0} dq \, q^2 J_6(q) + \frac{3 (13-19
\cos(2 \Omega) ) \cot^2(\Omega) \csc(\Omega) \log(\delta)}{8192 H^2}
+\mathcal{O}(\delta^0) \bigg]\,,
 \labell{cnflatx18}
 \eea
where
 \be
J_6(h)\,=\, \left(\frac{ \sin(\theta) \sqrt{1+h^2+
\dot{h}{}^2}}{\dot{h} h^5} + \frac{\sin(\Omega)}{h^5} - \frac{3
\cos(\Omega) \cot(\Omega)}{32 h^3} + \frac{3 (13-19 \cos(2 \Omega))
\cot^2(\Omega) \csc(\Omega)}{4096 h}  \right)\,.
 \ee

%%%%%%%%%%%%%%%%%%%%%%%%%
\subsection{EE for cone $c_2$ in CFT dual to Gauss-Bonnet gravity}
For five-dimensional cone $R^+\times S^2$ in CFT dual to the
Gauss-Bonnet gravity, the entanglement entropy turns out to be
 \bea
S_5 \big|_{c_2} &\,=\,& {8\,\pi^2 \tL^4 \ov \lp^4} \bigg[ \frac{H^3 \sin^2(\Omega) \left(1-4 \lambda  f_{\infty }\right)}{9 \delta ^3} - \frac{2 H \cos^2(\Omega) \left(2-7 \lambda  f_{\infty }\right)}{9 \delta }  \labell{gbx16} \\
& & - \left( \frac{\left(2\cos^2(\Omega) \left(2-7 \lambda  f_{\infty
}\right) h_0^2 - 3 \sin^2(\Omega) \left(1 - 4 \lambda  f_{\infty
}\right) \right) }{9 h_0^3} + \int^{h_0}_0 dh {\mathcal{L}_3 \ov
\mathcal{L}_4 } \right)\log(\delta/H)  + \mathcal{O}(\delta^0)
\bigg]\,, \nonumber
 \eea
where
 \bea
\mathcal{L}_3 &\,=\,& 3 \sin^4(\theta ) \left(1+h^2+ \dot{h}{}^2\right)^2 + 4 \lambda^2 f_{\infty }^2 \Big(h^4 \left(2+3 h^2+h^4\right) \sin^4(\theta) \nonumber \\
& & + 4 h^3 \left(2+3 h^2+h^4\right) \cos(\theta) \sin^3(\theta) \dot{h} + h^2 \left(4 + 6 h^2 + 3 h^4 + \left(8+9 h^2+3 h^4\right) \cos(2 \theta)\right) \sin^2(\theta ) \dot{h}{}^2 \nonumber \\
& & - h \left(4 - h^4 - \left(4+3 h^2+h^4\right) \cos(2 \theta) \right) \sin(2 \theta) \dot{h}{}^3 + \left(h^4 \cos^4 (\theta) - 3 h^2 \cos^2(\theta) \sin^2(\theta) + 6 \sin^4(\theta) \right) \dot{h}{}^4 \Big) \nonumber \\
& & - 2 \lambda  \sin^3(\theta) f_{\infty } \left(1 + h^2 + \dot{h}{}^2\right) \left(-4 h \cos(\theta) \dot{h} + \sin(\theta) \left(6+5 h^2+9 \dot{h}{}^2\right)\right) \nonumber \\
\mathcal{L}_4 &\,=\,& 3 h^4  \dot{h} \sqrt{1+h^2 + \dot{h}{}^2} \Big(\sin^2(\theta) \left(1+h^2+\dot{h}{}^2\right)+ 2 \lambda  f_{\infty } \Big(\left(2+3 h^2+h^4\right) \sin^2(\theta) \nonumber \\
&&+ h \left(2+h^2\right) \sin(2 \theta) \dot{h} + \left(h^2
\cos^2(\theta) - \sin^2(\theta) \right) \dot{h}{}^2 \Big)\Big)\,.
 \labell{gbx17}
 \eea

%%%%%%%%%%%%%%%%%%%%%%%%%%%%%%%
\subsection{EE for cone $c_3$ in CFT dual to Gauss-Bonnet gravity}
For cone $R^+\times S^3$ in $d=6$ dimensional CFT, which is dual to
Gauss-Bonnet gravity, the EE is given by
 \bea
S_6 \big|_{c_3} &\,=\,& {4\,\pi^3 \tL^5 \ov \lp^5} \bigg[\frac{H^4 \sin^3(\Omega) \left(3-10 \lambda  f_{\infty }\right)}{48 \delta ^4} - \frac{H^2 \cos^2(\Omega) \sin(\Omega) \left(27 - 86 \lambda  f_{\infty }\right)}{128 \delta ^2} \nonumber \\
& & + \frac{3 \cos(\Omega) \cot(\Omega) \left(\left(93-190 \lambda  f_{\infty}\right)-\left(3-2 \lambda  f_{\infty }\right) \cos(2 \Omega) \ri)}{8192} \log(\delta/H)^2 \nonumber \\
& & + \frac{1}{12288 h_0^4}\Big[-27 h_0^4 \cos(\Omega) (31-\cos(2 \Omega)) \cot(\Omega) \log\left(h_0\right) - 5184 h_0^2 \cos^2(\Omega) \sin(\Omega) \nonumber \\
&& + 3072 \sin^3(\Omega) - 2 \lambda f_{\infty } \Big(9 h_0^4 \cos(\Omega) (-95+\cos(2 \Omega)) \cot(\Omega) \log\left(h_0\right)  \labell{gbx18} \\
&&- 8256 h_0^2 \cos^2(\Omega) \sin(\Omega) + 5120 \sin^3(\Omega) \Big) + \int^{h_0}_0 dh\, (\mathcal{L}_5-\mathcal{L}_5')
\Big] \log(\delta/H) +\mathcal{O}(\delta^0)   \bigg]\,. \nonumber
 \eea
Here $\mathcal{L}_5$ and $\mathcal{L}_5'$ are terms which come from making the $h$ integrand
in original expression of EE finite and these are given by
\bea
\mathcal{L}_5 &\,=\,& {\mathcal{L}_{5N} \ov \mathcal{L}_{5D}} \qquad \textrm{with} \labell{l5} \\
\mathcal{L}_{5N} &\,=\,& \sin\theta\, \bigg(12 \, \lambda^2 f_{\infty}^2 h^8 \sin^4\theta +48 \, \lambda^2 f_{\infty }^2 \cos\theta\, h^7 \sin^3\theta\, \dot{h}+24\, \lambda  f_{\infty} \cos\theta\, \sin^3\theta\, h\, \dot{h} \left(1+\left(1-5 \lambda f_{\infty }\right) \dot{h}^2\right) \nonumber \\
&& +6 \, \lambda\,f_{\infty }  h^5 \sin(2 \theta)\, \dot{h} \left( \sin^2\theta+2 \, \lambda\,  f_{\infty } \left(5 \, \sin^2\theta + 2 \cos^2 \theta \, \dot{h}^2 \right)\right) + \lambda\,f_{\infty }  h^3 \sin^2\theta \,  \dot{h} \bigg(6 \sin^2\theta \, (3+\dot{h}^2) \notag \\
&& +2 \, \lambda \,  f_{\infty } \left(18\, \sin^2\theta - (1+17 \cos(2 \theta)) \dot{h}^2 \right)\bigg) + 2\, \lambda \,f_{\infty }  h^6 \left(3 \, \sin^4\theta + \lambda\, f_{\infty} \left(14 \, \sin^4\theta + 9\, \sin^2(2 \theta)\, \dot{h}^2 \right)\right) \notag \\
&& +2 \,\sin^4\theta \, \left(20 \lambda ^2 f_{\infty }^2 \dot{h}^4 + 3\, (1+\dot{h}^2)^2-8\, \lambda\, f_{\infty } (1+3 \, \dot{h}^2 + 2\, \dot{h}^4 )\right) + h^2 \sin^2\theta \, \bigg(12 \,\sin^2\theta \, (1+\dot{h}^2) \notag \\
&& +4 \, \lambda^2 f_{\infty }^2 \dot{h}^2 \big(7+17\, \cos(2 \theta)-9\, \cos^2\theta \, \dot{h}^2 \big)-\lambda\,  f_{\infty } \Big(24 \, \sin^2\theta + (17-23\, \cos(2 \theta)) \dot{h}^2 + 6 \, \cos^2\theta \, \dot{h}^4 \Big) \bigg) \notag \\
&& +h^4 \Big(6 \, \sin^4\theta + \lambda\, f_{\infty } \sin^2\theta \, \big(1+\cos(2 \theta)+6 \dot{h}^2\big) +4 \, \lambda^2 f_{\infty }^2 \big(4 \,\sin^4\theta + 3\, (5+8 \cos(2 \theta)) \sin^2\theta \, \dot{h}^2 \notag \\
&& + 3\, \cos^4\theta \, \dot{h}^4 \big)\Big)\bigg)\,, \notag \\
\mathcal{L}_{5D} &\,=\,& 6\, h^5 \dot{h} \sqrt{1+h^2+\dot{h}^2} \bigg(3\, \lambda\,f_{\infty }  h^4 \sin^2\theta +9\, \lambda\,f_{\infty }\cos\theta\, \sin\theta\, h\, \dot{h}+6\, \lambda\,f_{\infty } \cos\theta \, \sin\theta \, h^3 \dot{h} \notag \\
&& + \sin^2\theta \Big(1+\dot{h}^2 + 2\, \lambda \, f_{\infty } \big(2-\dot{h}^2 \big) \Big) + h^2 \left(\sin^2\theta+\lambda  f_{\infty } \left(7\,\sin^2\theta + 3\, \cos^2\theta \, \dot{h}^2 \right)\right)\bigg) \notag \\
\textrm{and} && \notag \\
\mathcal{L}_5' &\,=\,& - \frac{ \sin^3\Omega \, \left(3-10 \lambda\,  f_{\infty }\right)}{3 \, h^5} + \frac{\cos^2\Omega \, \sin\Omega\, \left(27-86 \lambda  f_{\infty }\right)}{32 h^3} \notag \\
&& \qquad \qquad \qquad \qquad- \frac{3 \cos\Omega \, \cot\Omega \, \left(93-3 \cos(2 \Omega) + 2\, \lambda \, (-95 + \cos(2 \Omega)) f_{\infty }\right)}{4096\, h} \,.
\eea
Note that in the asymptotic limit, $h\to 0$ and $\mathcal{L}_5 = \mathcal{L}_5'+\mathcal{O}(h)$.

%%%%%%%%%%%%%%%%%%%%%%%%%
\subsection{EE for $k \times S^2$ in CFT dual to Gauss-Bonnet gravity}
The values of $\hat{\mathcal{L}}$'s in eqn.~\eqref{gbx24} are
following:
 \bea
\hat{\mathcal{L}}_0 &\,=\,& \frac{1}{ h^2 \sqrt{1+h^2+\dot{h}^2} \left(\left(1-2 f_{\infty } \lambda \right) \dot{h}^2 + \left(1 - 2 f_{\infty } \lambda \right) h^2+ 4 f_{\infty } \lambda +1  \right)}  \Big(\left(1-6 f_{\infty } \lambda +8 f_{\infty }^2 \lambda ^2\right) \dot{h}^4 \nonumber \\
&& + 2 \left(1-5 f_{\infty } \lambda +\left(1-6 f_{\infty } \lambda +8 f_{\infty }^2 \lambda ^2\right) h^2\right) \dot{h}^2  \nonumber \\
&& \qquad \qquad \qquad \qquad \qquad \qquad + \left(1-6 f_{\infty } \lambda +8 f_{\infty }^2 \lambda ^2\right) h^4 + \left(2-10 f_{\infty } \lambda \right) h^2 -4 f_{\infty } \lambda+ 1 \Big) \,, \nonumber \\
\hat{\mathcal{L}}_N &\,=\,& \frac{2 f_{\infty } \lambda  \sqrt{1+h^2+\dot{h}^2}}{3 h^2} \nonumber \\
\hat{\mathcal{L}}_f &\,=\,& {-1 \ov 72 h^2 \left(1+h^2\right) \left(1+h^2+\dot{h}^2\right){}^{3/2} \left(\left(1-2 f_{\infty } \lambda \right) \dot{h}^2 + \left(1-2 f_{\infty } \lambda \right) h^2 + 4 f_{\infty } \lambda + 1 \right)} \times \nonumber \\
&& \bigg(5 \left(3-10 f_{\infty } \lambda +8 f_{\infty }^2 \lambda ^2\right) \left(1+h^2\right) \dot{h}^6 +\Big( 42-26 f_{\infty } \lambda -56 f_{\infty }^2 \lambda ^2+3 \left(29-80 f_{\infty } \lambda +64 f_{\infty }^2 \lambda ^2\right) h^2 \nonumber \\
&& \quad  +15 \left(3-10 f_{\infty } \lambda +8 f_{\infty }^2 \lambda ^2\right) h^4 \Big) \dot{h}^4 +\Big( 39+8 f_{\infty } \lambda +48 f_{\infty }^2 \lambda ^2+\left(123-172 f_{\infty } \lambda +64 f_{\infty }^2 \lambda ^2\right) h^2 \nonumber \\
&& \quad +3 \left(43-110 f_{\infty } \lambda +88 f_{\infty }^2 \lambda ^2\right) h^4+15 \left(3-10 f_{\infty } \lambda +8 f_{\infty }^2 \lambda ^2\right) h^6 \Big) \dot{h}^2 + 5 \left(3-10 f_{\infty } \lambda +8 f_{\infty }^2 \lambda ^2\right) h^8 \nonumber \\
&& \quad + \left(57-140 f_{\infty } \lambda +112 f_{\infty }^2 \lambda ^2\ri) h^6 + \left( 81-146 f_{\infty } \lambda +120 f_{\infty }^2 \lambda ^2\ri) h^4 \nonumber \\
&& \qquad \qquad \qquad \qquad \qquad \qquad \qquad \qquad + \left( 51-72 f_{\infty } \lambda +48 f_{\infty }^2 \lambda ^2 \right) h^2 -16 f_{\infty } \lambda + 12    \bigg) \nonumber \\
\hat{\mathcal{L}}_B &\,=\,& z\left(-\mathcal{L}_2' \rh +\mathcal{L}_3 \drh - \dot{\mathcal{L}_3} \rh + \mathcal{L}_5 \rh \right)  \nonumber \\
&\,=\,& -\frac{2 f_{\infty } \lambda  h \dot{g}_3 + 2 f_{\infty }
\lambda  h^3 \dot{g}_3 + \left(1-2 f_{\infty } \lambda \right) g_3 h^2
\dot{h} + g_3 \dot{h} \left(1+6 f_{\infty } \lambda + \left(1-4
f_{\infty } \lambda \right) \dot{h}^2\right)}{
\left(1+h^2+\dot{h}^2\right){}^{3/2}}\,.
 \labell{gbx25}
 \eea

%%%%%%%%%%%%%%%%%%%%%%%%%%%%%%%%%%%%%%%%%%%%


\begin{thebibliography}{99}


\bibitem %[wenx]
 {wenx} See, for example:\\
M.~Levin and X.~G.~Wen,
  ``Detecting Topological Order in a Ground State Wave Function,''
  Phys.\ Rev.\ Lett.\  {\bf 96}, 110405 (2006)
  %%CITATION = PRLTA,96,110405;%%
 [arXiv:cond-mat/0510613];\\
A.~Kitaev and J.~Preskill,
  ``Topological entanglement entropy,''
  Phys.\ Rev.\ Lett.\  {\bf 96}, 110404 (2006)
  [arXiv:hep-th/0510092];\\
  %%CITATION = PRLTA,96,110404;%%
A.~Hamma, R.~Ionicioiu and P.~Zanardi, ``Ground state entanglement and
geometric entropy in the Kitaev's model,'' Phys.\ Lett.\ A {\bf 337},
22 (2005) [arXiv:quant-ph/0406202].

\bibitem %[cardy0]
 {cardy0} See, for example:\\
P.~Calabrese and J.~L.~Cardy,
  ``Entanglement entropy and quantum field theory,''
  J.\ Stat.\ Mech.\  {\bf 0406}, P002 (2004)
  [arXiv:hep-th/0405152];\\
  %%CITATION = JSTAT,0406,P002;%%
P.~Calabrese and J.~L.~Cardy, ``Entanglement entropy and quantum field
theory: A non-technical introduction,''
  Int.\ J.\ Quant.\ Inf.\  {\bf 4}, 429 (2006)
  [arXiv:quant-ph/0505193].
  %%CITATION = 00445,4,429;%%

\bibitem %[fradkin]
 {fradkin} B.~Hsu, M.~Mulligan, E.~Fradkin and E.A.~Kim,
``Universal entanglement entropy in 2D conformal quantum critical
points,'' Phys.\ Rev.\ B {\bf 79}, 115421 (2009) [arXiv:0812.0203].
  %%CITATION = ARXIV:0812.0203;%%

\bibitem %[ryu1]
  {ryu1}
  S.~Ryu and T.~Takayanagi,
  ``Holographic derivation of entanglement entropy from AdS/CFT,''
  Phys.\ Rev.\ Lett.\  {\bf 96}, 181602 (2006)
  [arXiv:hep-th/0603001].
  %%CITATION = PRLTA,96,181602;%%

\bibitem %[ryu2]
 {ryu2} S.~Ryu and T.~Takayanagi,
  ``Aspects of holographic entanglement entropy,''
  JHEP {\bf 0608}, 045 (2006)
  [arXiv:hep-th/0605073].
  %%CITATION = JHEPA,0608,045;%%

\bibitem %[ryu3]
 {ryu3} T.~Nishioka, S.~Ryu and T.~Takayanagi,
  ``Holographic Entanglement Entropy: An Overview,''
  J.\ Phys.\ A  {\bf 42}, 504008 (2009)
  [arXiv:0905.0932 [hep-th]];
  %%CITATION = JPAGB,A42,504008;%%
T.~Takayanagi,
  ``Entanglement Entropy from a Holographic Viewpoint,''
  arXiv:1204.2450 [gr-qc].
  %%CITATION = ARXIV:1204.2450;%%


\bibitem %[igor0]
 {igor0} See, for example:\\
I.~R.~Klebanov, D.~Kutasov and A.~Murugan,
  ``Entanglement as a Probe of Confinement,''
  Nucl.\ Phys.\  B {\bf 796}, 274 (2008)
  [arXiv:0709.2140 [hep-th]];\\
  %%CITATION = NUPHA,B796,274;%%
T.~Nishioka and T.~Takayanagi, ``AdS Bubbles, Entropy and Closed String
Tachyons,'' JHEP {\bf 0701}, 090 (2007)
  [hep-th/0611035];\\
  %%CITATION = HEP-TH/0611035;%%
N.~Ogawa and T.~Takayanagi, ``Higher Derivative Corrections to
Holographic Entanglement Entropy for AdS Solitons,''
  JHEP {\bf 1110}, 147 (2011)
  [arXiv:1107.4363 [hep-th]];\\
  %%CITATION = ARXIV:1107.4363;%%
A.~Lewkowycz, ``Holographic Entanglement Entropy and Confinement,''
  arXiv:1204.0588 [hep-th].
  %%CITATION = ARXIV:1204.0588;%%


\bibitem %[mvr]
 {mvr} M.~Van Raamsdonk,
  ``Comments on quantum gravity and entanglement,''
  arXiv:0907.2939 [hep-th];\\
  %%CITATION = ARXIV:0907.2939;%%
M.~Van Raamsdonk,
  ``Building up spacetime with quantum entanglement,''
  Gen.\ Rel.\ Grav.\  {\bf 42}, 2323 (2010)
  [arXiv:1005.3035 [hep-th]].
  %%CITATION = GRGVA,42,2323;%%

\bibitem %[vm]
 {vm} V.~E.~Hubeny and M.~Rangamani,
  ``Causal Holographic Information,''
  arXiv:1204.1698 [hep-th];\\
  %%CITATION = ARXIV:1204.1698;%%
 B.~Czech, J.~L.~Karczmarek, F.~Nogueira and M.~Van Raamsdonk,
  ``The Gravity Dual of a Density Matrix,''
  arXiv:1204.1330 [hep-th].
  %%CITATION = ARXIV:1204.1330;%%

\bibitem %[tarun]
 {tarun} T.~Grover, A.M.~Turner and A.~Vishwanath,
  ``Entanglement Entropy of Gapped Phases and Topological Order in Three dimensions,''
  Phys.\ Rev.\ B {\bf 84}, 195120 (2011)
  [arXiv:1108.4038 [cond-mat.str-el]].
  %%CITATION = ARXIV:1108.4038;%%

\bibitem %[liu]
  {liu}
  H.~Liu and M.~Mezei,
  ``A Refinement of entanglement entropy and the number of degrees of freedom,''
  arXiv:1202.2070 [hep-th].
  %%CITATION = ARXIV:1202.2070;%%

\bibitem %[bomb]
 {bomb} L.~Bombelli, R.~K.~Koul, J.~Lee and R.~D.~Sorkin,
  ``A Quantum Source of Entropy for Black Holes,''
  Phys.\ Rev.\ D {\bf 34}, 373 (1986).
  %%CITATION = PHRVA,D34,373;%%

\bibitem %[mark]
 {mark} M.~Srednicki,
  ``Entropy and area,''
  Phys.\ Rev.\ Lett.\  {\bf 71}, 666 (1993)
  [hep-th/9303048].
  %%CITATION = HEP-TH/9303048;%%

\bibitem %[frank]
 {frank} M.~P.~Hertzberg and F.~Wilczek,
  ``Some Calculable Contributions to Entanglement Entropy,''
  Phys.\ Rev.\ Lett.\  {\bf 106}, 050404 (2011)
  [arXiv:1007.0993 [hep-th]].
  %%CITATION = ARXIV:1007.0993;%%

\bibitem %[janet]
  {janet}
  L.~-Y.~Hung, R.~C.~Myers and M.~Smolkin,
  ``Some Calculable Contributions to Holographic Entanglement Entropy,''
  JHEP {\bf 1108}, 039 (2011)
  [arXiv:1105.6055 [hep-th]].
  %%CITATION = ARXIV:1105.6055;%%

\bibitem %[solodukhin]
  {solodukhin}
  S.~N.~Solodukhin,
  ``Entanglement entropy, conformal invariance and extrinsic geometry,''
  Phys.\ Lett.\  {\bf B665}, 305-309 (2008).
  [arXiv:0802.3117 [hep-th]].

\bibitem %[log1]
  {log1} H.~Casini and M.~Huerta,
  ``Universal terms for the entanglement entropy in 2+1 dimensions,''
  Nucl.\ Phys.\ B {\bf 764}, 183 (2007)
  [hep-th/0606256].
  %%CITATION = HEP-TH/0606256;%%

\bibitem %[log3]
 {log3} H.~Casini, M.~Huerta and L.~Leitao,
``Entanglement entropy for a Dirac fermion in three dimensions: Vertex
contribution,''
  Nucl.\ Phys.\ B {\bf 814}, 594 (2009)
  [arXiv:0811.1968 [hep-th]].
  %%CITATION = ARXIV:0811.1968;%%

\bibitem %[hirata]
  {hirata} T.~Hirata, T.~Takayanagi,
  ``AdS/CFT and strong subadditivity of entanglement entropy,''
  JHEP {\bf 0702}, 042 (2007).
  [hep-th/0608213].

\bibitem %[log2]
  {log2} E.~Fradkin and J.~E.~Moore,
``Entanglement entropy of 2D conformal quantum critical points: hearing
the shape of a quantum drum,''
  Phys.\ Rev.\ Lett. {\bf 97}, 050404 (2006)
  [cond-mat/0605683].

\bibitem %[adscft]
  {adscft}
  J.~M.~Maldacena,
  ``The Large N limit of superconformal field theories and supergravity,''
  Adv.\ Theor.\ Math.\ Phys.\  {\bf 2}, 231-252 (1998).
  [hep-th/9711200].

\bibitem %[ent1]
  {ent1}
  L.~Y.~Hung, R.~C.~Myers and M.~Smolkin,
  ``On Holographic Entanglement Entropy and Higher Curvature Gravity,''
  JHEP {\bf 1104}, 025 (2011)
  [arXiv:1101.5813 [hep-th]].
  %%CITATION = JHEPA,1104,025;%%

\bibitem %[new]
  {new}
  I.~R.~Klebanov, T.~Nishioka, S.~S.~Pufu and B.~R.~Safdi,
  ``On Shape Dependence and RG Flow of Entanglement Entropy,''
  arXiv:1204.4160 [hep-th].
  %%CITATION = ARXIV:1204.4160;%%

\bibitem %[ben]
  {ben} 
  B.~R.~Safdi,
  ``Exact and Numerical Results on Entanglement Entropy in (5+1)-Dimensional CFT,''
  arXiv:1206.5025 [hep-th].
  %%CITATION = ARXIV:1206.5025;%%

\bibitem %[head1]
 {head1} M.~Headrick,
  ``Entanglement Renyi entropies in holographic theories,''
  Phys.\ Rev.\  D {\bf 82}, 126010 (2010)
  [arXiv:1006.0047 [hep-th]].
  %%CITATION = PHRVA,D82,126010;%%

\bibitem %[casini9]
 {casini9} H.~Casini, M.~Huerta and R.C.~Myers,
  ``Towards a derivation of holographic entanglement entropy,''
  JHEP {\bf 1105}, 036 (2011)
  [arXiv:1102.0440 [hep-th]].
  %%CITATION = JHEPA,1105,036;%%

\bibitem %[ent2]
  {ent2} J.~de Boer, M.~Kulaxizi and A.~Parnachev,
  ``Holographic Entanglement Entropy in Lovelock Gravities,''
  JHEP {\bf 1107}, 109 (2011)
  [arXiv:1101.5781 [hep-th]].
  %%CITATION = ARXIV:1101.5781;%%

\bibitem %[lovel]
 {lovel} D.~Lovelock, ``The Einstein tensor and its generalizations,''
  J.\ Math.\ Phys.\  {\bf 12}, 498 (1971);\\
  %%CITATION = JMAPA,12,498;%%
D.~Lovelock, ``Divergence-free tensorial concomitants,'' Aequationes
Math. {\bf 4}, 127 (1970).

\bibitem %[jacobson]
  {jacobson} T.~Jacobson and R.~C.~Myers,
  ``Black hole entropy and higher curvature interactions,''
  Phys.\ Rev.\ Lett.\  {\bf 70}, 3684 (1993)
  [hep-th/9305016].
  %%CITATION = HEP-TH/9305016;%%

\bibitem %[renyi]
  {renyi}
  L.-Y.~Hung, R.~C.~Myers, M.~Smolkin and A.~Yale,
  ``Holographic Calculations of Renyi Entropy,''
  JHEP {\bf 1112}, 047 (2011)
  [arXiv:1110.1084 [hep-th]].
  %%CITATION = ARXIV:1110.1084;%%

\bibitem %[EtasGB]
 {EtasGB} M.~Brigante, H.~Liu, R.~C.~Myers, S.~Shenker and
    S.~Yaida, ``Viscosity Bound Violation in Higher Derivative Gravity,''
  Phys.\ Rev.\ D {\bf 77} (2008) 126006
  [arXiv:htp-th/0712.0805];\\
  %%CITATION = PHRVA,D77,126006;%%
M.~Brigante, H.~Liu, R.~C.~Myers, S.~Shenker and S.~Yaida, ``The
Viscosity Bound and Causality Violation,''
  Phys.\ Rev.\ Lett.\  {\bf 100}, 191601 (2008)
  [arXiv:0802.3318 [hep-th]];\\
  %%CITATION = PRLTA,100,191601;%%
A.~Buchel and R.~C.~Myers,
  ``Causality of Holographic Hydrodynamics,''
  JHEP {\bf 0908}, 016 (2009)
  [arXiv:0906.2922 [hep-th]];\\
  %%CITATION = ARXIV:0906.2922;%%
D.~M.~Hofman, ``Higher Derivative Gravity, Causality and Positivity of
Energy in a UV complete QFT,''
  Nucl.\ Phys.\  B {\bf 823}, 174 (2009)
  [arXiv:0907.1625 [hep-th]];\\
  %%CITATION = NUPHA,B823,174;%%
X.~H.~Ge and S.~J.~Sin, ``Shear viscosity, instability and the upper
bound of the Gauss-Bonnet coupling constant,''
  JHEP {\bf 0905}, 051 (2009)
  [arXiv:0903.2527 [hep-th]];\\
  %%CITATION = JHEPA,0905,051;%%
R.~G.~Cai, Z.~Y.~Nie and Y.~W.~Sun,
  ``Shear Viscosity from Effective Couplings of Gravitons,''
  Phys.\ Rev.\  D {\bf 78}, 126007 (2008)
  [arXiv:0811.1665 [hep-th]];\\
  %%CITATION = PHRVA,D78,126007;%%
R.~G.~Cai, Z.~Y.~Nie, N.~Ohta and Y.~W.~Sun, ``Shear Viscosity from
Gauss-Bonnet Gravity with a Dilaton Coupling,''
  Phys.\ Rev.\  D {\bf 79}, 066004 (2009)
  [arXiv:0901.1421 [hep-th]];\\
  %%CITATION = PHRVA,D79,066004;%%
J.~de Boer, M.~Kulaxizi and A.~Parnachev,
  ``$AdS_7/CFT_6$, Gauss-Bonnet Gravity, and Viscosity Bound,''
  JHEP {\bf 1003}, 087 (2010)
  [arXiv:0910.5347 [hep-th]];\\
  %%CITATION = JHEPA,1003,087;%%
X.~O.~Camanho and J.~D.~Edelstein, ``Causality constraints in AdS/CFT
from conformal collider physics and Gauss-Bonnet gravity,''
  arXiv:0911.3160 [hep-th].\\
  %%CITATION = ARXIV:0911.3160;%%
M.~Fujita,
  ``Holographic Entanglement Entropy for d=4 N=2 SCFTs in F-theory,''
  arXiv:1112.5535 [hep-th].
  %%CITATION = ARXIV:1112.5535;%%

\bibitem %[highc]
 {highc} S.~Nojiri and S.D.~Odintsov,
``On the conformal anomaly from higher derivative gravity in AdS/CFT
correspondence,''
  Int.\ J.\ Mod.\ Phys.\  A {\bf 15}, 413 (2000)
  [arXiv:hep-th/9903033];\\
  %%CITATION = IMPAE,A15,413;%%
M.~Blau, K.S.~Narain and E.~Gava, ``On subleading contributions to the
AdS/CFT trace anomaly,''
  JHEP {\bf 9909}, 018 (1999)
  [arXiv:hep-th/9904179].
  %%CITATION = JHEPA,9909,018;%%

\bibitem %[cthem1]
 {cthem1} R.~C.~Myers and A.~Sinha,
  ``Holographic c-theorems in arbitrary dimensions,''
  JHEP {\bf 1101}, 125 (2011)
  [arXiv:1011.5819 [hep-th]].
  %%CITATION = JHEPA,1101,125;%%

\bibitem %[cthem2]
 {cthem2} R.~C.~Myers and A.~Sinha,
  ``Seeing a c-theorem with holography,''
  Phys.\ Rev.\  D {\bf 82}, 046006 (2010)
  [arXiv:1006.1263 [hep-th]].
  %%CITATION = PHRVA,D82,046006;%%

\bibitem %[finn]
 {finn} C.~Holzhey, F.~Larsen and F.~Wilczek,
  ``Geometric and renormalized entropy in conformal field theory,''
  Nucl.\ Phys.\  B {\bf 424}, 443 (1994)
  [arXiv:hep-th/9403108].
  %%CITATION = NUPHA,B424,443;%%

\bibitem %[fefferman]
  {fefferman} C.~Fefferman and C.~R.~Graham, ``Conformal Invariants,''
in Elie Cartan et les
Math\'ematiques d'aujourd hui (Ast\'erisque, 1985) 95;\\
C.~Fefferman and C.~R.~Graham, ``The Ambient Metric,'' arXiv:0710.0919
[math.DG].

\bibitem %[sthem]
  {sthem}
  R.~C.~Myers and A.~Singh,
  ``Comments on Holographic Entanglement Entropy and RG Flows,''
  arXiv:1202.2068 [hep-th].
  %%CITATION = ARXIV:1202.2068;%%

\bibitem %[casini6]
 {casini6} H.~Casini and M.~Huerta, ``On the RG running
 of the entanglement entropy of a circle,''
  arXiv:1202.5650 [hep-th].
  %%CITATION = ARXIV:1202.5650;%%

\bibitem %[anselmi2]
 {anselmi2} D.~Anselmi, J.~Erlich, D.Z.~Freedman and A.A.~Johansen,
  ``Positivity constraints on anomalies in supersymmetric gauge theories,''
  Phys.\ Rev.\  D {\bf 57}, 7570 (1998)
  [arXiv:hep-th/9711035].
  %%CITATION = PHRVA,D57,7570;%%

\bibitem %[next]
 {next} R.~C.~Myers and M.~Smolkin, ``Some Relevant
Contributions to Entanglement Entropy,'' in perpetual preparation.

\bibitem %[brian]
 {brian} B.~Swingle,``Mutual information and the structure of
entanglement in quantum field theory,'' arXiv:1010.4038 [quant-ph].
  %%CITATION = ARXIV:1010.4038;%%

\bibitem %[cusp]
 {cusp} D.~Correa, J.~Maldacena and A.~Sever, ``The quark anti-quark potential
and the cusp anomalous dimension from a TBA equation,''
  arXiv:1203.1913 [hep-th].
  %%CITATION = ARXIV:1203.1913;%%

\bibitem %[cusp2]
 {cusp2} N.~Drukker, D.~J.~Gross and H.~Ooguri,
  ``Wilson loops and minimal surfaces,''
  Phys.\ Rev.\ D {\bf 60}, 125006 (1999)
  [hep-th/9904191].
  %%CITATION = HEP-TH/9904191;%%

\bibitem %[cusp3]
 {cusp3} D.~E.~Berenstein, R.~Corrado, W.~Fischler and J.~M.~Maldacena,
``The Operator product expansion for Wilson loops and surfaces in the
large N limit,''
  Phys.\ Rev.\ D {\bf 59}, 105023 (1999)
  [hep-th/9809188].
  %%CITATION = HEP-TH/9809188;%%



\end{thebibliography}
\end{document}